\documentclass[aps,prx,preprint,onecolumn,citeautoscript,superscriptaddress,footinbib,
eqsecnum]{revtex4-1}  
\usepackage{amsmath,amssymb,bm,bbm} 
\usepackage{graphicx}  
\usepackage{color} 
\usepackage[dvipsnames]{xcolor}
\usepackage[papersize={8.5in,11in}]{geometry}
\usepackage[colorlinks=true]{hyperref}  
\usepackage{ulem}
\usepackage[mathscr]{euscript}
\usepackage[section]{placeins}
\hypersetup{
    bookmarks=true,         
    unicode=false,          
    pdftoolbar=true,        
    pdfmenubar=true,        
    pdffitwindow=false,     
    pdfstartview={FitH},    
    pdfsubject={},   
    pdfcreator={},   
    pdfproducer={}, 
    pdfkeywords={} {} {}, 
    pdfnewwindow=true,      
    colorlinks=true,       
    linkcolor=magenta, 
    citecolor=blue,        
    filecolor=magenta,      
    urlcolor=blue           
} 

\usepackage{amsfonts}
\usepackage{upgreek}
\usepackage{slashed}
\usepackage{latexsym}

\newcommand{\beq}{\begin{equation}}
\newcommand{\eeq}{\end{equation}}
\def\bea{\begin{eqnarray}}
\def\eea{\end{eqnarray}}
\newcommand{\nn}{\nonumber \\}

\newcommand{\equref}[1]{Eq.~(\ref{#1})}
\newcommand{\equsref}[2]{Eqs.~(\ref{#1}) and (\ref{#2})}
\newcommand{\secref}[1]{Sec.~\ref{#1}}
\newcommand{\figref}[1]{Fig.~\ref{#1}}
\newcommand{\refcite}[1]{Ref.~\onlinecite{#1}}

\newcommand{\tableref}[1]{Table~\ref{#1}}

\newcommand{\pdagger}{{\phantom{\dagger}}}
\newcommand{\diff}{\mathrm{d}}

\usepackage{braket}
\usepackage{comment}
\usepackage{slashed}
\usepackage{subcaption}

\newcommand{\bvec}[1]{\boldsymbol{#1}}
\renewcommand{\vec}[1]{\boldsymbol{#1}}

\begin{document}

\preprint{\href{https://arxiv.org/abs/1811.04930}{arXiv:1811.04930}}

\title{Gauge theory for the cuprates near optimal doping}

\author{Subir Sachdev}
\affiliation{Department of Physics, Harvard University, Cambridge MA 02138, USA}
\affiliation{Perimeter Institute for Theoretical Physics, Waterloo, Ontario, Canada N2L 2Y5}

\author{Harley D. Scammell}
\affiliation{Department of Physics, Harvard University, Cambridge MA 02138, USA}

\author{Mathias S. Scheurer}
\affiliation{Department of Physics, Harvard University, Cambridge MA 02138, USA}

\author{Grigory Tarnopolsky}
\affiliation{Department of Physics, Harvard University, Cambridge MA 02138, USA}

\date{\today
\\
\vspace{0.4in}}

\begin{abstract}
We describe the phase diagram of a 2+1 dimensional SU(2) gauge theory of fluctuating incommensurate spin density waves for the hole-doped cuprates. Our primary assumption is that all low energy fermionic excitations are gauge neutral and electron-like, while the spin density wave order is fractionalized into Higgs fields transforming as adjoints of the gauge SU(2).
The confining phase of the gauge theory is a conventional Fermi liquid with a large Fermi surface (and its associated $d$-wave superconductor). There is a quantum phase transition to a Higgs phase describing the `pseudogap' at lower doping. Depending upon the quartic terms in the Higgs potential, the Higgs phase exhibits one or more of charge density wave, Ising-nematic, time-reversal odd scalar spin chirality, and $\mathbb{Z}_2$ topological orders. It is notable that the emergent broken symmetries in our theory of fluctuating spin density waves co-incide with those observed in diverse experiments. 
For the electron-doped cuprates, the spin density wave fluctuations are at wavevector $(\pi,\pi)$, and then the corresponding SU(2) gauge theory only has a crossover between the confining and Higgs regimes, with an exponentially large confinement scale deep in the Higgs regime. On the Higgs side, for both the electron- and hole-doped cases, and at scales shorter than the confinement scale (which can be infinite when $\mathbb{Z}_2$ topological order is present), the electron spectral function has a `fractionalized Fermi liquid (FL*)' form with small Fermi surfaces. We also describe the deconfined quantum criticality of the Higgs transition in the limit of a large number of Higgs flavors, and perturbatively discuss its coupling to fermionic excitations.
\end{abstract}

\maketitle
\tableofcontents

\section{Introduction}
\label{sec:intro}

Decades of intense experimental study of the cuprate superconductors have demonstrated that there is a fundamental transformation in the nature of the electronic state near optimal doping \cite{CPLT18}. Among the many experimental studies \cite{CPLT18}, we highlight a few of relevance to us in the hole-doped cuprates: ({\it i\/}) scanning tunnelling microscopy (STM) observations indicate nematic and charge density orders which vanish near optimal doping \cite{Fujita14}; ({\it ii\/}) STM observations also indicate that the electron spectral functions transforms from `Fermi arcs' to a full Fermi surface across optimal doping \cite{He14,Fujita14}; ({\it iii\/}) the specific heat exhibits a well-defined peak at optimal doping \cite{Michon18}; ({\it iv\/}) there are indications of time-reversal symmetry breaking which disappear across optimal doping \cite{Xia08}.

This paper presents a 2+1 dimensional SU(2) gauge theory of fluctuating incommensurate spin density wave (SDW) fluctuations which can unify these and other disparate observations. We note that there is considerable numerical evidence that antiferromagnetic spin fluctuations are responsible for the `pseudogap' in the underdoped state at intermediate 
temperatures \cite{Tremblay06,Jarrell06,Ferrero09,Tremblay12,Gunn15,Wu17}.
Our central idea is to fractionalize the SDW order parameter (see Eq.~(\ref{e1}) below), 
while keeping the low energy fermionic excitations as non-fractionalized electrons (we note evidence for electron-like bound states in recent ultracold atom studies of the doped Hubbard model \cite{Greiner18a,Bloch18,Greiner18b}). Similar ideas have been examined in a number of earlier studies \cite{Zaanen01,Demler02,Morinari02,Zaanen02A,Zaanen02B,Mross12,Mross12PRB}. In our case, we will fractionalize the SDW order by transforming to a rotating reference frame in spin space \cite{Shraiman88,Schulz90,Schrieffer2004,SS09,DCSS15b,CSS17,SCWFGS,WSCSGF,SS18}. 

\begin{figure}[tb]
\begin{center}
\includegraphics[height=3.75in]{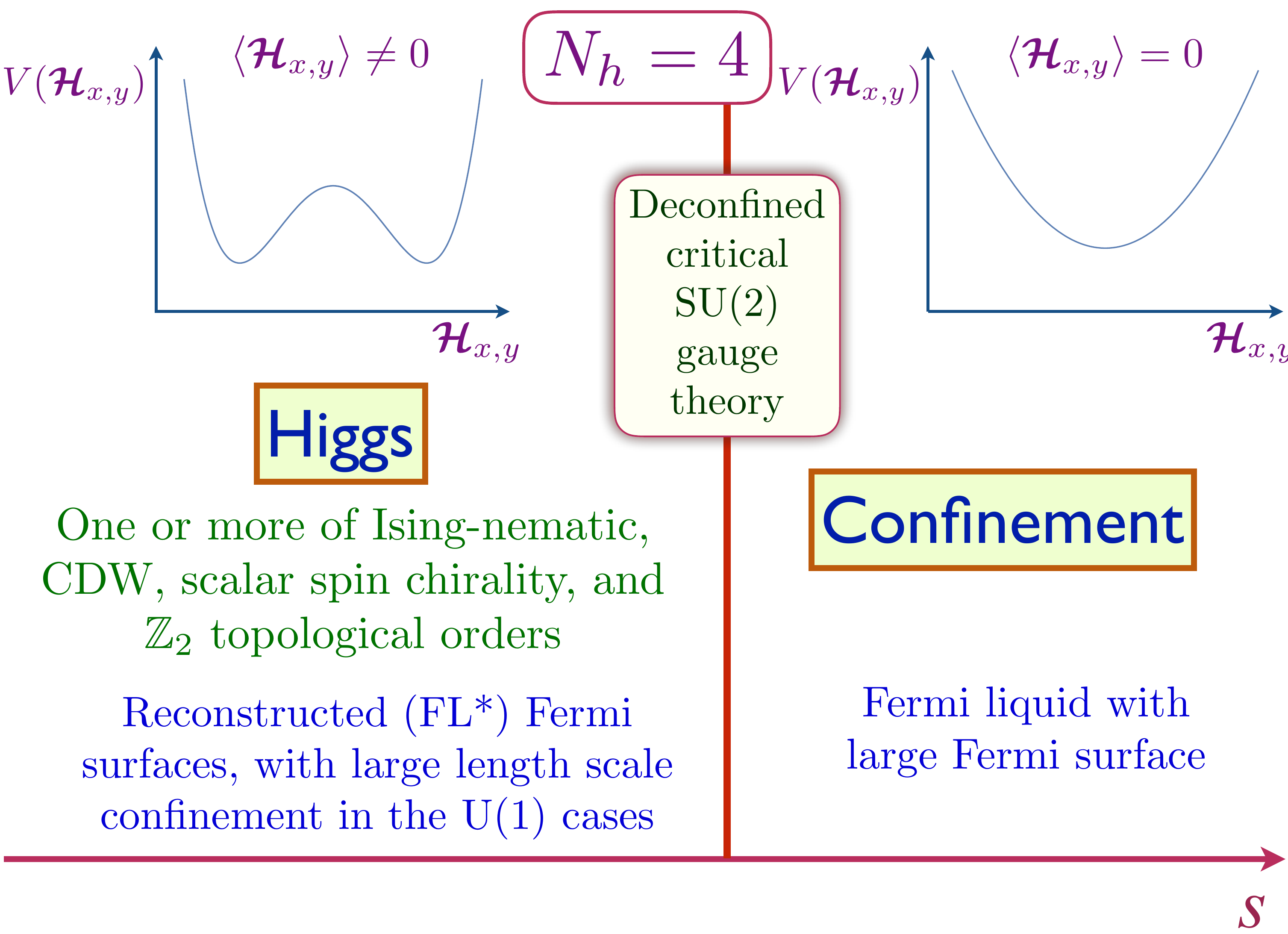} 
\end{center}
\caption{Schematic phase diagram of SU(2) gauge theory with adjoint Higgs fields for the hole-doped cuprates. The number of real adjoint Higgs fields is denoted by $N_h$. The ``Higgs'' region maps onto to the underdoped pseudogap regime, while the ``Confinement'' region maps onto the overdoped side. Details of the phase diagram within the Higgs region appear in Fig.~\ref{MFPhasediagram}. The deconfined critical theory is described in Section~\ref{sec:dcp} at large $N_h$.}
\label{fig:phasediag_h}
\end{figure}

Readers have likely heard of the example of quantum chromodynamics (QCD), a non-Abelian gauge theory which is confining {\it i.e.\/} all physical excitations are gauge neutral. Correspondingly, our SU(2) gauge theory also has a confining phase: this is just the ordinary Fermi liquid, and maps to the overdoped cuprates; see Fig.~\ref{fig:phasediag_h}.
However, unlike QCD and the Weinberg-Salam theory of weak interactions, it is also possible for non-Abelian gauge theories to display `Higgs' phases in which certain discrete gauge charges can be deconfined. In the Weinberg-Salam theory, condensation of the Higgs scalar field breaks the SU(2) gauge symmetry completely, and we obtain a low energy theory of massive fermions interacting weakly via exchange of massive gauge bosons: there is no `fractionalization' here. However, more subtle types of Higgs phases are also possible in which a discrete subgroup of the non-Abelian gauge group remains unbroken \cite{FradkinShenker,NRSS91,Wen91,Bais92}: then the matter fields can carry discrete gauge charges and be fractionalized. See Ref.~\cite{SS18} for a recent review of this phenomenon in a condensed matter context. We shall exploit this phenomenon here, and map a Higgs phase to the underdoped cuprates. An ancillary phenomenon for certain configurations of Higgs fields is that global symmetries can be broken in the Higgs phase: this will also be useful to us in the mapping to the underdoped cuprates. Finally, there can also be Higgs-confining quantum phase transitions, or just crossovers, and we will map such phenomena to optimal doping criticality.

There have been earlier attempts to describe optimal doping using Higgs criticality in a SU(2) gauge theory \cite{DCSS15b}. However these involved active fractionalized excitations on the Fermi surface, in contrast to a key feature of the present approach already highlighted above: all Fermi surface excitations  are electron-like. Theories with low energy fractionalized excitations have a number of subtle features, including non-trivial Berry phases associated with topological gauge excitations. We argue here that such effects can be sidestepped in the present approach, allowing significant progress into the structure of both the Higgs and confined phases, and of the phase transition between them.

We will find that the Higgs phase in our approach has a rich emergent structure, with numerous possibilities for broken symmetries, which can also co-exist with the $\mathbb{Z}_2$ topological order of the `toric code' \cite{NRSS91,Wen91,Bais92,Kitaev03}. Remarkably, the broken symmetries which emerge from the Higgs phase are essentially those observed in the hole-doped cuprates.
(An exception is the pair density wave \cite{SeamusPDW}; this can be treated in a parallel manner in a different SU(2) gauge theory obtained by transforming to a rotating reference frame in pseudospin space \cite{CXSS10,OurLoopCurrentsPaper,SS18}, and will be described in a forthcoming paper.) A phenomenologically attractive candidate is the novel state we denote $(F)_\theta$: this has uni-directional charge density wave (CDW) order co-existing with time-reversal breaking, modulated scalar spin chirality (the scalar spin chirality is symmetry equivalent to spontaneous orbital currents \cite{OurLoopCurrentsPaper}). This combination of orders in the $(F)_\theta$ state is induced by short-range, collinear, incommensurate spin correlations along one spatial direction, and short-range, spiral, incommensurate spin correlations along the other spatial direction.

We will also consider the case of commensurate N\'eel SDW fluctuations, which is expected to apply to the electron-doped cuprates. Here, the SU(2) gauge theory does not exhibit a Higgs-confining phase transition, only a crossover between physically rather different regimes. In the Higgs regime, the confinement length scale can become exponentially large, and so fractionalized behavior can appear at intermediate scales.

We transform to a spacetime-dependent rotating reference frame by writing the electron spin magnetic moment $\bvec{S}_i$ as 
\beq
\bvec{\sigma}\cdot \bvec{S}_i = R_i \bvec{\sigma} R_i^\dagger \cdot \bvec{H}_i \label{e1}
\eeq
where $\bvec{\sigma}$ are the Pauli matrices, $R_i$ is a spacetime dependent
SU(2) rotation matrix, and $\bvec{H}_i$ is the spin magnetic moment in the rotated reference frame; Eq.~(\ref{e1}) is the key expression fractionalizing spin excitations into the emergent fields $R$ and $H$.  (Note that we are taking the path integral point of view here, and all fields in Eq.~(\ref{e1}) are to be viewed as insertions in such a path integral, and not as canonical operators.) The physical idea is that the primary fluctuations in the SDW order are orientational, and the $R_i$ rotation transforms to a frame of reference in which certain momentum space components of $\bvec{H}_i$ vary slowly.
The transformation in Eq.~(\ref{e1}) introduces a SU(2) gauge 
invariance \cite{SS09} under which
\beq
R_i \rightarrow R_i V_i^\dagger \quad, \quad \bvec{\sigma} \cdot \bvec{H}_i \rightarrow V_i \bvec{\sigma} V_i^\dagger \cdot \bvec{H}_i
\eeq
where $V_i$ is a spacetime-dependent SU(2) matrix generating the gauge transformation. So $\bvec{H}$ transforms as an adjoint of SU(2), and it is our Higgs field (see Maldacena's perspective \cite{Maldacena14} on Higgs fields resulting from transformations to rotated reference frames). 
To capture the spatial form of the spin density wave, we parameterize the Higgs field as 
\begin{equation}
\bvec{H}_i = \text{Re}\left[ \bvec{\mathcal{H}}_x e^{i\vec{K}_x\cdot\vec{r}_i} + \bvec{\mathcal{H}}_y e^{i\vec{K}_y\cdot\vec{r}_i} \right], \label{Parametrization}
\end{equation}
where $\vec{K}_x=\pi(1-\delta,1)$, $\vec{K}_y=\pi(1,1-\delta)$ with irrational $\delta$ are the incommensurate wavevectors (throughout, we set the lattice constant to unity). In some of the cuprates, it may be that CDW/SDW fluctuations are commensurate with $\delta = 1/4$: this will modify the effective action by higher order terms which are expected to be irrelevant, and essentially all of our results will continue to apply. Our theory will be formulated assuming slow spacetime variations of the complex Higgs fields $\bvec{\mathcal{H}}_{x,y}$. So we will consider a SU(2) gauge theory with $N_h = 4$ real, adjoint Higgs scalars (the number of real adjoint Higgs scalars will be denoted by $N_h$).

From Eq.~(\ref{e1}), we note that the wavevectors of the dominant spin fluctuations can be shifted from $\vec{K}_{x,y}$ by the dominant wavevectors of $R$ fluctuations. The latter can be at zero, but this need not necessarily be the case \cite{CSS17,SCWFGS}. Formally, we define $\vec{K}_{x,y}$ so that $2 \vec{K}_{x,y}$ are the wavevectors of the CDW orders defined below in Eq.~(\ref{ScalarHiggsChriality}).

We will assume that the $R_i$ are gapped across the transition, and (as stated above) that the only low energy fermionic excitations are non-fractionalized electrons $c_i$. The $c_i$ are invariant under SU(2) gauge transformations, and this strongly constrains the nature of their coupling to the $\bvec{\mathcal{H}}_{x,y}$ Higgs fields. As we will describe in Section~\ref{sec:CouplingToElectrons}, the simplest allowed  coupling to the electrons is quartic, and a more disruptive cubic Yukawa coupling is not permitted \cite{CS93,SS94,Morinari02,TGTS10,Mross12,Mross12PRB}. 
We will examine the influence of the electronic excitations perturbatively.

For the electron-doped cuprates, we follow a similar approach, and replace Eq.~(\ref{Parametrization}) by
\begin{equation}
\bvec{H}_i =  \bvec{\mathcal{H}} e^{i\vec{K}\cdot\vec{r}_i} \label{Parametrization2}
\end{equation}
where $\vec{K}=(\pi,\pi)$ and $\bvec{\mathcal{H}}$ is real, and so there is only $N_h = 1$ real adjoint Higgs field. The Higgs sector of this theory is called the Georgi-Glashow model \cite{GG72,tH74,Polyakov74} in the particle physics literature, and we are interested in this model in 2+1 spacetime dimensions \cite{Kibble00,Dunne00,Kibble02}.

We will present our theory for optimally doped cuprates in Section~\ref{sec:su2}. We describe the phase diagram of the Higgs side, where the SU(2) gauge invariance is broken down either to $\mathbb{Z}_2$ or U(1).
The $\mathbb{Z}_2$ topological order is stable, but the U(1) topological order is ultimately unstable due to the proliferation of monopoles. But the monopole density becomes exponentially small as we move deeper into the Higgs side, as we review in Appendix~\ref{app:monopole}.
In the $N_h=4$ theory we find that the Higgs phase always has gauge-invariant bilinears of $\bvec{\mathcal{H}}_{x,y}$ which break a global symmetry. 

The pattern of symmetry breaking also distinguishes different phases on the Higgs side, and these considerations lead to the phase diagrams in Figs.~\ref{fig:phasediag_h}, \ref{MFPhasediagram}, and \ref{fig:phasediag_e}. The symmetry breaking on the hole-doped side 
is characterized by 6 different order parameters (symmetry 
properties are defined in Table~\ref{RepresentationOfSymmetries}):
the Ising nematic order, $\phi$, the CDW order parameters $\Phi_{x,y}$ at wavevectors $2 \vec{K}_{x,y}$, the CDW order parameters $\Phi_{\pm}$ at wavevectors $\vec{K}_x \pm \vec{K}_y$, and the scalar spin chiralities $\bvec{\chi}$ (see Eq.~(\ref{defchi}); $\bvec{\chi}$ also measures the presence of spontaneous orbital currents \cite{OurLoopCurrentsPaper}). These order parameters can be expressed gauge-invariant non-linear functions of the Higgs fields:
\bea
\phi=|\bvec{\mathcal{H}}_x|^2-|\bvec{\mathcal{H}}_y|^2 \,, \quad
&&\Phi_x = \bvec{\mathcal{H}}_x \cdot \bvec{\mathcal{H}}_x\,, \quad
\Phi_y = \bvec{\mathcal{H}}_y \cdot \bvec{\mathcal{H}}_y\,, \quad
\Phi_+ = \bvec{\mathcal{H}}_x \cdot \bvec{\mathcal{H}}_y\,, \quad
\Phi_- =   \bvec{\mathcal{H}}_x \cdot \bvec{\mathcal{H}}_y^*\nn
&&  \chi_{ijk} = \bvec{H}_i \cdot (\bvec{H}_j \times \bvec{H}_k) \label{ScalarHiggsChriality}
\eea
So the Higgs fields can be viewed as the democratic fractionalizations of all these order parameters. Moreover, as noted in Eq.~(\ref{e1}), upon including the $R_i$, the Higgs fields also fractionalize the SDW order parameter. We note that this unification of the order parameters via fractionalization is quite distinct from unifications in which order parameters with different symmetries are put in a common multiplet of a larger group \cite{Pepin18}.

The connection between the Higgs fields and the CDW/Ising-nematic order parameters in Eq.~(\ref{ScalarHiggsChriality}) parallels
that between the SDW order and the CDW/Ising-nematic order parameters in Landau theories of `stripe' orders \cite{Zachar98,KFE98}. In these Landau theories, appearance of the SDW broken symmetry induces the secondary CDW/Ising-nematic broken symmetries. In our case, we have `gauged' the SDW order, and so there is no SDW broken symmetry in the Higgs phase; nevertheless, the gauge-invariant CDW/Ising-nematic orders do appear upon condensation of the Higgs field.

Section~\ref{sec:flstar} will turn to a consideration of the electron spectral function within the Higgs phase. Unlike previous approaches with fractionalized order parameters \cite{Zaanen01,Demler02,Morinari02,Zaanen02A,Zaanen02B,Mross12,Mross12PRB}, our SU(2) gauge theory does allow for reconstructed 
Fermi surfaces in a `fractionalized Fermi liquid (FL*)' regime. The nature of such a FL* state has been discussed in previous works \cite{SCWFGS,WSCSGF}, and here we will include additional computations which clarify the relationship to the theory presented in Section~\ref{sec:su2}.

Section~\ref{sec:dcp} will examine the quantum critical point describing the onset of the Higgs phase for $N_h > 1$. The critical theory is generally strongly coupled. In the expansion in $\epsilon=4-D$, where $D$ is spacetime dimensionality, a suitable fixed point does not exist for small values of $N_h$, and there is runaway flow similar to that in the well-known U(1) gauge theory \cite{HLM}. However, a controlled fixed point can be found in the limit of large $N_h$, while ignoring the coupling to the fermionic excitations. We will present a description of the large $N_h$ deconfined quantum critical point in a model with O($N_h$) global flavor symmetry. The scaling dimensions of operators at this quantum critical point allows us to estimate the importance of the electronic Fermi surface excitations in a metal, or the nodal excitations in a $d$-wave superconductor: this analysis appears in Section~\ref{sec:CouplingToElectrons}.

Section~\ref{sec:impvis} will turn to saddle points of the action which involve textures around a single spatial point. Section~\ref{sec:imp} will consider point-like impurities, which can induce textures of sub-dominant order parameters in the Higgs phase.
Section~\ref{sec:vison} will describe the structure of $\mathbb{Z}_2$ vortex (`visons' \cite{SenthilFisher}) solutions in the Higgs regimes with $\mathbb{Z}_2$ topological order. An interesting feature of these visons is that they are accompanied by non-trivial textures in the order parameters of the broken symmetries. 
It is important to note that these visons, involving gauge rotations in spin space, are distinct from visons studied earlier \cite{SS92,TSMPAF01} and searched for experimentally \cite{Moler01,Moler02}; the latter  involve gauge rotations in Nambu pseudospin space \cite{OurLoopCurrentsPaper}. 

\section{SU(2) gauge theory}
\label{sec:su2}
Our primary theory will be an effective action for $\bvec{\mathcal{H}}_{x,y}$ on the hole-doped cuprates, and for $\bvec{\mathcal{H}}$ on the electron-doped cuprates. We will begin with the discussion for the hole-doped cuprates,
and defer the electron-doped case to Section~\ref{sec:electron}.

The SU(2) gauge invariance demands that we introduce a real SU(2) gauge field $\bvec{A}_\mu$, where $\mu$ is a spacetime index. Then we consider the Lagrangian
\begin{subequations}
\beq
\mathcal{L}_{\mathcal{H}} = \frac{1}{4 g^2} \bvec{F}_{\mu\nu} \cdot
 \bvec{F}_{\mu\nu} + \left| \partial_\mu \bvec{\mathcal{H}}_x - \bvec{A}_\mu \times \bvec{\mathcal{H}}_x \right|^2 + 
\left| \partial_\mu \bvec{\mathcal{H}}_y - \bvec{A}_\mu \times \bvec{\mathcal{H}}_y \right|^2 + V(\bvec{\mathcal{H}}_{x,y}) \label{LH1}
\eeq
where the SU(2) field strength is
\beq
\bvec{F}_{\mu\nu} = \partial_\mu \bvec{A}_\nu - \partial_{\nu} \bvec{A}_{\mu} - \bvec{A}_\mu \times \bvec{A}_{\nu}.
\eeq


\newcommand{\extrasp}{\hspace{0.7em} \,}
\begin{table}[bt]
\begin{center}
\caption{Representation of the square-lattice symmetries and time-reversal ($\Theta$) on the complex Higgs fields $\bvec{\mathcal{H}}_{x,y}$ introduced in \equref{Parametrization} and on four different (gauge-invariant) bi-linear order parameters. Here $T_{\bvec{\eta}}$ denotes translation by square-lattice vector $\bvec{\eta}$, $C_{4}$ four-fold rotation along the $z$ axis, and $\sigma_x$ reflection at the $xz$ plane. Although these symmetry operations generate the full symmetry group of the system, we have also added two-fold rotation $C_2$ and the diagonal reflection $\sigma_d$ (with $i_x \leftrightarrow i_y$) for convenience of the reader.}
\label{RepresentationOfSymmetries}
 \begin{tabular}{ccccccc} \hline \hline
 Fields  & $\Theta$ & $T_{\bvec{\eta}}$  &   $C_4$ & $\sigma_x$ & $C_2$ & $\sigma_d$    \\ \hline
$\bvec{\mathcal{H}}_x$ & \extrasp$-\bvec{\mathcal{H}}_x$\extrasp & $e^{i\bvec{K}_x\cdot{\bvec{\eta}}}\bvec{\mathcal{H}}_x$ & \extrasp$\bvec{\mathcal{H}}_y$\extrasp & \extrasp$\bvec{\mathcal{H}}_x$ \extrasp & \extrasp $\bvec{\mathcal{H}}_x^*$\extrasp & \extrasp $\bvec{\mathcal{H}}_y$ \extrasp \\
$\bvec{\mathcal{H}}_y$ & $-\bvec{\mathcal{H}}_y$ & $e^{i\bvec{K}_y\cdot{\bvec{\eta}}}\bvec{\mathcal{H}}_y$ & $\bvec{\mathcal{H}}_x^*$ & $\bvec{\mathcal{H}}_y^*$ & $\bvec{\mathcal{H}}_y^*$ & $\bvec{\mathcal{H}}_x$  \\ 
\extrasp$ \phi=|\bvec{\mathcal{H}}_x|^2-|\bvec{\mathcal{H}}_y|^2$\extrasp  & $\phi$ & $\phi$ & $-\phi$ & $\phi$ & $\phi$ & $-\phi$ \\ 
$\Phi_x = \bvec{\mathcal{H}}_x \cdot \bvec{\mathcal{H}}_x$ & $\Phi_x$ & $e^{2i\bvec{K}_x\cdot{\bvec{\eta}}}\Phi_x$ & $\Phi_y$ & $\Phi_x$ & $\Phi_x^*$ & $\Phi_y$ \\
$\Phi_y = \bvec{\mathcal{H}}_y \cdot \bvec{\mathcal{H}}_y$ & $\Phi_y$ & $e^{2i\bvec{K}_y\cdot{\bvec{\eta}}}\Phi_y$ & $\Phi_x^*$ & $\Phi_y^*$ & $\Phi_y^*$ & $\Phi_x$ \\ 
$\Phi_+ = \bvec{\mathcal{H}}_x \cdot \bvec{\mathcal{H}}_y$ & $\Phi_+$ & $e^{i(\bvec{K}_x+\bvec{K}_y)\cdot{\bvec{\eta}}}\Phi_+$ & $\Phi_-^*$ & $\Phi_-$ & $\Phi_+^*$ & $\Phi_+$ \\
$\Phi_- =   \bvec{\mathcal{H}}_x \cdot \bvec{\mathcal{H}}_y^*$ & $\Phi_-$ & \extrasp$e^{i(\bvec{K}_x-\bvec{K}_y)\cdot{\bvec{\eta}}}\Phi_-$\extrasp & $\Phi_+$ & $\Phi_+$ & $\Phi_-^*$ & $\Phi_-^*$ \\ \hline \hline
 \end{tabular}
\end{center}
\end{table}

The form of the effective Higgs potential $V(\bvec{\mathcal{H}}_{x,y})$ is constrained by the symmetries of the system. The action of time-reversal and the square-lattice symmetries, translation $T_{\vec{\eta}}$ by lattice vector $\vec{\eta} = (n,m)$, $n,m\in\mathbb{Z}$, and the elements of the point group $C_{4v}$, follow directly from the parametrization in \equref{Parametrization} and are summarized in \tableref{RepresentationOfSymmetries}. Note that, by virtue of $\vec{K}_{x}$ being incommensurate, the phases $\{e^{i\vec{K}_{x}\cdot \vec{\eta}}|\vec{\eta} = (n,0), \,n \in \mathbb{Z}\}$ densely cover $\text{U}(1)$ and the same holds for $\vec{K}_y$. Consequently, translational invariance effectively becomes a $\text{U}(1)\times \text{U}(1)$ symmetry with action $(\bvec{\mathcal{H}}_{x},\bvec{\mathcal{H}}_{y}) \longrightarrow (e^{i\varphi_1}\bvec{\mathcal{H}}_{x},e^{i\varphi_2}\bvec{\mathcal{H}}_{y})$, $0\leq \varphi_{1,2} \leq 2\pi$. Gauge and translational invariance together with the additional discrete symmetries in \tableref{RepresentationOfSymmetries} strongly constrain the possible terms in the Higgs potential; up to quartic order in $\bvec{\mathcal{H}}_{x,y}$, there are only five independent terms that can be written as \cite{Demler02,Vicari06,Vicari08}
\bea
&& V(\bvec{\mathcal{H}}_{x,y}) = s\left( \bvec{\mathcal{H}}^*_x \cdot \bvec{\mathcal{H}}^{\phantom{*}}_x + \bvec{\mathcal{H}}^*_y \cdot \bvec{\mathcal{H}}_y^{\phantom{*}} \right) + u_0\left( \bvec{\mathcal{H}}^*_x \cdot \bvec{\mathcal{H}}^{\phantom{*}}_x + \bvec{\mathcal{H}}^*_y \cdot \bvec{\mathcal{H}}^{\phantom{*}}_y \right)^2 \nn
&& +\frac{u_1}{4} (\bvec{\mathcal{H}}^*_x \cdot \bvec{\mathcal{H}}^{\phantom{*}}_x - \bvec{\mathcal{H}}^*_y \cdot \bvec{\mathcal{H}}^{\phantom{*}}_y)^2 
+ \frac{u_2}{2}\left( | \bvec{\mathcal{H}}_x \cdot \bvec{\mathcal{H}}_x |^2 + | \bvec{\mathcal{H}}_y \cdot \bvec{\mathcal{H}}_y |^2 \right) +  u_3 \left(|\bvec{\mathcal{H}}_x \cdot \bvec{\mathcal{H}}_y|^2 + |\bvec{\mathcal{H}}^{\phantom{*}}_x \cdot \bvec{\mathcal{H}}^*_y|^2 \right) \nn
&& \qquad \qquad = s\left( \bvec{\mathcal{H}}^*_x \cdot \bvec{\mathcal{H}}^{\phantom{*}}_x + \bvec{\mathcal{H}}^*_y \cdot \bvec{\mathcal{H}}_y^{\phantom{*}} \right) + u_0\left( \bvec{\mathcal{H}}^*_x \cdot \bvec{\mathcal{H}}^{\phantom{*}}_x + \bvec{\mathcal{H}}^*_y \cdot \bvec{\mathcal{H}}^{\phantom{*}}_y \right)^2 \nn
&&\qquad \qquad \qquad + \frac{u_1}{4} \phi^2
  + \frac{u_2}{2}\left( | \Phi_x |^2 + | \Phi_y |^2 \right) + u_3 \left( |\Phi_+|^2 + |\Phi_-|^2 \right)
\label{HiggsPotential}
\eea \label{FullHiggsAction}\end{subequations}
using the bi-linear combinations $\phi$, $\Phi_{x,y}$, and $\Phi_{\pm}$ defined in Eq.~(\ref{ScalarHiggsChriality}) and \tableref{RepresentationOfSymmetries}. The gauge-invariant bi-linears $\phi$, $\Phi_{x,y}$, and $\Phi_{\pm}$ are not only useful to parametrize the Higgs potential but will also serve as order parameters for nematic order, charge density wave (CDW) order with wavevector $2\bvec{K}_{x,y}$, and with wavevector $\bvec{K}_{x}\pm \bvec{K}_{y}$, respectively.

We note that the potential in Eq.~(\ref{HiggsPotential}) is the same as that appearing in previous works \cite{Demler02,Vicari06} on incommensurate SDW order. Here we have `gauged' the SDW order parameters to the Higgs fields $\bvec{\mathcal{H}}_{x,y}$, whose condensation therefore preserves spin rotation symmetry, and the gradient terms in Eq.~(\ref{LH1}) are covariant gradients with a SU(2) gauge field. The previous works \cite{Demler02,Vicari06}
used a more general potential with separate couplings for the two terms associated with $u_3$: these separate couplings are not needed because they are required to be equal by the $\sigma_x$ symmetry.

For the case of commensurate SDW with $\delta = 1/4$, symmetry allows new terms in $V(\bvec{\mathcal{H}}_{x,y})$ which are eighth order in $\bvec{\mathcal{H}}_{x,y}$, {\it e.g.\/} $(\bvec{\mathcal{H}}_{x} \cdot \bvec{\mathcal{H}}_{x})^4$. These are likely irrelevant at the critical point, and do not significantly modify any of our results.


\subsection{Coupling to electrons}\label{sec:CouplingToElectrons}
Besides the Higgs field, our proposed minimal gauge theory near optimal doping also contains low-energy electron-like quasiparticles that carry both charge and spin and exhibit a large Fermi surface in the absence of a Higgs condensate. 
We will also consider the case where the electrons pair into a conventional $d$-wave superconductor, in which case the large Fermi surface reduces to 4 nodal points.
We assume that the high-energy electrons and the gapped spinons have already been integrated out and discuss the structure of the effective theory based on symmetries. 

Using $c_{is}$ to denote the field operator of an electron-like quasiparticle at site $i$ with spin $s=\uparrow,\downarrow$, the associated bare Lagrangian for the metallic large Fermi surface state of the $c_{is}$ reads as 
\begin{equation}
  \mathcal{L}_{c} =   c_{is}^\dagger (\partial_\tau - \mu) c^\pdagger_{is} - \sum_{\bvec{\eta},s}t_{\bvec{\eta}} c_{is}^\dagger c^\pdagger_{i+\bvec{\eta}s}, \label{BareElectronLagr}
 \end{equation} 
where $t_{\bvec{\eta}}$ denotes the (renormalized) hopping matrix element along the bond $\bvec{\eta}$ and $\mu$ the chemical potential. Due to SU(2) gauge invariance, a conventional Yukawa term of the form $c^\dagger c H$ is not allowed and the lowest order (in fields and gradients/bond length) coupling between electrons and the Higgs fields is quartic,
\begin{equation}
    \mathcal{L}_{\text{int}} = \lambda_1 \, \bvec{H}_i^2 c_{is}^\dagger c_{is}^\pdagger + \dots \, . \label{CouplingToElectrons}    
\end{equation}
We can also consider terms which involve the gauge-invariant combinations of 
Higgs fields on nearby sites: these will require suitable factors of the SU(2) gauge connection to make them gauge-invariant. The most important of such terms reduce to a coupling of the Ising-nematic order parameter $\phi$, defined in Eq.~(\ref{ScalarHiggsChriality}), to the electrons
\beq
\mathcal{L}_{\phi} = \lambda_2 \, \phi_i \left( c_{is}^\dagger c_{i+\hat{x},s}^\pdagger + c_{is}^\dagger c_{i-\hat{x},s}^\pdagger - c_{is}^\dagger c_{i+\hat{y},s}^\pdagger - c_{is}^\dagger c_{i-\hat{y},s}^\pdagger \right)\,. \label{Lphi}
\eeq
As a consequence of the additional power of the Higgs fields, these couplings are less relevant than a Yukawa coupling, and will be treated perturbatively. In \secref{sec:flstar}, where we start from the spin-fermion model, we will see more explicitly how such couplings can be generated upon integrating out spinons and high-energy electrons. There can also be further coupling terms containing higher powers in fields and gradients. For instance, there are terms of the form $c^\dagger c H^3$ where the ``scalar Higgs chiralities'',
$\chi_{ijk}$ in Eq.~(\ref{ScalarHiggsChriality})
couple to (a suitably chosen combintation of) bond currents $J_{jk} = i \sum_s c^\dagger_{js} c^\pdagger_{ks} + \text{H.c.}$. As will become important in the next subsection, this means that Higgs phases that break time-reversal  symmetry (as indicated by $\braket{\chi_{ijk}}\neq 0$) can lead to time-reversal-symmetry-breaking signatures, such as bond currents, in the electronic properties.

Near the critical point to be discussed in Section~\ref{sec:dcp}, we can study the influence of electrons in the metallic state by following Hertz \cite{hertz} and integrating the electrons out perturbatively in powers of $\lambda_{1,2}$.
After combining with the parametrization in Eq.~(\ref{Parametrization}), we find that the leading effect at order $\lambda_{1,2}^2$ is a renormalization of the quartic $u_i$ couplings in Eq.~(\ref{HiggsPotential}) via $u_i \rightarrow u_i - c_i \lambda_{1,2}^2$, where $i=0,1,2,3$ and the $c_i$ are positive couplings dependent upon the nature of the Fermi surface. This will enhance the possibility of obtaining the corresponding $\phi$, $\Phi_{xy}$, $\Phi_{\pm}$ orders in the Higgs phase. This mechanism is similar to that discussed in Refs.~\onlinecite{Comin14,DCSS14} for the enhancement of certain CDW orders.

Going beyond the static renormalization of couplings, the electrons also induce new dynamic terms in the action. Unlike the cases considered by Hertz \cite{hertz}, the dynamic renormalization is now in the {\it quartic\/} terms of the action (\ref{LH1}), and so of less importance. The simplest such renormalization of the action involves the modulus squared of all the Higgs fields. Such a modulus couples to long-wavelength fluctuations of the electron density, which are suppressed by the long-range Coulomb interactions. However, in a Hubbard-like model with only short-range interactions, the 
electron density fluctuations contribute a term
of the form \cite{Morinari02} 
\beq
\mathcal{S}_{d0} = d_0 \int_{\vec{q},\omega} 
\left[\bvec{\mathcal{H}}^*_x \cdot \bvec{\mathcal{H}}^{\phantom{*}}_x + \bvec{\mathcal{H}}^*_y \cdot \bvec{\mathcal{H}}_y^{\phantom{*}}\right]_{\vec{q},\omega} \frac{|\omega|}{q} \left[\bvec{\mathcal{H}}^*_x \cdot \bvec{\mathcal{H}}^{\phantom{*}}_x + \bvec{\mathcal{H}}^*_y \cdot \bvec{\mathcal{H}}_y^{\phantom{*}}\right]_{-\vec{q},-\omega} 
\label{eq:d0}
\eeq
The scaling dimension of this non-local term can be determined from a knowledge of the scaling dimension of the Higgs modulus squared, which is in turn related to the tuning parameter $s$ across the transition in Eq.~(\ref{HiggsPotential}). Using the fact that $\mbox{dim}[s] = 1/\nu$, where $\nu$ is the correlation exponent, we obtain \cite{Morinari02}
\beq
\mbox{dim}[d_0] = -D + \frac{2}{\nu}\,,
\eeq
where $D=3$ is the spacetime dimensionality. The correlation length exponent $\nu$ will be computed in the large $N_h$ limit in Section~\ref{sec:dcp}, and we find that $\nu =1$ at $N_h = \infty$, while $1/N_h$ corrections reduce $\nu$ to a value below 1 (see Eqs.~(\ref{nu}) and (\ref{nuN})). We conclude that the scaling dimension of $d_0$ is likely to be close to zero. But we reiterate that the $d_0$ term is not present in the presence of long-range Coulomb interactions, when Eq.~(\ref{eq:d0}) can be disregarded \cite{Morinari02}.

The most important dynamic term \cite{Morinari02} appears from the coupling of the electrons to the Ising-nematic order parameter, $\phi$ in Eq.~(\ref{Lphi}). In this case we obtain
\beq
\mathcal{S}_{d1} = d_1 \int_{\vec{q},\omega} 
\phi (\vec{q}, \omega)  \frac{|\omega|}{q} \phi(-\vec{q},-\omega) \,, \label{Sd1}
\eeq
and 
\beq
\mbox{dim}[d_1] = D - 2 \,\mbox{dim}[\phi] \label{eq:d1}
\eeq
In the analysis of the O($N_h$) invariant model of
Section~\ref{sec:dcp}, the $\phi$ order parameter is part of the multiplet of the $Q_{\ell m}$ traceless symmetric tensors in
Eq.~(\ref{defQlm}). These have the scaling dimension obtained in Eq.~(\ref{dimQ}) in the $\epsilon$ expansion, and in Eq.~(\ref{dimQN}) in the $1/N_h$ expansion.
We also note that the corresponding computation of the scaling dimension of $\phi$ in the model without the SU(2) gauge field \cite{Vicari08,Vicari02}---they obtained $\mbox{dim}[\phi] \approx 1$, in which case the coupling to the Fermi surface is relevant.

Finally, we consider the dynamic terms associated with the CDW order parameters. These are at non-zero wavevectors, and so the influence of the Fermi surface is significantly weaker than that for zero wavevector orders, similar to results in the original analysis of Hertz \cite{hertz}. So for the $\Phi_{x,y}$ orders we find
\beq
\mathcal{S}_{d2} = d_2 \int_{\vec{q},\omega} \left[
\Phi_x (\vec{q}, \omega) \, |\omega|\, \Phi_x (-\vec{q},-\omega) + \Phi_y (\vec{q}, \omega) \, |\omega|\, \Phi_y (-\vec{q},-\omega) \right] \,, 
\eeq
and 
\beq
\mbox{dim}[d_2] = D -1 - 2 \,\mbox{dim}[\Phi_x]\,.
\eeq
The CDW order parameters are also part of the $Q_{\ell m}$ operators in Eq.~(\ref{defQlm}), and so
comparing with Eq.~(\ref{eq:d1}) we conclude that $d_2$ is likely irrelevant.

We can extend the above analysis for the confinement-Higgs transition to the case where the electrons are in a $d$-wave superconductor. The large Fermi surface is replaced by 4 nodal points, and the analysis parallels that in Ref.~\onlinecite{Vicari08}, with the Higgs field replacing the spin density wave order parameter. As in Ref.~\onlinecite{Vicari08}, we expect the most important coupling to be between a fermion bilinear and the Ising nematic order parameter, and scaling dimension of the associated coupling is $1 - \mbox{dim}[\phi]$. This is more likely to be irrelevant than that in Eq.~(\ref{Sd1}).

\subsection{Phase diagram for the hole-doped cuprates}
\label{sec:phase}
To see which phases are possible in the Higgs theory of incommensurate SDW fluctuations outlined above, we begin with a mean-field minimization of the Higgs potential in \equref{HiggsPotential}.
A similar minimization of such a potential was carried out by De Prato {\it et al.} \cite{Vicari06} in the context of a theory with $\vec{A}_\mu=0$ describing SDW ordering. Our results are consistent with theirs, but the physical interpretations of the states are quite different: because we have gauged the SDW order into the Higgs fields, the Higgs condensate does not yield SDW order. Our attention is focused on gauge-invariant combinations of the Higgs fields which lead to other broken symmetries.

We take $s<0$ to obtain minima with non-zero Higgs fields and further assume that $u_0 > 0$ has been chosen sufficiently large to ensure stability of the potential (otherwise higher order terms are required which we will not consider here). Under these assumptions, the phase diagram can be compactly plotted as a function of, say, $u_1/|u_3|$, $u_2/|u_3|$, and the sign of $u_3$. 
As summarized in \figref{MFPhasediagram}, there are seven distinct phases denoted by (A), (B), (C), (D), (E), (F)$_\theta$, and (G)$_\theta$ that differ either by the symmetries they break or by the invariant gauge group, i.e., whether there is a residual U(1) gauge invariance or whether all SU(2) gauge fields are gapped and there is $\mathbbm{Z}_2$ topological order.
In the following, we discuss all phases one-by-one:

\begin{figure}[tb]
\begin{center}
\begin{subfigure}[t]{0.55\textwidth}
        \centering
        \includegraphics[height=3.35in]{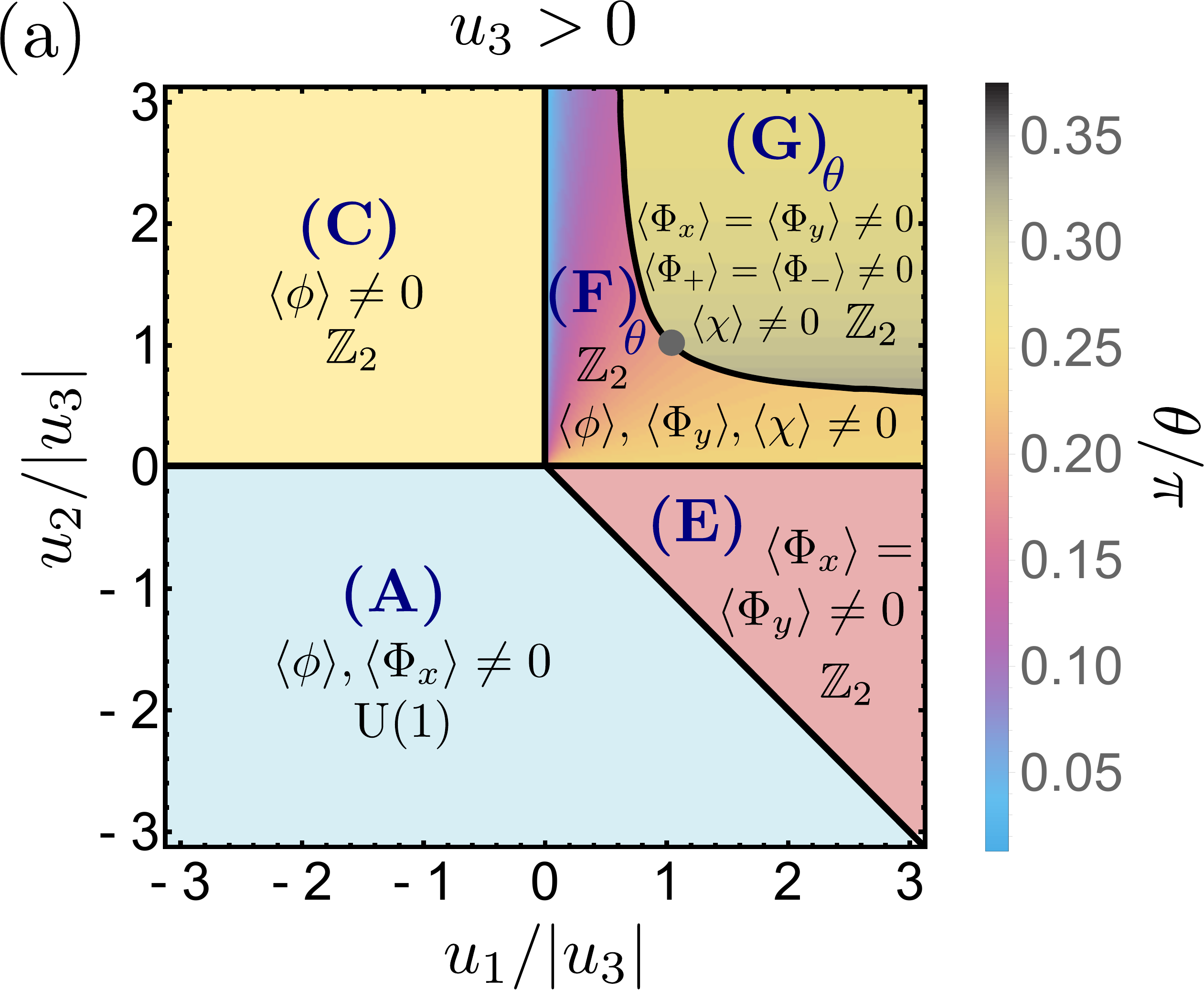}
\end{subfigure}%
~~
\begin{subfigure}[t]{0.55\textwidth}
        \centering
        \includegraphics[height=3.35in]{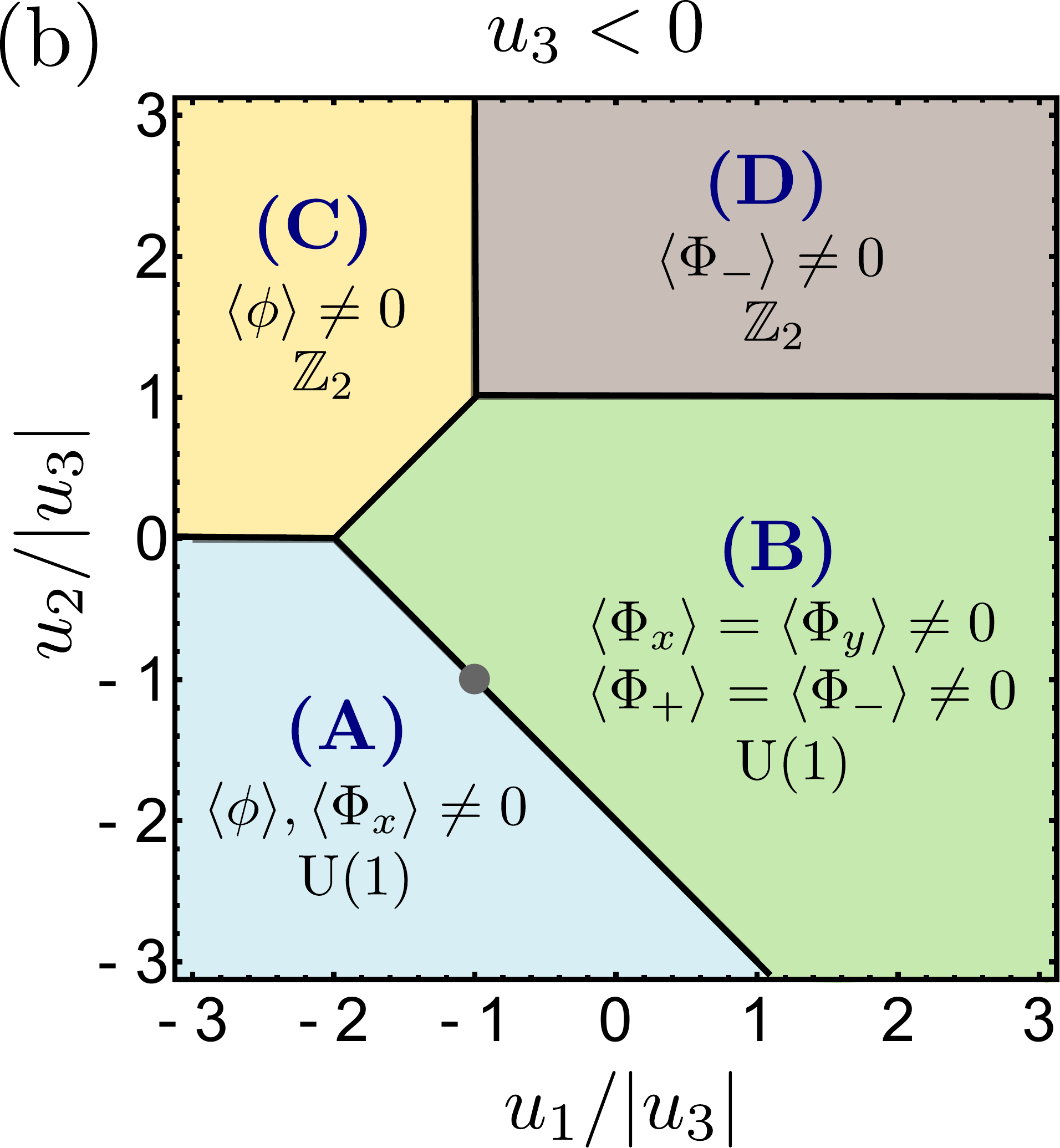}
\end{subfigure}
\end{center}
\caption{Higgs-field phase diagram for the hole-doped cuprates using the parametrization of the Higgs potential given in \equref{HiggsPotential} for (a) $u_3>0$ and (b) $u_3 < 0$. While the phases (A), (B), (C), (D), and (E) are defined by discrete configurations of the complex Higgs fields $\bvec{\mathcal{H}}_{x,y}$ (given in the main text), the saddle point values continuously vary as a function of $u_{1,2,3}$ for the phases (F)$_\theta$ and (G)$_\theta$. As indicated by the subscript, this variation is described by a single parameter $\theta$, defined in \equsref{DefinitionOfFTheta}{DefinitionFTheta}, and shown in color code in the region of the phase diagram where (F)$_\theta$ and (G)$_\theta$ appear. For all phases, we indicate whether the invariant gauge group is U(1) or $\mathbb{Z}_2
$ and which of the time-reversal-invariant order parameters $\phi$, $\Phi_{x,y}$, and $\Phi_{\pm}$ (defined in \tableref{RepresentationOfSymmetries}) and whether the time-reversal-odd scalar Higgs chiralities $\chi$ [see \equref{ScalarHiggsChriality}] are non-zero. The latter operators indicate broken time-reversal symmetry and can couple to the spin chiralities and orbital currents (see main text). In all phases, we only show one possible order parameter configuration while all other symmetry related states (domains) follow from \tableref{RepresentationOfSymmetries}. The gray dots indicate the $O(4)$ invariant points (with $u_1=u_2=u_3$), which we will focus on in the large-$N_h$ analysis of \secref{sec:dcp}.}
\label{MFPhasediagram}
\end{figure}

\begin{description}
\item[Phase (A)] This phase is defined by
\begin{equation}
\bvec{\mathcal{H}}_x = H_0(1,0,0)^T, \quad \bvec{\mathcal{H}}_y = \bvec{0}, \quad \text{or} \quad  \bvec{H}_i = H_0\left(\cos(\vec{K}_x \cdot \vec{r}_i),0,0\right)^T,
\end{equation}
or any symmetry-equivalent description. 
Note that the Higgs condensate is invariant under U(1) gauge transformations along the $(1,0,0)^T$ direction
, and so SU(2) has been broken down to U(1). The U(1) gauge field is ultimately unstable to confinement at the longest length scales, driven by the proliferation of monopoles, as described in Appendix~\ref{app:monopole}.
However, the associated confinement length-scale can be exponentially large deep in the Higgs phase. As can be seen from $\bvec{H}_i^2 = H_0^2 \cos^2(\pi \delta i_x)$, which is gauge invariant and directly couples to the electronic charge density, see \equref{CouplingToElectrons}, this phase induces unidirectional CDW modulations. The residual symmetry group is $C_{2v} \, \otimes \, \Theta \, \otimes \, \mathscr{T}_{y}$ where $\mathscr{T}_{y} = \{T_{0,m}|m\in\mathbb{Z}\}$ is the group containing all lattice translations along the $y$ direction. This is also reflected in $\braket{\phi}=\braket{\Phi_x}=H_0^2$ and $\braket{\Phi_y}=\braket{\Phi_{\pm}}=0$.

\item[Phase (B)] Here we have
\begin{equation}
\bvec{\mathcal{H}}_x = \bvec{\mathcal{H}}_y = \frac{H_0}{\sqrt{2}}(1,0,0)^T, \quad \text{or} \quad  \bvec{H}_i = \frac{H_0}{\sqrt{2}}\left(\cos(\vec{K}_x \cdot \vec{r}_i)+\cos(\vec{K}_y \cdot \vec{r}_i),0,0\right)^T,
\end{equation}
again with a U(1) gauge field in the Higgs phase. In accordance with $\braket{\phi}=0$ and $\braket{\Phi_{x,y}}=\braket{\Phi_{\pm}}=H_0^2/2$, this phase leads to bi-directional CDW modulations,
\begin{equation}
\bvec{H}_i ^2 = \frac{H_0^2}{2}\left[ \cos(\pi\delta i_x) + \cos(\pi\delta i_y) \right]^2,
\end{equation}
and has a residual symmetry group $C_{4v} \otimes \Theta$ (no lattice translational symmetry is preserved).

\item[Phase (C)] The Higgs field is a planar spiral,
\begin{equation}
\bvec{\mathcal{H}}_x = \frac{H_0}{\sqrt{2}}(1,i,0)^T, \quad \bvec{\mathcal{H}}_y = \bvec{0}, \quad \text{or} \quad  \bvec{H}_i = \frac{H_0}{\sqrt{2}}\left(\cos(\vec{K}_x \cdot \vec{r}_i),\sin(\vec{K}_x \cdot \vec{r}_i),0\right)^T. \label{SaddlePointPhaseC}
\end{equation}
This Higgs texture has been studied before in the context of the SU(2) gauge theory \cite{CSS17,OurLoopCurrentsPaper}. Now the Higgs condensate is invariant only under a $\mathbb{Z}_2$ gauge transformation of $-1$, and so this phase has $\mathbb{Z}_2$ topological order \cite{FradkinShenker,NRSS91,Wen91,Bais92} intertwined with Ising nematic order; the residual symmetry group is given by $C_{2v} \otimes \Theta \otimes \mathscr{T}_{x,y}$ where $\mathscr{T}_{x,y}=\{T_{n,m}|n,m\in\mathbb{Z}\}$ is the full square-lattice-translation group. Due to the preserved translation symmetry, it does not induce CDW modulations. These observations are also reflected in $\braket{\phi}=H^2_0$ and $\braket{\Phi_{x,y}}=\braket{\Phi_{\pm}}=0$.

\item[Phase (D)] In this case, we have
\begin{equation}
\bvec{\mathcal{H}}_x = \bvec{\mathcal{H}}_y = \frac{H_0}{2}(1,i,0)^T, \quad \text{or} \quad  \bvec{H}_i = \frac{H_0}{2}\begin{pmatrix}\cos(\vec{K}_x \cdot \vec{r}_i)+\cos(\vec{K}_y \cdot \vec{r}_i) \\ \sin(\vec{K}_x \cdot \vec{r}_i)+\sin(\vec{K}_y \cdot \vec{r}_i) \\ 0\end{pmatrix},
\end{equation}
which leads to $\mathbb{Z}_2$ topological order intertwined with CDW modulations, 
\begin{equation}
\bvec{H}^2_i = \frac{H_0^2}{2} \left[ 1 + \cos(\pi\delta(i_x-i_y)) \right],
\end{equation}
and rotational symmetry breaking: the residual symmetry group is given by $C_{2v} \otimes \Theta \otimes \mathscr{T}_{x+y}$ where $\mathscr{T}_{x+y}=\{T_{n,n}|n\in\mathbb{Z}\}$ is the group of simultaneous translations along $x$ and $y$. Again, this is consistenly seen in $\braket{\phi}=\braket{\Phi_{x,y}}=\braket{\Phi_+}=0$ and $\braket{\Phi_-}=H_0^2/2$.

\item[Phase (E)] Here it holds
\begin{equation}
\bvec{\mathcal{H}}_x = \frac{H_0}{\sqrt{2}}(1,0,0)^T,\quad \bvec{\mathcal{H}}_y = \frac{H_0}{\sqrt{2}}(0,1,0)^T\quad \text{or} \quad  \bvec{H}_i = \frac{H_0}{\sqrt{2}}\begin{pmatrix}\cos(\vec{K}_x \cdot \vec{r}_i) \\ \cos(\vec{K}_y \cdot \vec{r}_i) \\ 0\end{pmatrix},
\end{equation}
which is quite similar to phase (B) except for exhibiting $\mathbb{Z}_2$ topological order. There are CDW modulations,
\begin{equation}
\bvec{H}_i ^2 = \frac{H_0^2}{2}\left[ \cos^2(\pi\delta i_x) + \cos^2(\pi\delta i_y) \right],
\end{equation}
breaking all lattice-translation symmetries and the residual symmetry group is given by $C_{4v} \otimes \Theta$. It holds $\braket{\Phi_{x,y}}= H_0^2/2$ and $\braket{\phi}=\braket{\Phi_\pm}=0$.

\item[Phase (F)$_\theta$] This phase is defined by
\begin{subequations}\begin{equation}
\bvec{\mathcal{H}}_x = \frac{H_0}{\sqrt{2}}\cos(\theta)\,(1,i,0)^T,\quad \bvec{\mathcal{H}}_y = H_0 \sin (\theta) \,(0,0,1)^T, 
\end{equation}
or, equivalently,
\begin{equation}
 \bvec{H}_i = \frac{H_0}{\sqrt{2}}\begin{pmatrix}\cos(\theta)\cos(\vec{K}_x \cdot \vec{r}_i) \\ \cos(\theta)\sin(\vec{K}_x \cdot \vec{r}_i) \\ \sqrt{2} \sin(\theta) \cos(\vec{K}_y \cdot \vec{r}_i) \end{pmatrix}, \label{FthetaParam}
\end{equation}\label{DefinitionOfFTheta}\end{subequations}
which, as opposed to all other phases discussed so far, contains a parameter, $\theta \in (0,\pi/2)$, that varies continuously in the phase diagram (at the O(4) invariant point, we have $\theta = \cot^{-1} (\sqrt{2}) \simeq 0.62$). This variation is indicated in color code in \figref{MFPhasediagram}(a).
Note that \equref{DefinitionOfFTheta} contains phase (C) and phase (A) in the limits $\theta=0$ and $\theta=\pi/2$, respectively. For angles in between, $\theta \in (0,\pi/2)$, however, this is a distinct state: as opposed to all states we discussed so far, it breaks time-reversal symmetry. More precisely, its residual symmetry group is generated by $\sigma_x$, $\Theta T_{1,0}$, and $\Theta C_2$; the associated three-dimensional magnetic space group can be written as $m'mm \otimes \mathscr{T}'_x$ with $\mathscr{T}'_{x} = \{\Theta T_{n,0}|n\in\mathbb{Z}\}$. In accordance with these symmetries, we find $\braket{\phi}=\cos 2\theta$ and $\braket{\Phi_y}=\sin^2\theta $ while $\braket{\Phi_x}=\braket{\Phi_{\pm}}=0$. On top of inducing CDW modulations,
\begin{equation}
 \bvec{H}_i^2 = \frac{H_0^2}{2} \left[ \cos^2(\theta) + 2 \sin^2(\theta) \cos^2(\pi\delta i_y) \right],
\end{equation}
time-reversal symmetry is broken in this phase. Recalling that spin-rotation invariance is preserved, we expect that the Higgs field texture in \equref{DefinitionOfFTheta} will induce non-zero and spatially modulated scalar spin chiralities $\braket{\chi^S_{ijk}} = \braket{\bvec{S}_i \cdot (\bvec{S}_j \times \bvec{S}_k)}$ and bond currents $\braket{J_{ij}} \propto \text{Im}\braket{c^\dagger_{is}c^\pdagger_{js}}$. To see this more explicitly, we recall from \secref{sec:CouplingToElectrons} that the lowest-order gauge-invariant, time-reversal-odd order parameters in terms of Higgs fields are the scalar Higgs chiralities in \equref{ScalarHiggsChriality} which can couple to $\chi^S_{ijk}$ and $J_{ij}$. Focusing on the smallest triangles on the square lattice, consider (cf.~the analogous discussion of spin chirality and orbital currents in \refcite{OurLoopCurrentsPaper})
\begin{equation}
\chi_{s,s'}(i) := ss'\, \bvec{H}_i \cdot \left(\bvec{H}_{i+s\vec{e}_x} \times \bvec{H}_{i+s'\vec{e}_y} \right), \qquad s,s'=\pm. \label{ScalarSpinChirality}
\end{equation}
To connect to the analysis of \refcite{OurLoopCurrentsPaper}, we decompose $\bvec{\chi} = (\chi_{+,+},\chi_{-,+},\chi_{-,-},\chi_{+,-})^T$ locally into the four different symmetry-breaking patterns A--D, summarized in \figref{PatternDefinition} along with the corresponding orbital currents. We find 
\begin{equation}
\braket{\bvec{\chi}(i)} = -(-1)^{i_x+i_y}\sqrt{2} \sin \pi\delta\sin\theta\cos^2\theta\left[ (1-\cos\pi\delta ) \cos (\pi\delta i_y) \bvec{v}_A + \sin\pi\delta \sin (\pi\delta i_y) \bvec{v}_D \right],
\label{defchi}
\end{equation}
where $\bvec{v}_A=(1,1,-1,-1)^T$ and $\bvec{v}_D=(1,1,1,1)^T$, i.e., the induced spin-chirality and orbital-current modulations change character from local A-like to local D-like form as a function of $i_y$. Note that, as expected from the symmetries derived above, we have purely pattern A for $i_y=0$.

\begin{figure}[tb]
\begin{center}
\includegraphics[width=0.8\linewidth]{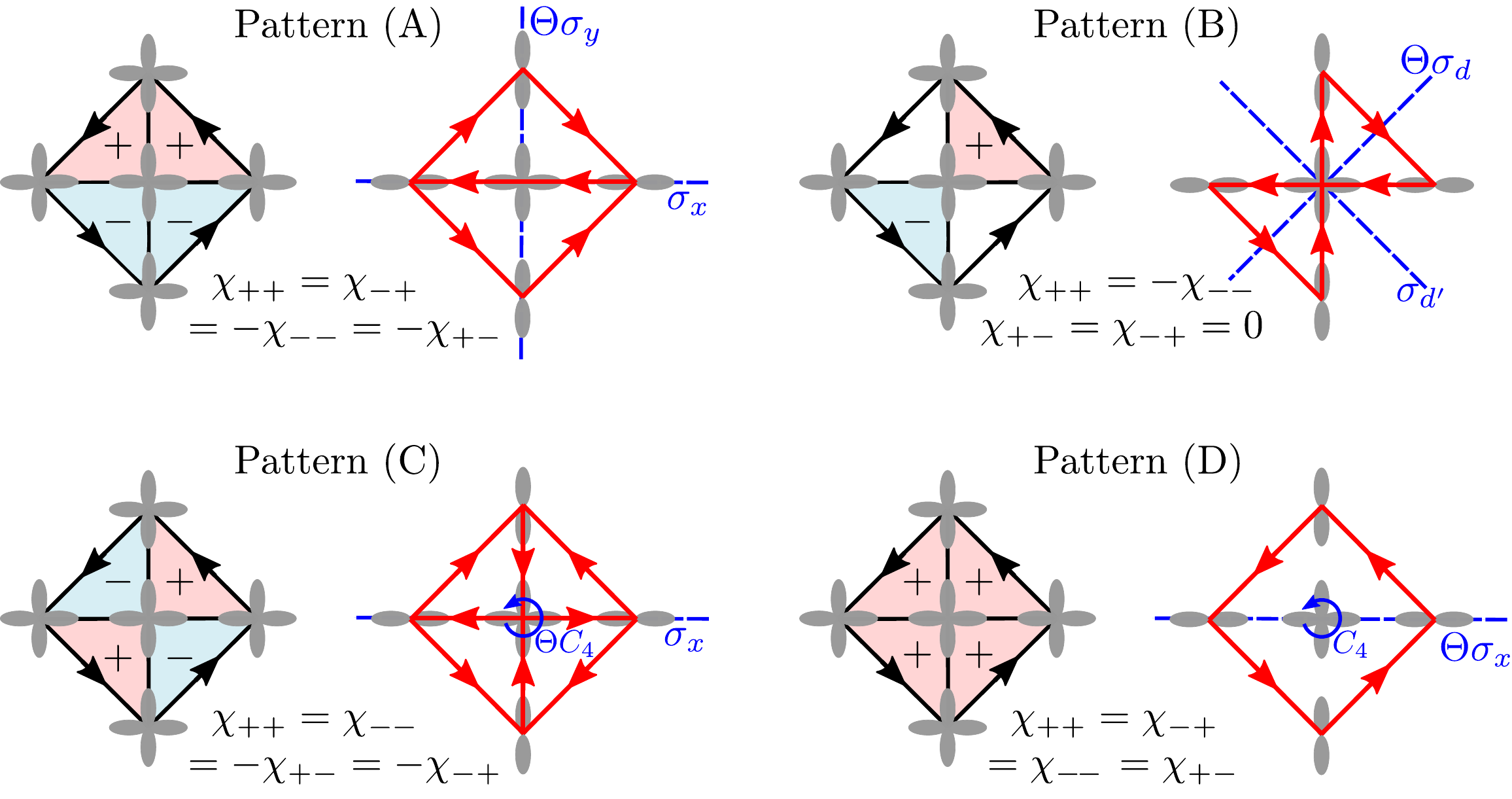}
\caption{The modulated scalar spin chirality (\ref{ScalarSpinChirality}) can locally be decomposed into the four different patterns A--D in the classification of Ref.~\onlinecite{OurLoopCurrentsPaper}. We also indicate the generators of the residual symmetries (blue) and the resulting orbital-current pattern (red). Note that pattern C and D do not allow for (homogeneous) currents on the square lattice which is why we show the loop currents in the three-orbital model \cite{OurLoopCurrentsPaper}.}
\label{PatternDefinition}
\end{center}
\end{figure}

\item[Phase (G)$_\theta$] Finally, this phase is defined by
\begin{subequations}\begin{equation}
\bvec{\mathcal{H}}_x = \frac{H_0}{\sqrt{2}}\begin{pmatrix}\cos(\theta)\\ i\sin(\theta) \\ 0\end{pmatrix},\qquad \bvec{\mathcal{H}}_y = \frac{H_0}{\sqrt{2}}\begin{pmatrix}\cos(\theta)\\ 0 \\  i\sin(\theta)\end{pmatrix}, 
\end{equation}
or
\begin{equation}
 \bvec{H}_i = \frac{H_0}{\sqrt{2}}\begin{pmatrix}\cos(\theta) \left[ \cos(\vec{K}_x \cdot \vec{r}_i) +  \cos(\vec{K}_y \cdot \vec{r}_i) \right] \\ \sin(\theta)\sin(\vec{K}_x \cdot \vec{r}_i) \\ \sin(\theta)\sin(\vec{K}_y \cdot \vec{r}_i) \end{pmatrix}, \label{GtheraParam}
\end{equation}\label{DefinitionFTheta}\end{subequations}
which also involves a parameter, $\theta \in (0,\pi/2)$, that varies continuously in the phase diagram and is indicated in color code in \figref{MFPhasediagram}. This phase approaches phase (B) in the limit $\theta \rightarrow 0$ and phase (E) for $\theta\rightarrow \pi/2$. At the O(4) invariant point, $\theta = \tan^{-1} (\sqrt{2}) \simeq 0.96$. For phase (G)$_\theta$, we obtain $\braket{\phi}=0$ and $\braket{\Phi_{x,y}}=H_0^2\cos(2\theta)/2$, and $\braket{\Phi_\pm} = H_0^2\cos^2(\theta)/2$ indicating the presence of CDW modulations,
\begin{equation}
 \bvec{H}_i^2 = \frac{H_0^2}{2} \left[ \cos^2\theta \left(\cos(\pi\delta i_x) + \cos(\pi\delta i_y)\right)^2 + \sin^2\theta \left(\sin^2(\pi\delta i_x) + \sin^2(\pi\delta i_y)\right) \right].
\end{equation}
The residual symmetry group is generated by $C_4$ and $\Theta\sigma_x$ and, hence, the associated 3D magnetic space group is given by $\frac{4}{m}m'm'$ (no translations are preserved). Again, due to the broken time-reversal and translational symmetry, we expect spatially modulated scalar spin chiralities, $\braket{\chi^S_{ijk}}\neq 0$, and orbital currents, $\braket{J_{ij}} \neq 0$ which can be diagnosed by the order parameters defined in \equref{ScalarSpinChirality}. We find
\begin{align}\begin{split}
\braket{\bvec{\chi}(i)} &= (-1)^{i_x+i_y}\sin^2 \pi\delta\sin^2\theta\cos\theta\Bigl[ \left( \cos (\pi\delta i_x) +  \cos (\pi\delta i_y) \right)\bvec{v}_D \\ & \qquad + \tanh(\pi \delta/2) (\sin (\pi\delta i_x)-\sin (\pi\delta i_y))\bvec{v}_A - 2 \tanh(\pi \delta/2) \sin (\pi\delta i_x) \bvec{v}_B   \Bigr],
\end{split}\end{align}
where $\bvec{v}_B = (1,0,-1,0)^T$ showing that we obtain a spatially varying combination of spin chiralities/orbital currents of type A, B, and D of \figref{PatternDefinition}. In accordance with the symmetry arguments presented above, there is only pattern D at $(i_x,i_y)=(0,0)$.

\end{description}

\subsection{Electron-doped cuprates}
\label{sec:electron}

\begin{figure}[tb]
\begin{center}
\includegraphics[height=3.75in]{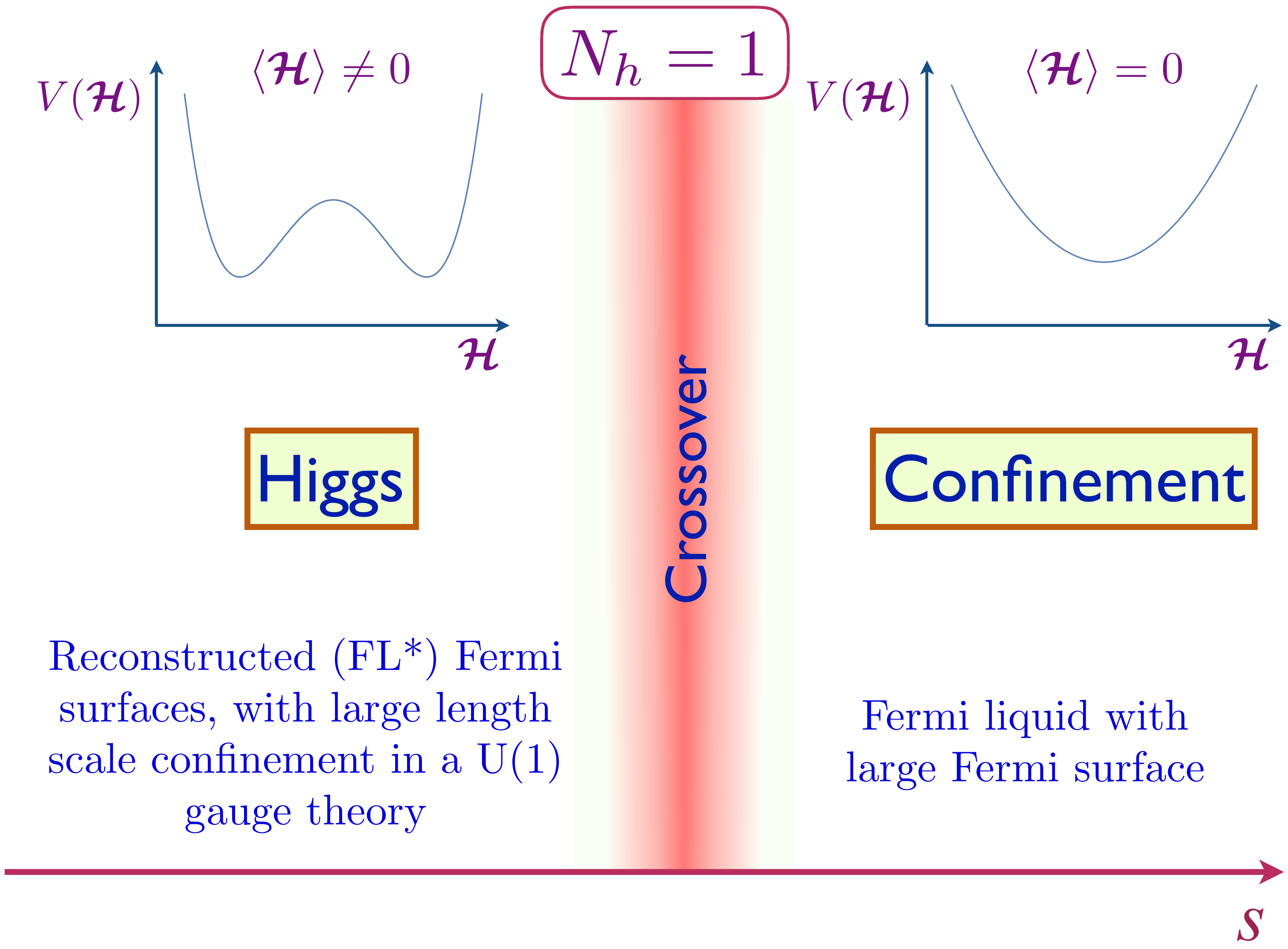}
\end{center}
\caption{Schematic phase diagram of SU(2) gauge theory for $N_h=1$ with a real, adjoint Higgs field for the electron-doped cuprates; compare to Fig.~\ref{fig:phasediag_h} for the hole-doped cuprates. Now there is only
a crossover between the Higgs and confinement regimes at low and high doping respectively. The crossover is associated with an increase in the confinement scale to an exponentially large value within the Higgs region.}
\label{fig:phasediag_e}
\end{figure}
As discussed in Section~\ref{sec:intro}, we assume commensurate, two sublattice, antiferromagnetic fluctuations for the electron-doped side of the phase diagram, and replace the parametrization (\ref{Parametrization}) by that in Eq.~(\ref{Parametrization2})
\beq
\bvec{H}_{i} = (-1)^{i_x+i_y} \bvec{\mathcal{H}}(\vec{r}_i),
\eeq
with only $N_h = 1$ real adjoint Higgs field $\bvec{\mathcal{H}}$, so that Eq.~(\ref{FullHiggsAction}) reduces to the Georgi-Glashow model \cite{GG72,tH74,Polyakov74}. It is easily seen that the most general quartic Higgs potential consistent with SU(2) gauge invariance and time-reversal automatically conserves all square-lattice symmetries, and has the usual Landau-Ginzburg form
\beq
V  = s \, \bvec{\mathcal{H}} \cdot \bvec{\mathcal{H}} + u (\bvec{\mathcal{H}} \cdot \bvec{\mathcal{H}})^2, 
\eeq
where $u>0$ to ensure stability.
For $s>0$ we can ignore the Higgs field, and then the SU(2) gauge theory is in a confining phase.
For $s<0$, we anticipate a Higgs condensate in which $\langle \bvec{\mathcal{H}} \rangle = H_0 (0,0,1)$. 
Such a condensate leaves an unbroken U(1) gauge field
and is, consequently, also confining in $2+1$ dimensions from the proliferation of monopoles, as discussed in Appendix~\ref{app:monopole}. The simplest assumption is that there is no phase transition, and theory is always confining. 
Indeed, this is what was found in Monte Carlo simulations \cite{Kibble00,Kibble02}. See the phase diagram in Fig.~\ref{fig:phasediag_e}.

However, deep in the Higgs phase, the monopoles can be very dilute (see Appendix~\ref{app:monopole}), and the confinement length scale is exponentially large.
Then, at length scales smaller than the confinement scale, we have a deconfined U(1) gauge theory and the Fermi surface can reconstruct, but the large Fermi surface eventually reappears at the longest scales.
This is our candidate theory for the electron-doped cuprates, with no confinement-induced broken symmetry in the underdoped region.

\section{FL* spectral function}
\label{sec:flstar}

It is useful to begin the discussion of the electron spectral function in the pseudogap by recalling the approach taken in recent work on the SU(2) gauge theory \cite{SCWFGS,WSCSGF}. Along with the transformation of the spin magnetic moment in Eq.~(\ref{e1}), this work also rotates the electron operator in spin space to a fermionic `chargon' $\psi_i$
\beq
c_i = R_i \psi_i \,.
\eeq
The chargon carries SU(2) gauge charge, but is invariant under the global spin rotation symmetry. The transformations of all fields under the global and gauge SU(2)'s is summarized in Table~\ref{tab:fields}.
\begin{table}[h]
    \centering
    \begin{tabular}{|c|c|c|c|c|c|}
\hline
Field & Symbol & Statistics & SU(2) gauge & SU(2) spin & U(1) charge\\
\hline 
\hline
Electron & $c$ & fermion & ${\bm 1}$ & ${\bm 2}$ & -1\\
SDW order & $\boldsymbol{S}$ & boson & ${\bm 1}$ & ${\bm 3}$ & 0 \\
Chargon & $\psi$ & fermion & ${\bm 2}$ & ${\bm 1}$ & -1 \\
Spinon & $R$ & boson & $\bar{\bm 2}$ & $ {\bm 2}$ & 0 \\
Higgs & $\boldsymbol{H}$ & boson & ${\bm 3}$ & ${\bm 1}$ & 0 \\
\hline
\end{tabular}
    \caption{Fields and their quantum numbers. 
The transformations under the SU(2)'s are labelled by the dimension
of the SU(2) representation, while those under the electromagnetic U(1) are labeled by the U(1) charge. Earlier works \cite{SCWFGS,WSCSGF} described the pseudogap using $\bvec{H}$, $\psi$, $R$. Section~\ref{sec:su2} uses $\bvec{H}$ and $c$, and Section~\ref{sec:flstar} uses $\bvec{H}$, $c$, and $R$.}
    \label{tab:fields}
\end{table}

\begin{figure}
\begin{center}
\includegraphics[height=6cm]{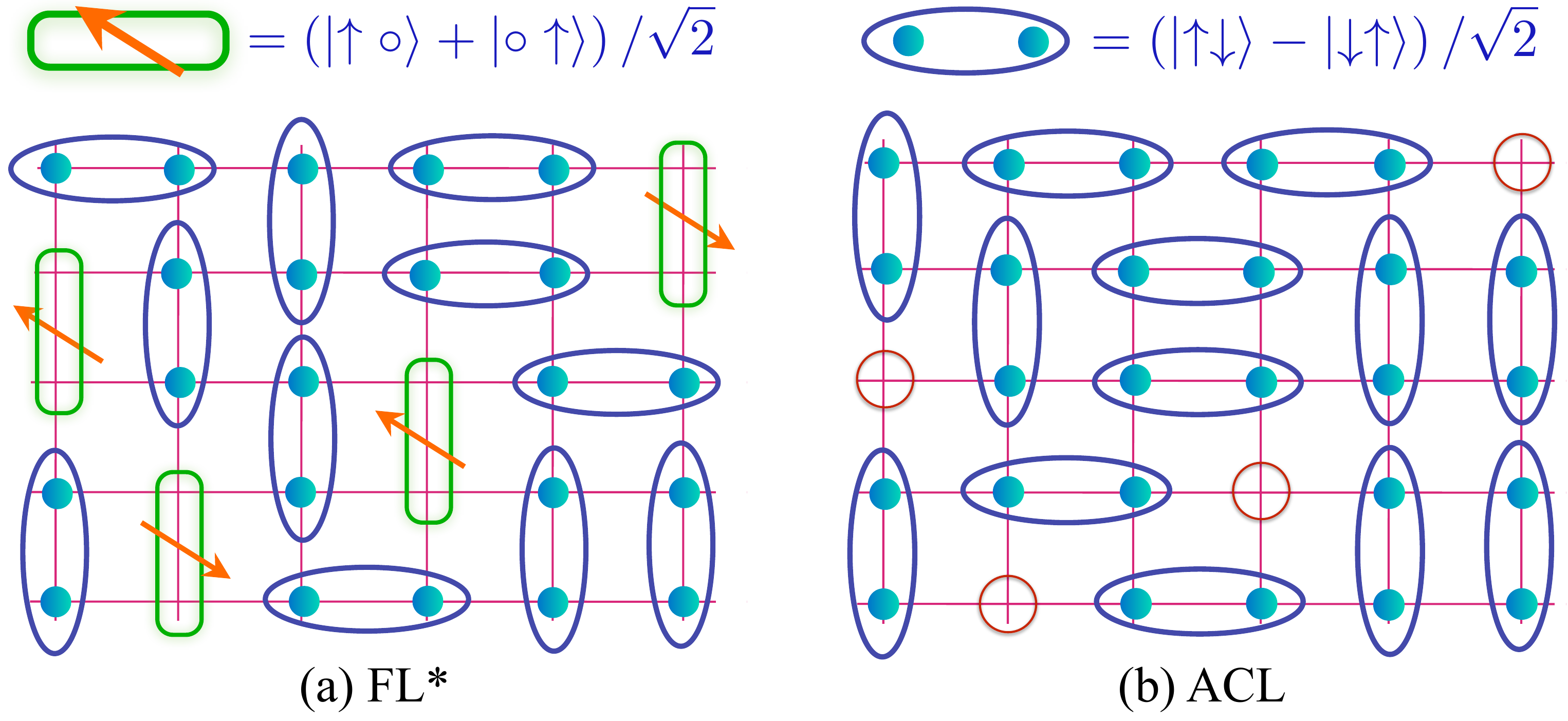}
\end{center}
\caption{(a) A component of a resonating bond wavefunction for FL* in a single-band model on the square lattice \cite{Punk15}. 
The density of the green
bonds is $p$, and these are fermions which form `reconstructed' Fermi surfaces of volume $p$ with electron-like quasiparticles.
(b) A component of a wavefunction for an ACL. The vacancies are the `holons', or more generally, the `chargons'; they are 
assumed to be fermions which form a Fermi surface of 
spinless quasiparticles of charge $e$. The blue resonating dimers represent the deconfined $\mathbb{Z}_2$ or U(1) gauge fields.}
\label{fig:nambu}
\end{figure}

The analysis of Refs.~\onlinecite{SCWFGS,WSCSGF} proceeds by assuming that the Higgs field condenses, and this allows the $\psi$ and $R$ to deconfine. The deconfinement is possible if the Higgs condensation breaks SU(2) down to $\mathbb{Z}_2$, and at length scales smaller than the possible large confinement scale when SU(2) is broken down to U(1). The Higgs condensate also reconstructs the $\psi$ Fermi surface to small electron and/or hole pockets, yielding a metallic state with chargon Fermi surfaces: such a metal is an `algebraic charge liquid (ACL)' and is illustrated in Fig.~\ref{fig:nambu}(b).
The electron spectral function of the ACL is computed as a convolution of the $\psi$ and $R$ propagators, and this was the procedure followed in Refs.~\onlinecite{SCWFGS,WSCSGF}. There are strong residual attractive interactions between the $\psi$ and $R$ arising from the underlying electron hopping terms, and these interactions can lead to the formation of electron-like quasiparticles, as was shown in the simplified model of Ref.~\onlinecite{SBCS16} (recall the evidence for such electron-like bound states in ultracold atom studies \cite{Greiner18a,Bloch18,Greiner18b}). 
When all the low energy fermions become electron-like, we obtain the FL* state shown schematically in Fig.~\ref{fig:nambu}(a). 

The underlying assumption of the analysis in Section~\ref{sec:su2} was that the low energy fermionic excitations remain electron-like across optimal doping. And the resulting theory for the
confining to Higgs transition was formulated in terms of the $\bvec{H}$ and $c$ fields of Table~\ref{tab:fields}. Here, we wish to describe the FL* state also using the $\bvec{H}$ and $c$ fields. We will show below that this is possible, but we will also need to include the gapped $R$ spinons in such a theory. The theory we present here is similar in spirit to earlier analysis using a `spinon-dopon' approach \cite{Ribeiro06,Punk12}, in which the dopons were electron-like fermions, and the spinons where fermionic \cite{Ribeiro06} or bosonic \cite{Punk12} analogs of $R$.

For completeness, we also note an approach to the pseudogap physics based upon quasi-one-dimensional physics \cite{Tsvelik17,Tsvelik17b}. This can be considered a version of a FL* state because it has poles in the electron Green's function, along with neutral spin-1/2 excitations.

\subsection{Model and effective electron-electron interaction}
As outlined above, we start with electrons, $c_{is}$, on the square lattice described by the Lagrangian $\mathcal{L}_{c}$ in \equref{BareElectronLagr} which are coupled to collective spin fluctuations $\bvec{\Phi}_i$ via the usual spin-fermion coupling $\mathcal{L}_{\text{int}} = \lambda_0 \sum_{s,s'}c_{is}^\dagger \bvec{\sigma}_{ss'} c^\pdagger_{is'}  \cdot \bvec{\Phi}_i$. To describe orientational fluctuations of SDW order, we transform to a rotating reference frame according to \equref{e1} and rewrite the spin-fermion coupling in terms of spinons $R_i$ and the Higgs field $\bvec{H}_i$,
\begin{equation}
    \mathcal{L}_{\text{int}} = \sum_{s,s'} c_{is}^\dagger (R^\pdagger_i\bvec{\sigma} R_i^\dagger)^\pdagger_{s,s'} c^\pdagger_{is'}  \cdot \bvec{H}_i \label{Coupling}.
\end{equation}
To connect to our previous discussion in \secref{sec:CouplingToElectrons}, we note in passing that a coupling of the form of \equref{CouplingToElectrons} will be generated by \equref{Coupling} upon integrating over the spinons $R_i$ and a shell of high-energy electrons $c_i$. 

The total Lagrangian, $\mathcal{L}=\mathcal{L}_c+\mathcal{L}_{\text{int}} + \mathcal{L}_R + \mathcal{L}_{\mathcal{H}}$, also contains a spinon part, $\mathcal{L}_R$, and a Higgs contribution, $\mathcal{L}_{\mathcal{H}}$ which we will address next. 
The general form of $\mathcal{L}_{\mathcal{H}}$ has been discussed at length in \secref{sec:su2} above.
In this section, we are interested in the spectral properties and Fermi surface reconstruction of the electron-like quasiparticles $c_{is}$ in the Higgs phase and not in the additional symmetry breaking that occurs in the incommensurate phases analyzed in \secref{sec:phase}. For this reason, we will focus on the simplest, commensurate Higgs condensate with $\braket{\bvec{H}_i} = (-1)^{i_x+i_y} (H_0,0,0)^T$ which we had proposed for the electron-doped cuprates. For sufficiently small, but non-zero, deformations $\delta \ll 1$ of the wavevectors of the Higgs texture, we can describe the incommensurate, symmetry-breaking states relevant to the hole-doped side while retaining approximately the same spectral properties.

To parameterize the spinons $R_i \in \text{SU}(2)$ in this Higgs phase, we write
\begin{equation}
R_i = \begin{pmatrix} z_{i\uparrow} & -z^*_{i\downarrow} \\ z_{i\downarrow} & z^*_{i\uparrow}  \end{pmatrix}, \quad z_i^\dagger z_i^\pdagger =1, \label{eq:Rz}
\end{equation}
where $z_i$ are complex bosons related to the $\mathbb{C}\mathbb{P}^1$ bosons for fluctuating antiferromagnetism \cite{CSS17,SCWFGS}.
Since we do not know its precise form, we will take the phenomenolgical ansatz 
\begin{equation}
    \mathcal{S}_R = \frac{1}{g}\int_q z^\dagger_q \left( \Omega_n^2 + E^2_{\bvec{q}}\right) z_q, \label{SpinonAction}
\end{equation}
for the spinon action $\mathcal{S}_R$. Here $\int_q \dots \equiv T\sum_{\Omega_n} \int_{\text{BZ}}\frac{\diff^2 \bvec{q}}{(2\pi)^2} \dots$ with bosonic Matsubara frequencies $\Omega_n$ and integration of $\bvec{q}$ over the first Brillouin zone (BZ), $z_{q}\equiv z_{\Omega_n,\bvec{q}}$ denotes the Fourier transform of $z_i$, $g$ is a parameter determining the strength quantum fluctuations, and $E_{\bvec{q}}$ is the spinon dispersion. We treat the constraint $z_i^\dagger z^\pdagger_i =1$ on average,
\begin{equation}
\braket{z_i^\dagger z^\pdagger_i} = 1, \label{ConstraintOnAv}
\end{equation}
and the associated Lagrange multiplier determines the gap $E_{\bvec{q}=0}=\Delta$ of the spinons. Solving \equref{ConstraintOnAv} for the Lagrange multiplier or, equivalently, for the spinon gap $\Delta$ yields $\Delta=\Delta(g)$. For our purposes, it will be simpler and more insightful to specify $\Delta$ and solve for $g=g(\Delta)$. To be consistent with the conventional continuum $\mathbb{C}\mathbb{P}^1$ model, we demand that $E^2_{\bvec{q}}=\Delta^2 + v^2 \bvec{q}^2$ as $|\bvec{q}|\rightarrow 0$ with some constant $v$ and that $E_{\bvec{q}}$ be minimal at $\bvec{q}=0$. Since we perform the calculation on the lattice, we will further have to make sure that $E_{\bvec{q}}$ is periodic on the Brillouin zone and is invariant under the point symmetries of $C_{4v}$. Below we will use
\begin{equation}
  E_{\vec{q}} = \sqrt{ 2v^2(2-\cos(q_x)-\cos(q_y)) + \Delta^2}, \label{SimplestSpectrum}   
\end{equation}
which satisfied all of the above constraints.

Deep in the Higgs phase and at length scales much shorter than the confinement length, we can ignore both the Higgs-field and gauge-field fluctuations. Using the identity $R^\pdagger_i \sigma_z R_i^\dagger = z_i^\dagger \bvec{\sigma} z^\pdagger_i \cdot \bvec{\sigma}$, we obtain the simplified action
\begin{equation}
\mathcal{S}= \int_k  c_k^\dagger (-i\omega_n - \xi_{\bvec{k}}) c_k  + H_0 \int_{k,q,\Delta q} (c^\dagger _k \bvec{\sigma} c^\pdagger_{k+\Delta q}) \cdot (z^\dagger_{q + \Delta q+\bvec{K}} \bvec{\sigma} z^\pdagger_q) + \mathcal{S}_R, \label{FullAction}
\end{equation}
where spin indices have been suppressed, $c_k \equiv c_{\omega_n,\bvec{k}}$ with fermionic Matsubara frequencies $\omega_n$ are the Fourier transforms of the electron operators, $\bvec{K}=(\pi,\pi)^T$ and $q+\bvec{K} \equiv (i \Omega_n,\bvec{q}+\bvec{K})$.
For the electronic dispersion $\xi_{\bvec{k}}$, we focus on nearest, $t$, and next-to-nearest-neighbor hopping amplitudes, $t'$, i.e.,
\begin{equation}
\xi_{\bvec{k}} = -2 t(\cos(k_x)+\cos(k_y) ) - 4 t' \cos(k_x)\cos(k_y) - \mu.
\end{equation}
In all numerical plots, we will use $t$ as unit of energy (setting $t=1$).

\begin{figure}[tb]
\begin{center}
\includegraphics[width=0.45\linewidth]{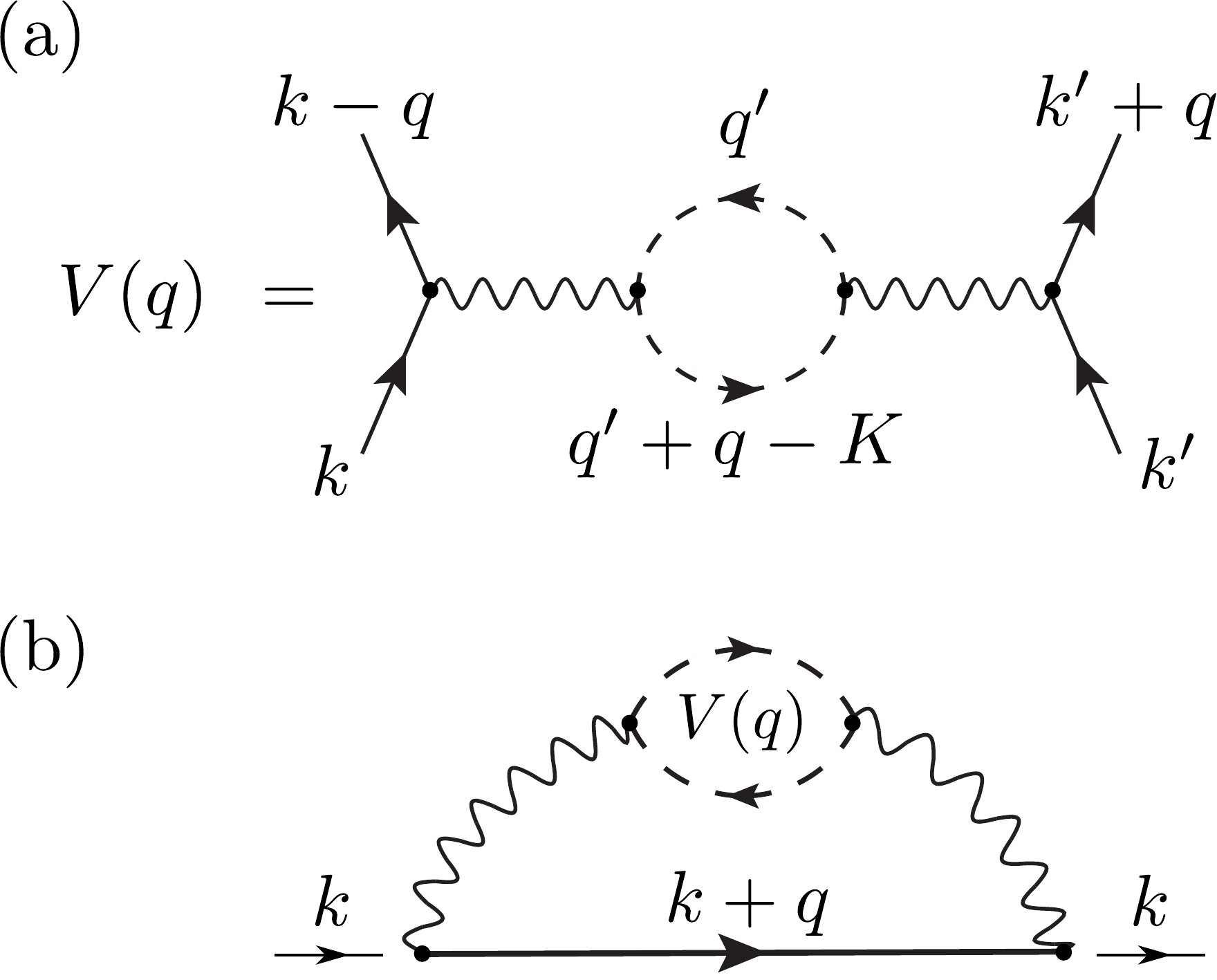}
\caption{Diagrammatic representation of (a) effective electron-electron interaction (\ref{EffecInteraction}) mediated by spinon fluctuations and of (b) the leading-order electronic self-energy (\ref{LeadingOrderSE}). We use solid and dashed lines to represent electron and spinon Green's functions. The curvy lines denote the spinon-electron interaction ($\propto H_0$) in \equref{FullAction}.}
\label{SpinonInducedInteraction}
\end{center}
\end{figure}

To obtain an effective interaction for the electrons, we integrate out the spinons and expand the resulting action of the fermions to leading non-trivial order (quartic) in the electronic fields. This yields the following effective action 
\begin{subequations}\begin{equation}
\mathcal{S}_c^{\text{eff}}= \int_k  c_k^\dagger (-i\omega_n + \xi_{\bvec{k}}) c^\pdagger_k - \int_q \bvec{S}_q \cdot \bvec{S}_{-q} V(q), \qquad \bvec{S}_q = \int_k c^\dagger_{k+q} \bvec{\sigma} c^\pdagger_{k}, \label{EffectiveElectronAction}
\end{equation}
for the electron-like quasiparticles.
The crucial ``form factor'' $V(q)$ of the effective interaction is given by
\begin{equation}
V(\bvec{q},i\Omega_n) = V_0 \int_{q'} \frac{1}{\Omega_{n'}^2+E_{\bvec{q}'}^2}\frac{1}{(\Omega_{n'}+\Omega_{n})^2+E_{\bvec{q}'+\bvec{q}-\bvec{K}}^2}, \qquad V_0 = H_0^2 g^2, \label{EffectiveInteraction}
\end{equation}\label{EffecInteraction}\end{subequations}
and illustrated diagrammatically in \figref{SpinonInducedInteraction}(a). From the structure of \equref{EffectiveInteraction}, we can already see that the interaction is peaked for momentum transfers $\bvec{q} \simeq \bvec{K}$ and, as such, can give rise to pseudogap-like reconstruction of the Fermi surface in the vicinity of the antiferromagnetic zone boundary.
Notice that $g=g(\Delta)$ and $H_0$ only enter in form of the pre-factor $V_0=H_0^2 g^2$ and not independently. We, thus, do not have to solve \equref{ConstraintOnAv} explicitly and can, without loss of generality, work at fixed $V_0$ and $\Delta$.

\subsection{Spectral function in leading order}
We are only interested in the qualitative impact of the spinon-mediated interaction in \equref{EffecInteraction} on the spectral function of the electrons and, hence, restrict the analysis to the first-order contribution to the electronic self-energy $\Sigma$.
The corresponding diagram is shown in \figref{SpinonInducedInteraction}(b) and reads algebraically as
\begin{equation}
    \Sigma(k) = -3 \int_q V(q) G_0(k+q), \label{LeadingOrderSE}
\end{equation}
where $G_0(k)=(i\omega_n - \xi_{\bvec{k}})^{-1}$ is the bare electronic Green's function. 
This approximation for the electron self-energy is similar to that used in spin-fluctuation theories for the pseudogap \cite{Kampf90a,Kampf90b,Tremblay97,Eremin01,Chubukov10}, with the crucial difference that the 2-spinon propagator $V(q)$ is replaced by a Landau-damped spin-fluctuation propagator. In the latter case, the Landau damping restores the large Fermi surface at low energies. In our case, it is important to note that the fractionalized spinons do not undergo Landau damping from a gauge neutral Fermi surface.

Inserting the explicit form of $V(q)$ [see \equref{EffectiveInteraction}] into \equref{LeadingOrderSE}, performing the two Matsubara sums and the analytic continuation, the imaginary part of the leading-order retarded self-energy can be written as
\begin{align}\begin{split}
\text{Im} \Sigma^R(\omega,\bvec{k}) = -\frac{\pi V_0}{4} \int_{\text{BZ}}\frac{\diff^2 \bvec{q}}{(2\pi)^2}\frac{\diff^2 \bvec{q}'}{(2\pi)^2} \frac{1}{E_{\bvec{q}'}E_{\bvec{q}'+(\bvec{q}-\bvec{K})}} & \Bigl[ \Theta(\xi_{\bvec{k}+\bvec{q}}) \delta(\omega - (\xi_{\bvec{k}+\bvec{q}}+E_{\bvec{q}'}+E_{\bvec{q}'+(\bvec{q}-\bvec{K})})) \\ &+ \Theta(-\xi_{\bvec{k}+\bvec{q}}) \delta(\omega - (\xi_{\bvec{k}+\bvec{q}}-E_{\bvec{q}'}-E_{\bvec{q}'+(\bvec{q}-\bvec{K})})) \Bigr] \label{LeadingSimpleProp}
\end{split}\end{align}
while its real part simply follows from Kramers-Kronig. We immediately notice from this expression that
\begin{equation}
\text{Im} \Sigma^R(\omega,\bvec{k}) = 0 \qquad \text{for}\,\, |\omega| < 2\Delta  \label{VanishingSpectralWeight}
\end{equation}
as it should be due to the spinon gap: there is no decay into (pairs of) spinons for energies below the two-spinon gap $2\Delta$. 

\begin{figure}[tb]
\begin{center}
\includegraphics[width=\linewidth]{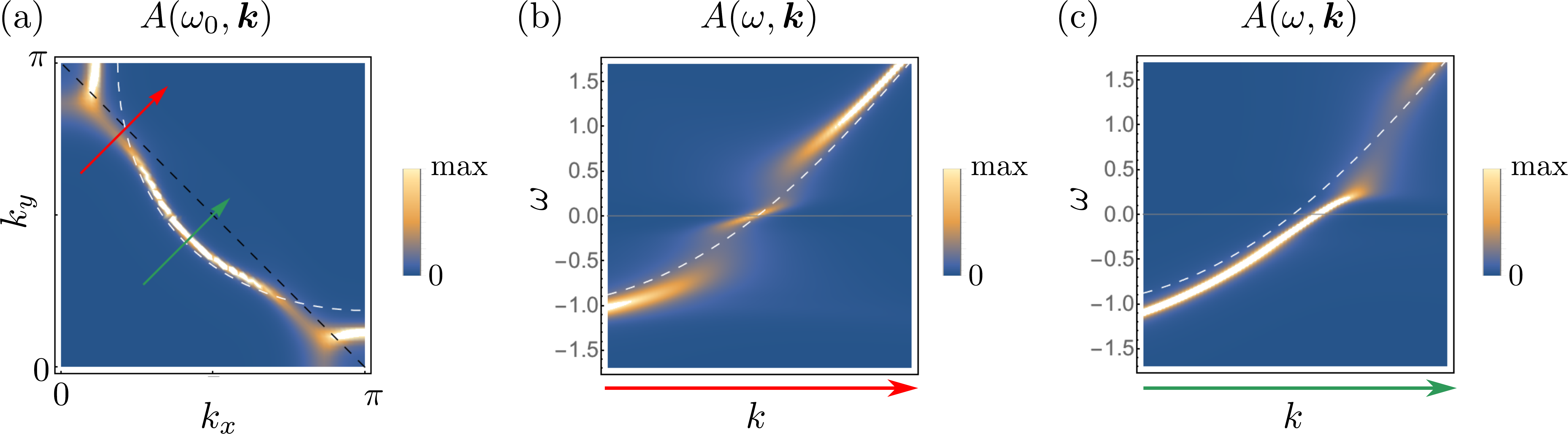}
\caption{Illustration of the spectral function $A(\omega,\vec{k})$ in first order in $V_0$. The momentum dependence of the spectral function at fixed low energy $\omega_0 = -0.1 < -2\Delta$ is shown in (a) where the black (white) dashed line indicates the antiferromagnetic zone boundary (the momenta where $\xi_{\vec{k}}=\omega_0$). In (b) and (c), the energy, $\omega$, and momentum dependence along the red and green arrows in (a) is shown. Here, the white dashed line is the bare electronic dispersion. These results are obtained by numerical integration of \equref{LeadingSimpleProp} using the spinon spectrum in \equref{SimplestSpectrum} and the cutoff procedure given in the main text. Here we have taken $v^2=0.1$, $\Delta=0.03$, $t'=-0.4$, $\mu=-0.9$, $V_0/(2\pi)^4=0.15$, $\eta=0.02$, and $\Lambda=2\Delta$.}
\label{SpectrumInFLStar}
\end{center}
\end{figure}

As readily follows from its structure, the main contribution of the integral in \equref{LeadingSimpleProp} comes from spinons with large momenta if we assume that $E_{\vec{q}}^2$ increases quadratically in $\vec{q}$. Since our current description of the FL$^*$ state is by design only valid at very low energies (below the energies associated with spinon-chargon bound state formation), \equref{LeadingSimpleProp} has to be supplemented with a cutoff. This can, for instance, be achieved by adding additional terms to the spinon dispersion in \equref{SimplestSpectrum} that increase more rapidly with $\vec{q}$ or by adding a momentum cutoff to \equref{LeadingSimpleProp}. Here we will use a soft-cutoff for the spinon energy by adding an extra factor of $\exp(-(E_{\vec{q}'}+E_{\vec{q}'+\vec{q}-\vec{K}}-2\Delta)/\Lambda)$ in the integral in \equref{LeadingSimpleProp}. The resulting spectral function is illustrated in \figref{SpectrumInFLStar} and clearly shows a strong reduction of the low-energy spectral weight in the vicinity of the hot-spots while leaving the part of the Fermi surface closer to $\vec{k}=(\pi/2,\pi/2)$ intact. Note, however, that there are additional weak peaks in the spectral function in an energy window of the size of the spinon gap centered around $\omega=0$, see \figref{SpectrumInFLStar}(b), which are related to \equref{VanishingSpectralWeight}. We leave the question whether the inclusion of higher-order diagrams, e.g., a self-consistent Hartree-Fock analysis, will remove these low-energy features and the necessity of introducing a cutoff for future work.

Instead we will try to understand the leading-order results shown in \figref{SpectrumInFLStar} more intuitively by investigating the limit of small $\Lambda$. In that case, we can effectively focus on the contribution of zero-momentum spinons to the integral in \equref{SpectrumInFLStar} which leads to  
\begin{equation}
\Sigma^R(\omega,\bvec{k}) \propto \frac{ V_0}{4\Delta^2} \Biggl[  \frac{\Theta(\xi_{\bvec{k}+\bvec{K}})}{\omega - (\xi_{\bvec{k}+\bvec{K}}+2\Delta) + i0^+}+ \frac{\Theta(-\xi_{\bvec{k}+\bvec{K}})}{\omega - (\xi_{\bvec{k}+\bvec{K}}-2\Delta) + i0^+} \Biggr].
\end{equation}
This expression strongly resembles the self-energy associated with long-range antiferromagnetic order, except for the region $-2\Delta < \omega < 2\Delta$, where $\text{Im} \Sigma^R(\omega,\bvec{k})$ vanishes; in fact, in the limit $\Delta \rightarrow 0$, it approaches $\frac{V_0}{4\Delta^2}   \frac{1}{\omega - \xi_{\bvec{k}+\bvec{Q}}+i0^+}$ which \textit{is} the self-energy of long-range antiferromagnetism. This naturally explains that the spectral function in \figref{SpectrumInFLStar} looks very similar to that in the presence of antiferromagnetic order.

\section{Deconfined criticality at large $N_h$}
\label{sec:dcp}

Our purpose here is to describe the quantum critical point associated with the onset of the Higgs phase for $N_h >1$ (recall that there is no phase transition for $N_h=1$).
We are interested in $N_h=4$ action for the hole-doped cuprates described in Section~\ref{sec:su2} with the Higgs potential in Eq.~(\ref{HiggsPotential}). To generalize this for arbitrary $N_h$, we will only consider the simpler case with global O($N_h$) flavor symmetry. De Prato {\it et al.} \cite{Vicari06} carried out an RG analysis of the theory without the SU(2) gauge field, but with the full potential in Eq.~(\ref{HiggsPotential}): they found that all stable fixed points had O(4) symmetry. We are assuming here that the same holds true in the presence of the SU(2) gauge field.

To write the action in an explicitly O($N_h$) invariant form, this section will not use the bold vector notation, and instead write
$\bvec{\mathcal{H}}_x = (\mathcal{H}_{1x}, \mathcal{H}_{2x}, \mathcal{H}_{3x})$ and similarly for $\bvec{\mathcal{H}}_y$ and the SU(2) gauge field $\bvec{A}_\mu$.
Further, we write the complex Higgs fields in terms of their real and imaginary parts
\beq
\mathcal{H}_{a x} = H_{a 1} + i H_{a 2} \quad , \quad \mathcal{H}_{a y} = H_{a 3} + i H_{a 4}\,.
\label{HReIm}
\eeq
So now we denote the Higgs fields as $H_{a\ell}$, where $a=1,2,3$ is a color index, and $\ell=1 \ldots 4$ is a flavor index. This has a natural generalization to a flavor index $\ell=1 \ldots N_h$.

The $N_h = 4$ Higgs potential in Eq.~(\ref{HiggsPotential}) turns out to be O(4) invariant when $u_1 = u_2 = u_3$. This follows from the identity
\bea
&& H_{a \ell} H_{a m} H_{b \ell} H_{b m} - \frac{1}{4} H_{a \ell} H_{a \ell} H_{b m} H_{b m} \nn
&&~~~~~ = \frac{1}{4} (\bvec{\mathcal{H}}^*_x \cdot \bvec{\mathcal{H}}^{\phantom{*}}_x - \bvec{\mathcal{H}}^*_y \cdot \bvec{\mathcal{H}}^{\phantom{*}}_y)^2 
+ \frac{1}{2}\left( | \bvec{\mathcal{H}}_x \cdot \bvec{\mathcal{H}}_x |^2 + | \bvec{\mathcal{H}}_y \cdot \bvec{\mathcal{H}}_y |^2 \right) +  |\bvec{\mathcal{H}}_x \cdot \bvec{\mathcal{H}}_y|^2 + |\bvec{\mathcal{H}}^{\phantom{*}}_x \cdot \bvec{\mathcal{H}}^*_y|^2  \nn
&&~~~~~ = \frac{1}{4} \left( \mathcal{H}_{ax}^\ast \mathcal{H}_{ax} - \mathcal{H}_{ay}^\ast \mathcal{H}_{ay} \right)^2 
+ \frac{1}{2} \left[ \left| \mathcal{H}_{ax} \mathcal{H}_{ax}\right|^2 + \left| \mathcal{H}_{ay} \mathcal{H}_{ay} \right|^2 \right] + 
  \left| \mathcal{H}_{ax} \mathcal{H}_{ay}\right|^2 + \left| \mathcal{H}_{ax} \mathcal{H}_{ay}^\ast \right|^2  \nn
&&~~~~~ = \frac{1}{4} \phi^2 + \frac{1}{2} \left[ |\Phi_x|^2 + |\Phi_y|^2 \right] +    |\Phi_+|^2 + |\Phi_-|^2 
  \,.
\label{Identity1}
\eea
We can, therefore, write down the O($N_h$) invariant generalization of the Lagrangian in Eq.~(\ref{FullHiggsAction})
\beq
\mathcal{L}_H = \frac{1}{4g^2} F_{a \mu\nu} F_{a \mu \nu} + \frac{1}{2} \left( \partial_\mu H_{a\ell} - \varepsilon_{abc} A_{b\mu}
H_{c\ell} \right)^2 + V(H) \label{LH2}
\eeq
with the field strength
\beq 
F_{a \mu\nu} = \partial_\mu A_{a \nu} - \partial_\nu A_{a \mu} - \varepsilon_{abc} A_{b\mu} A_{c \nu}\,,
\eeq
and the Higgs potential
\beq
V(H) =  \frac{s}{2} \, H_{a \ell} H_{a \ell}  + {u}_0 (H_{a \ell} H_{a \ell})^2
+ {u}_1 \left( H_{a \ell} H_{a m} H_{b \ell} H_{b m}
- \frac{1}{N_h} (H_{a \ell} H_{a \ell})^2 \right) \,.
\label{HiggsPotential2}
\eeq

\subsection{Mean field theory}
\label{sec:mft}

We begin with a mean field analysis of $\mathcal{L}_H$ analogous to that in Section~\ref{sec:phase}.
The Higgs phase will now be characterized by a gauge-invariant order parameter which transforms as a symmetric traceless second-rank tensor of O($N_h$):
\beq
Q_{\ell m} = H_{a\ell} H_{a m} - \frac{\delta_{\ell m}}{N_h} H_{an} H_{an}\,. \label{defQlm}
\eeq
All the order parameters in Section~\ref{sec:phase} correspond to different components of $Q_{\ell m}$, and they are treated here at an equal footing. Also note that the square of $Q_{\ell m}$ equals the $u_1$ quartic term in the Higgs potential
\beq
Q_{\ell m} Q_{\ell m} = H_{a \ell} H_{a m} H_{b \ell} H_{b m}
- \frac{1}{N_h} H_{a \ell} H_{a \ell} H_{b m} H_{b m}\,.
\label{Q2}
\eeq
The identity in Eq.~(\ref{Q2}) is the O($N_h$) invariant analog of the identity in Eq.~(\ref{Identity1}).

By the singular value decomposition theorem, any Higgs field can be written in the form
\beq
H_{a \ell} = O_{1,ab} W_{b m} O_{2,m\ell}
\eeq
where $O_1$ and $O_2$ are orthogonal matrices in color and flavor spaces respectively, and $W$ is a rectangular matrix with only 
\beq
p \equiv \mbox{min}(3, N_h)
\eeq
non-zero elements along its diagonal. We call the diagonal elements $w_{1\ldots p}$.
Then
\beq
V(H) = \frac{s}{2} \left(\sum_i^p w_i^2 \right) + \left(u_0 - \frac{u_1}{N_h} \right)\left(\sum_i^p w_i^2 \right)^2 + u_1 \left(\sum_i^p w_i^4 \right)
\eeq
The form of the minimum of $V(H)$ in the Higgs phase, $s<0$, depends upon the sign of $u_1$.
\subsubsection{$u_1> 0$}
The minimum has all the $w_i$ equal, $w_1 = \ldots = w_p$. This Higgs phase 
has $\mathbb{Z}_2$ topological order for all $N_h > 1$. For $N_h \leq 3$, there is no broken symmetry, and $\langle Q_{\ell m} \rangle = 0$.
But there is a broken symmetry for $N_h >3$, and the O($N_h$) symmetry is broken to O($N_h-3$)$\times$O($3$) with 
$\langle Q_{\ell m} \rangle \neq 0$; this broken symmetry co-exists with $\mathbb{Z}_2$ topological order.
\subsubsection{$u_1 < 0$}
Now only one of $w_{1\ldots p}$ is non-zero. The O($N_h$) symmetry is broken to O($N_h-1$), and there is a residual U(1) gauge invariance,
which ultimately leads to confinement. We always have $\left\langle Q_{\ell m} \right\rangle \neq 0$ for $N_h > 1$. The residual U(1) gauge invariance
is also present for $N_h=1$, and the absence of a broken symmetry implies the absence of a phase transition for this value of $N_h$ alone.

\subsection{$\epsilon$ expansion}
\label{sec:eps}

This section describes a renormalization group (RG) analysis of the Higgs-confinement critical point of the action in Eq.~(\ref{LH2}) to leading order in $\epsilon = 4-D$,
where $D$ is the spacetime dimensionality. Such an RG analysis has been carried out by Hikami \cite{Hikami:1980qk}
and others \cite{Vicari01,Vicari02,Kawamura}. We perform a similar analysis below, correcting a numerical error in Hikami's results, and obtain the scaling dimension of the $Q_{\ell m}$ order parameter.

To renormalize the Lagrangian (\ref{LH2}), we must include counterterms $Z_i$, gauge fixing ${\cal L}_{\xi}$ and ghost fields ${\cal L}_{C}$,
\begin{align}
\notag{\cal L}_{YM}&=\frac{1}{2g^2}Z_3 A_a^{\mu}(g_{\mu\nu}\partial^2-\partial_\mu\partial_\nu)A_a^{\nu}+\frac{1}{g^2}Z_{3g}\varepsilon_{abc}A_a^{\mu}A_b^{\nu}\partial_\mu A_{c\nu} - \frac{1}{4g^2}Z_{4g}\varepsilon_{abe}\varepsilon_{cde}A_a^{\mu}A_b^{\nu} A_{c\mu} A_{d\nu}\\
\notag{\cal L}_{\xi}&=\frac{1}{2}\xi^{-1}A_a^{\mu}\partial_\mu\partial_\nu A_a^{\nu}\\
\notag{\cal L}_{C}&=-\frac{1}{2}Z_{2c}\partial^\mu\bar{C}_a\partial_mu C_a - Z_{1c}\varepsilon_{abc}A_{c\mu}\partial^\mu\bar{C}_a C_b\\
\label{LHcounter}
{\cal L}_{H}&=\frac{1}{2}Z_2\left(\partial_\mu H_{bl}\right)^2-Z_1\varepsilon_{abc}A^\mu_{a} H_b \partial_\mu H_c+\frac{1}{2}Z_4\varepsilon_{abe}\varepsilon_{cde}A^\mu_{a}A_{b\mu}H_c H_d-V(H).
\end{align} 
Similarly we must include counterterms into the Higgs potential (\ref{HiggsPotential2}). Moreover, for this subsection it is convenient to rewrite the potential as follows,
\begin{align}
\label{VH1}
V(H)&=\frac{1}{2}Z_s s H_{bl}H_{bl} + Z_{v_0}v_0(H_{bl}H_{bl})^2  + Z_{v_1}v_1 \left[H_{bl}H_{bm}H_{cl}H_{cm} - (H_{bl}H_{bl})^2\right],
\end{align}
which returns to the form (\ref{HiggsPotential2}) upon the replacements $v_0=u_0-u_1(N_h-1)/N_h$; $v_1=u_1$.

Our task is to compute the counter-terms $Z_1, Z_2, Z_3, Z_{u_1}, Z_{u_2}, Z_{s}$ in an $\epsilon$-expansion and obtain the $\beta$-functions.
Since the $\beta$-functions (or equivalently, the counterterms $Z_i$) can be extracted from Machacek \cite{Machacek1983I} or Hikami \cite{Hikami:1980qk} (after correcting for the numerical error), we defer the technical details to Appendix~\ref{app:epsilon} and present just the final results. However, such details will become important for our computation of the scaling dimension of the $Q_{\ell m}$ order parameter, which has not been considered previously. Denoting $\alpha=g^2$, and rewriting the mass term as $s=t\mu^2$, where $\mu$ is the energy scale and $t\sim (g-g_c)$ is the dimensionless tuning parameter of the QPT, the $\beta$-functions are found to be,
\begin{align}
\notag\beta_\alpha&=\frac{\partial \alpha}{\partial\ln\mu}=\frac{1}{8 \pi ^2} \left(\frac{N_h}{3}-\frac{22}{3}\right)\alpha ^2-\alpha  \epsilon\\
\notag\beta_{v_0}&=\frac{\partial v_0}{\partial\ln\mu}=\frac{1}{8\pi^2}\left(\frac{3}{2}\alpha^2 +4(3N_h+8)v_0^2 -16 (N_h-1) v_0v_1+8 (N_h-1) v_1^2-12 v_0\alpha \right) -v_0 \epsilon\\
\notag\beta_{v_1}&=\frac{\partial v_1}{\partial\ln\mu}=\frac{1}{8\pi^2}\left(\frac{3}{4}\alpha ^2 +4 (N_h-5)v_1^2-12v_1\alpha+48 v_0
   v_1\right)-v_1 \epsilon\\
\beta_{t}&=\frac{\partial \ln t}{\partial\ln\mu}=\frac{1}{{8 \pi ^2}} \left(4(3N_h+2)v_0 -8(N_h-1)v_1 -6\alpha\right) - 2.
\end{align}
This set of $\beta$-functions admit one nontrivial stable fixed point for which all interaction couplings are non-zero. This fixed point cannot be expressed in closed analytic form for generic $N_h$. However, taking the large $N_h$ limit we obtain the simple expressions,
\begin{align}
\label{FP}
    \notag\alpha^*&= \left(\frac{24\pi^2}{N_h}+\frac{528\pi^2}{N_h^2} + O \big(N_h^{-3}\big) \right) \epsilon,\\
    \notag v_0^*&= \left(\frac{2\pi^2}{N_h}-\frac{56\pi^2}{N_h^2}+ O \big(N_h^{-3}\big)\right) \epsilon,\\ 
    v_1^*&= \left(\frac{2\pi^2}{N_h}+\frac{4\pi^2}{N_h^2}+ O \big(N_h^{-3}\big)\right) \epsilon. 
\end{align}
The $\alpha^*$ corrects the numerical error of Ref. \cite{Hikami:1980qk}. 
At this fixed point we evaluate the (gauge dependent) critical exponent $\nu$, 
\begin{align}
\label{nu}
    \frac{1}{\nu} &=-\beta_{t}(\alpha^*,v_0^*,v_1^*) =2-\left(1-\frac{102}{N_h}-\frac{448}{N_h^2} + O \big(N_h^{-3}\big)\right)\epsilon + \ldots. 
\end{align}
Importantly these results coincide with those obtained via the large $N_h$ expansion technique presented in Section \ref{sec:largeN}. Moreover, $\nu$ coincides with Ref.~\onlinecite{Hikami:1980qk} to $O \big(N_h^{-1}\big)$, despite the error in $\alpha^*$.

\begin{figure}[tb]
\begin{center}
\includegraphics[width=0.5\linewidth]{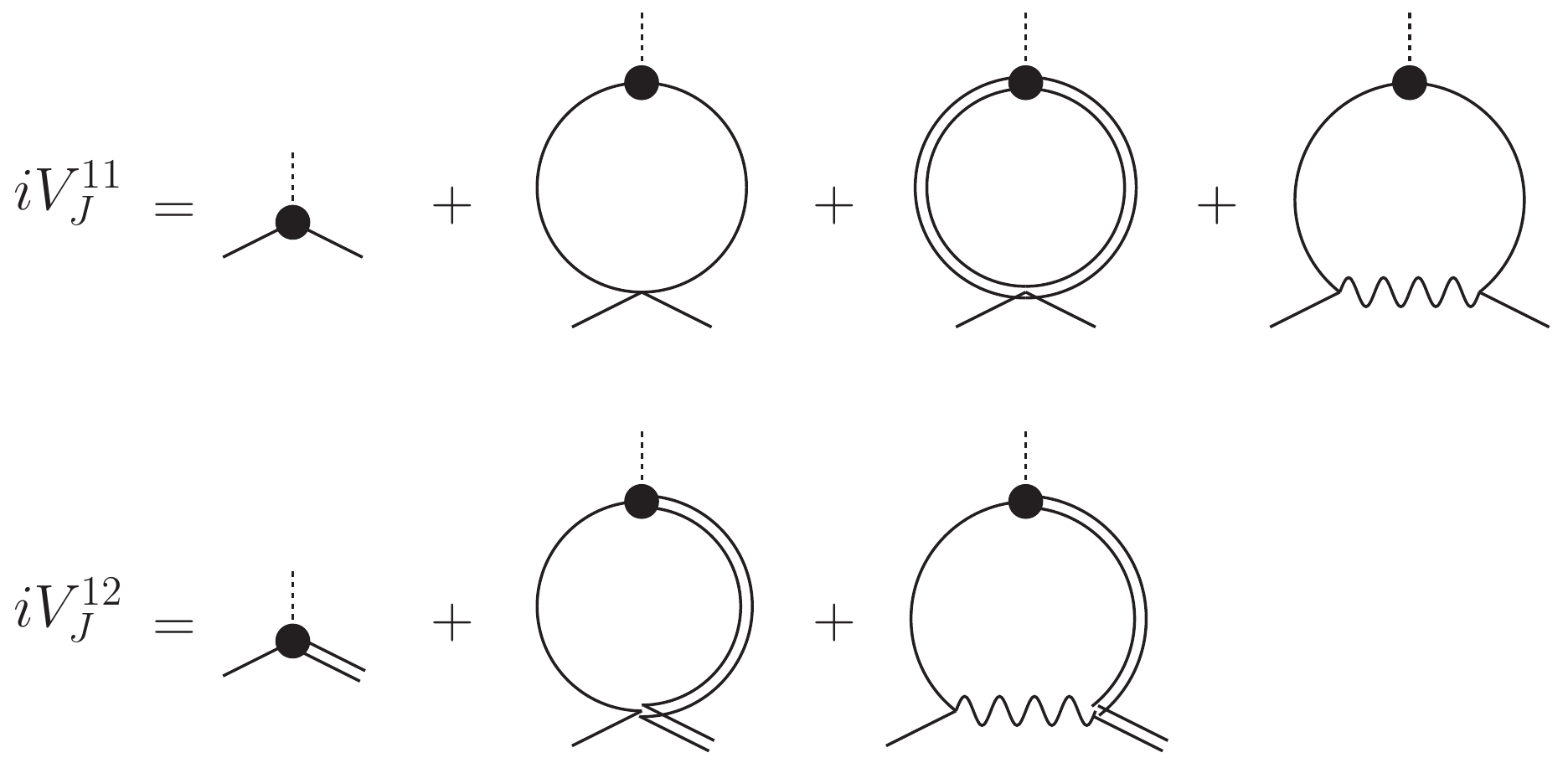}
\caption{First-order corrections to the vertices of the source term $J_{\ell m}$. We denote by $V_J^{\ell m}$ the series which contains only source term vertices $J_{\ell m}$. (a) Shows the series $V_J^{11}$, and (b) shows the series $V_J^{12}$. All other $V_J^{\ell m}$ are related to (a) or (b) by symmetry. The black dot represents source term vertices. All other vertices can be deduced from Eq. (\ref{LHcounter}), see Appendix~\ref{app:epsilon} for the corresponding Feynman rules. Single and double lines represent the propagator of the fields $H_{b1}$ and $H_{b2}$, respectively. Curvy line represents the gauge field $A_{b\mu}$.}
\label{SourceDiagrams}
\end{center}
\end{figure}

Let us now consider the scaling dimension of the $Q_{\ell m}$ order parameter. Since this has not been considered previously, we provide the details of the computation. To begin, we introduce a source-term ${\cal L}_J$ to the Lagrangian (\ref{LHcounter}),
\beq
{\cal L}_J =Z_J J_{\ell m}Q_{\ell m}
\eeq
where $Z_J$ is a counterterm needed to regularize $J_{\ell m}$. In Fig.~\ref{SourceDiagrams}(a) and (b), we establish a diagrammatic series for the components $J_{11}$ and $J_{12}$, respectively. All other components of $J_{\ell m}$ immediately follow. From such series we determine $Z_J$ (to first-order). It is important to verify that the single counterterm $Z_J$ is sufficient to cancel the $O \big(\epsilon^{-1}\big)$ divergences for all $J_{\ell m}$.
Explicitly including vertex and symmetry factors, the diagrammatic series of Fig.~\ref{SourceDiagrams}(a) and (b) give (in Euclidean metric),
\begin{align}
\label{VJ11}
\notag \frac{N_h}{N_h-1}iV_J^{11}&=2iZ_J J_{11}+(iZ_J J_{11})\{-i[24v_0-8v_0+16v_0-16(v_0-v_1)]\}\frac{1}{i}\int \frac{d^d l}{(2\pi)^d}\frac{1}{(l^2+s)((l+k)^2+s)}\\
\notag& \hspace{1.95cm}  + 4(iZ_J J_{11})(ig)^2\frac{1}{i^2}\int \frac{d^d l}{(2\pi)^d}\frac{l^\mu l^\nu P^{\mu\nu}(l)}{(l^2+s)^2 l^2}\\
&=2iZ_J J_{11} \left(1 - \frac{1}{8\pi^2\epsilon}\left(8(v_0+v_1) - 2g^2\right)\right) + O \big(\epsilon^{0}\big),\\
\notag iV_J^{12}&=iZ_J J_{12}+(iZ_J J_{12})(-i8(v_0+v_1))\frac{1}{i}\int \frac{d^d l}{(2\pi)^d}\frac{1}{(l^2+s)((l+k)^2+s)}\\
\notag& \hspace{1.85cm}  + 2(iZ_J J_{12})(ig)^2\frac{1}{i^2}\int \frac{d^d l}{(2\pi)^d}\frac{l^\mu l^\nu P^{\mu\nu}(l)}{(l^2+s)^2 l^2}\\
&=iZ_J J_{12} \left(1 - \frac{1}{8\pi^2\epsilon}\left(8(v_0+v_1) - 2g^2\right)\right) + O \big(\epsilon^{0}\big).
\end{align}
To obtain the final expressions we employ the Feynman gauge: $P^{\mu\nu}(l)=g^{\mu\nu}$. 
Demanding that all $O \big(\epsilon^{-1}\big)$ terms cancel and setting $g^2=\alpha$, we obtain the desired, unique expression for the counterterm
\beq
\label{Lsource}
Z_J =1 + \frac{1}{8\pi^2\epsilon}\left(8(v_0+v_1) - 2\alpha\right).
\eeq

Next, the bare source is related to the renormalized source via $J^0_{\ell m}=Z_J Z_2^{-1} J_{\ell m}$. $Z_2$ is the Higgs field-renormalization counterterm shown in Eq. (\ref{LHcounter}) -- it is given by Eq. (\ref{Z_2}). The engineering dimension of $J_{\ell m}$ is 2, while the anomalous dimension is computed via $\gamma_J=\frac{\partial}{\partial\log\mu}\log\left(Z_J Z_2^{-1}\right)$. Together we find the scaling dimension of the source term,
\begin{align}
    \mbox{dim}[J_{\ell m}]&=2-\gamma_J=2-\frac{1}{8\pi^2}(8(v_0^*+v_1^*)-6\alpha^*).
\end{align}
In the final equality we denote that the coupling constants are to be evaluated at the fixed point (\ref{FP}). Finally,
the scaling dimension of the order parameter $Q_{\ell m}$ is obtained by demanding that the Lagrangian density for the source-term ${\cal L}_J$ (\ref{Lsource}) has dimension $D$, which tells us that 
\begin{align}
   \notag \mbox{dim}[Q_{\ell m}]=D-\mbox{dim}[J_{\ell m}]&=2-\epsilon+\frac{1}{8\pi^2}(8(v_0^*+v_1^*)-6\alpha^*)
    \\
    &=2-\epsilon-\frac{14\epsilon}{N_h}\left(1+\frac{32}{N_h} + O \big(N_h^{-2}\big)\right) + \ldots.\label{dimQ}
\end{align}
In the last equality the scaling dimension is evaluated at the fixed point (\ref{FP}) in the large $N_h$ limit.

\subsection{Large $N_h$ expansion}
\label{sec:largeN}

In this section we consider a generalization of our model and now assume that the gauge group is $O(M)$.  This theory already was studied 
in \cite{Hikami:1980qk, Vasiliev:1984gn, Vasiliev:1984bf}. And here we merely repeat the main results of these papers and additionally compute the large $N_{h}$ anomalous dimension of the gauge invariant traceless operator $Q_{\ell m}=H_{a\ell}H_{am}-(\delta_{\ell m}/N_{h}) H_{an}H_{an}$. Our original theory is obtained by setting $M=3$. 
 As usual in the large $N$ computations it is convenient to start with 
a critical Euclidean action which has the form 
\begin{align}
S_{\textrm{crit}} = \frac{1}{2}\int d^{D}x\Big((\partial_{\mu}H_{a\ell}-i T^{A}_{ab}A^{A}_{\mu}H_{b\ell})^{2}+\lambda_{ab}H_{a\ell}H_{b\ell}\Big)\,,
\end{align}
where $a,b=1,\dots, M$,\;  $\ell=1,\dots,N_{h}$ and we introduced the Hubbard-Stratonovich matrix field $\lambda_{ab}(x)$ and dropped the bare non-abelian Maxwell term, since  in $D<4$
it is irrelevant at low momenta. The dynamics of the gauge field $A_{\mu}^{A}(x)$ and the field $\lambda_{ab}(x)$ is governed by induced propagators, so one can equivalently write the effective large $N_{h}$ action in the form 
\begin{align}
S_{\textrm{crit}}  =&  \frac{1}{2}\int d^{D}x\Big( A_{\mu}^{A}(x)\Pi^{AB}_{\mu\nu}(x-y)A_{\nu}^{B}(y)+\lambda_{ab}(x)K_{ab,cd}(x-y)\lambda_{cd}(y)+\partial_{\mu}H_{a\ell}\partial_{\mu}H_{a\ell}\notag\\
&\qquad\qquad +2iA_{\mu}^{A}J_{\mu}^{A}+(T^{A}T^{B})_{ab}A^{A}_{\mu}A^{B}_{\mu}H_{a\ell}H_{b\ell}+\lambda_{ab}H_{a\ell}H_{b\ell}\Big)\,, \label{critact2}
\end{align}
where $J_{\mu}^{A}(x) = \frac{1}{2}T^{A}_{ab}(H_{a\ell}\partial_{\mu}H_{b\ell}-\partial_{\mu}H_{a\ell}H_{b\ell})$ and one computes
\begin{align}
&\Pi^{AB}_{\mu\nu}(x) = \langle J_{\mu}^{A}(x) J_{\nu}^{B}(0)\rangle_{0} = -\delta^{AB}S_{2}\frac{N_{h}(D-2)\Gamma(\frac{D}{2}-1)^{2}}{(4\pi^{D/2})^{2}(x^{2})^{D-1}}\Big(\delta^{\mu\nu}-2 \frac{x^{\mu}x^{\nu}}{x^{2}}\Big)\,,  \notag\\
&K_{ab,cd}(x) =  -\frac{1}{4}\langle H_{a\ell}H_{b\ell}(x)H_{cm}H_{dm}(0)\rangle_{0} =-\frac{N_{h}\Gamma(\frac{D}{2}-1)^{2}}{4(4\pi^{D/2})^{2}}\frac{(\delta_{ac}\delta_{bd}+\delta_{ad}\delta_{cb}) }{(x^{2})^{D-2}}\,, \label{kernln}
\end{align}
where we defined $\textrm{Tr}(T^{A}T^{B})=S_{2}\delta^{AB}$ and  used that 
\begin{align}
\langle H_{a\ell}(x)H_{bm}(0)\rangle_{0} = \delta_{\ell m}\delta_{ab}G(x) = \delta_{\ell m}\delta_{ab} \frac{\Gamma(\frac{D}{2}-1)}{4\pi^{D/2}} \frac{1}{(x^{2})^{D/2-1}}\,.
\end{align}

 The critical action (\ref{critact2}) is the starting point for large $N_{h}$ diagrammatic  computations. This effective action is  gauge invariant as it should due to conservation of current.  After the gauge fixing procedure one has to add ghosts and gauge fixing term to the action. We note that ghosts do not contribute in the leading $1/N_{h}$ computations as explained in \cite{ Vasiliev:1984bf}.  
 
\begin{figure}[h!]
\includegraphics[width=0.93\textwidth]{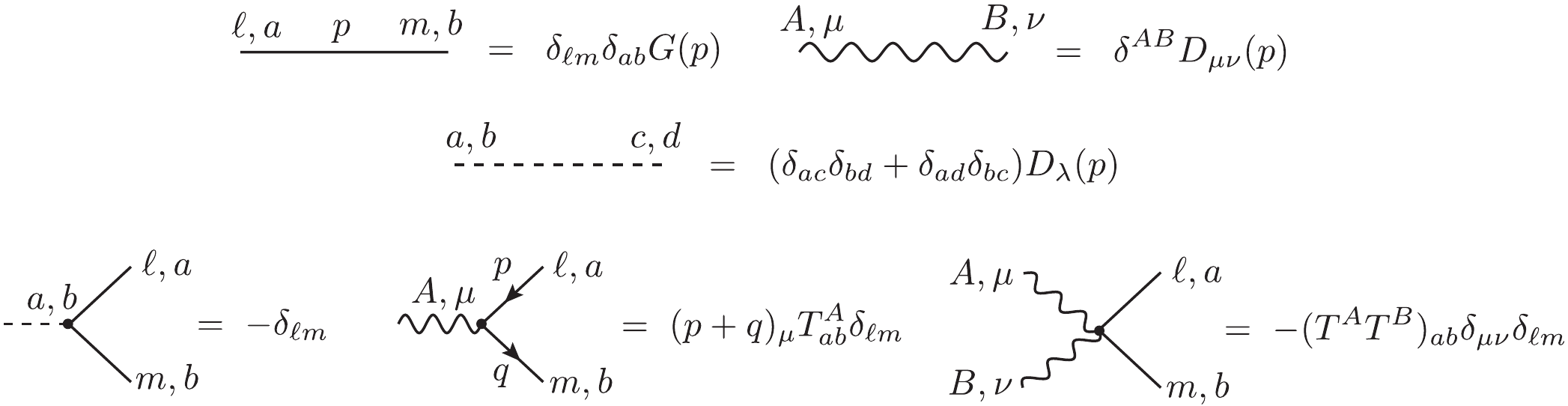}
\caption{\label{feynruleslargen} Feynman rules for the large $N_{h}$ effective action.}
\end{figure}

 The induced gauge and scalar propagators are obtained by inverting the non-local kernels in (\ref{kernln}), so we obtain in momentum space
\begin{align}
&\langle A^{A}_{\mu}(p)A_{\nu}^{B}(-p)\rangle = \delta^{AB}D_{\mu\nu}(p)\,,\quad \langle \lambda_{ab}(p) \lambda_{cd}(-p)\rangle = (\delta_{ac}\delta_{bd}+\delta_{ad}\delta_{cb}) D_{\lambda}(p)\,,
\end{align}
where 
\begin{align}
&D_{\mu\nu}(p) =  \frac{1}{S_{2}N_{h}} \frac{C_{A}}{(p^{2})^{D/2-1+\Delta}}\Big(\delta_{\mu\nu}-(1-\xi)\frac{p_{\mu}p_{\nu}}{p^{2}}\Big), \quad D_{\lambda}(p) =  \frac{1}{N_{h}}\frac{C_{\lambda}}{(p^{2})^{D/2-2+\Delta}} \,, \notag\\
&C_{A} =-2^{D+2} (4\pi) ^{\frac{D-3}{2}}\Gamma \Big(\frac{D+1}{2}\Big) \sin \big(\frac{\pi  D}{2}\big)\,, \quad C_{\lambda} =2^D (4 \pi )^{\frac{D-3}{2}}  \Gamma \Big(\frac{D-1}{2}\Big)\sin \big(\frac{\pi  D}{2}\big)\,,
\end{align}
 and we introduced  an arbitrary gauge parameter  $\xi$ and  the regulator $\Delta$ to handle divergences \cite{Vasiliev:1975mq,Vasiliev:1981yc, Vasiliev:1981dg, Derkachov:1997ch, Ciuchini:1999cv}, which should be sent to zero at the end of the calculation.  We summarize the Feynman rules in Fig.~\ref{feynruleslargen}.

\begin{figure}[h!]
\includegraphics[width=0.20\textwidth]{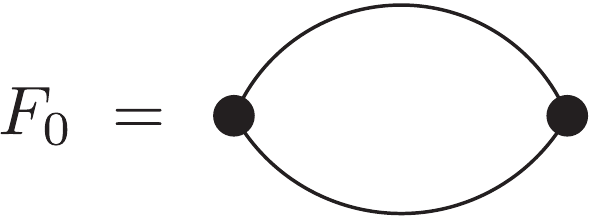}
\caption{\label{D0diag} The leading $N_{h}$ diagram $F_{0}(p)$ contributing to $\langle O^{\textrm{ren}}_{m}(p)  O^{\textrm{ren}}_{m}(-p) \rangle$.}
\end{figure}

\begin{figure}[h!]
\includegraphics[width=0.50\textwidth]{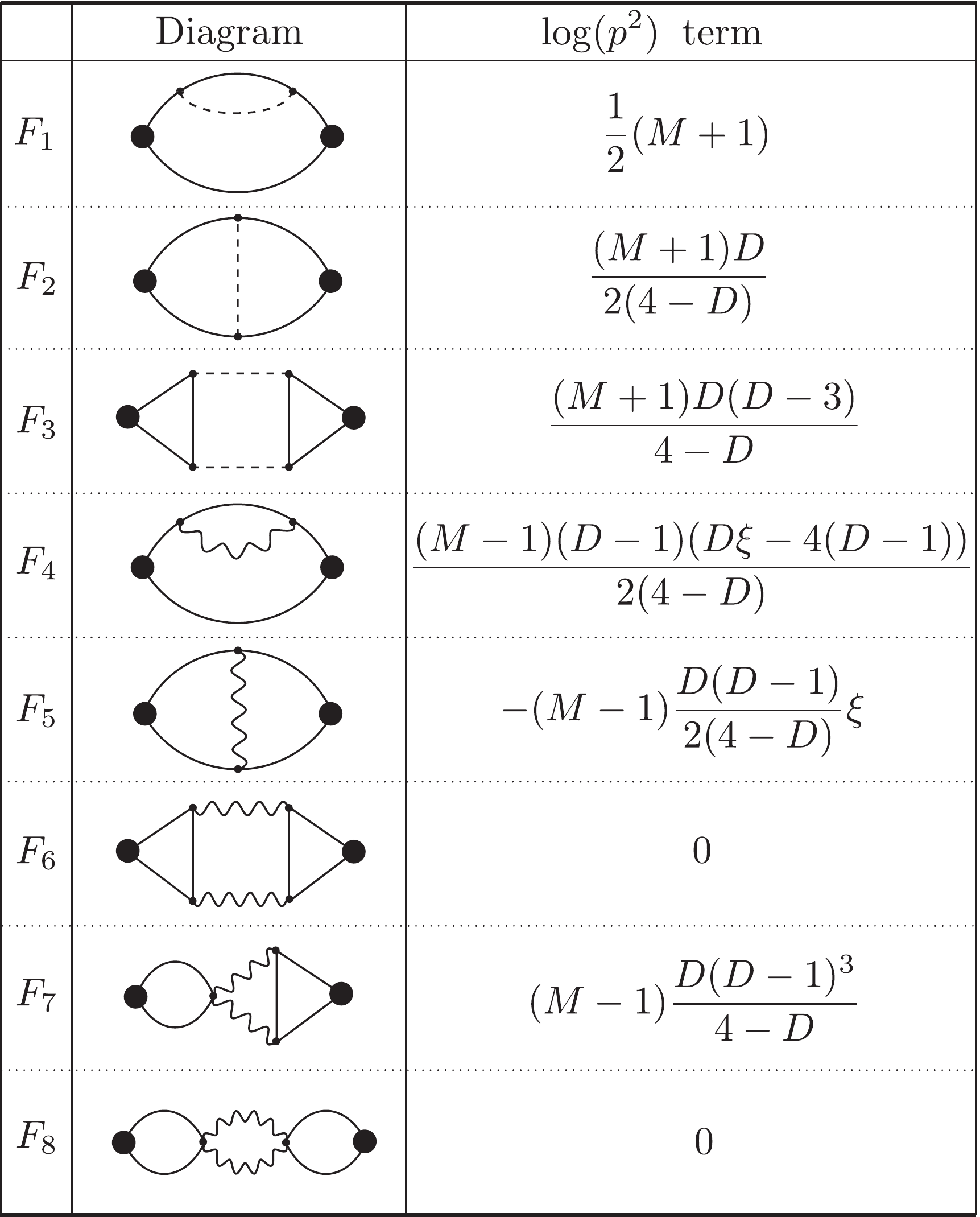}
\caption{\label{feyndiagrams} All Feynman diagrams contributing at $1/N_{h}$ order to $\gamma_{m}$.}
\end{figure}

We consider first the computation of the anomalous dimension of the operator $O_{m}= \frac{1}{2}H_{a\ell}H_{a\ell}$.  Let us define the renormalized operator as $O_{m}^{\textrm{ren}}= Z_{O_{m}}O_{m}$, where $Z_{O_{m}}$ is the renormalization constant. We have in momentum space
\begin{align}
\langle O_{m}^{\textrm{ren}}(p)  O_{m}^{\textrm{ren}}(-p)\rangle = \frac{C_{O_{m}}}{(p^{2})^{2-D/2-\gamma_{m}}}\,,
\end{align}
where the  dimension of the operator $O_{m}^{\textrm{ren}}$ reads $\Delta_{O_{m}}= D-2+\gamma_{m}$, and $\gamma_{m} = \gamma_{m1}/N_{h}+\mathcal{O}(1/N_{h}^{2})$. In order to find $\gamma_{m1}$ we have to calculate the diagrams represented in the table \ref{feyndiagrams}: 
\begin{align}
\langle O_{m}^{\textrm{ren}}(p)  O_{m}^{\textrm{ren}}(-p)\rangle =   Z_{O_{m}}\Big(F_{0}+\sum_{n=1}^{8}F_{n}\Big)\,.
\end{align}
More precisely we have to pick up $\log (p^{2})$ terms in each diagram. In the table \ref{feyndiagrams} we listed results for such terms in each diagram
in the units of $F_{0}\eta_{scal,1}/N_{h}$, where $\eta_{\textrm{scal},1}$ is the coefficient of $1/N$ term in the exponent $\eta$
 of the simple $\sigma$-model \cite{AbeHikami}
 \bea
\eta_{\textrm{scal},1} &=& \frac{2 (D-4) \sin \left(\frac{\pi D}{2}\right) \Gamma (D-1)}{\pi  d (\Gamma (D/2))^2}\nn
&=& \frac{8}{3 \pi^2} \quad, \quad D=3 \nn
&=& \frac{\epsilon^2}{2} + \ldots \quad , \quad D=4-\epsilon \,.
\eea
and $F_{0}$ is the leading one-loop diagram represented in figure \ref{D0diag} and given by 
 \begin{align}
F_{0}(p) = \frac{1}{2}MN_{h} \int \frac{d^{D}q}{(2\pi)^{D}}\frac{1}{q^{2}(p+q)^{2}} =-\frac{MN_{h}}{2 C_{\lambda}} \frac{1}{(p^{2})^{2-D/2}}\,.
\end{align}
\noindent
All diagrams are computed in momentum space using the well-known large $N$ methods (see for example \cite{Diab:2016spb, Giombi:2016fct}).
Also we used that $T^{A}_{ab}T^{A}_{bc}=C_{2}\delta_{bc}$, where $C_{2}=S_{2}(M-1)/2$ for $O(M)$ group.

We find for $\gamma_{m1}$
 \begin{align}
\gamma_{m1}= \eta_{\textrm{scal},1} (F_{1}+\dots +F_{8}) \,,
\end{align}
and here $F_{n}$ are exactly expressions given in the table \ref{feyndiagrams}. 
Therefore we obtain 
 \begin{align}
\gamma_{m1}= \eta_{\textrm{scal},1}\frac{(D-2) (D-1)}{4-D} (2+ (M-1)D^2) \,.
\end{align}
We are interested in the case $M=3$, thus for the exponent $\nu$ we get in $1/N$ order \cite{Hikami:1980qk}
\bea
\frac{1}{\nu} &=& D-2+ \frac{2\eta_{\textrm{scal},1}}{N_{h}}\frac{(D-2) (D-1)}{4-D} (1+ D^2) \nn
&=& 1 + \frac{320}{3 \pi^2 N_h} \,, \quad D=3
\,, \label{nuN}
\eea
and this agrees with Eq.~(\ref{nu}) when $\epsilon$ is small.

In order to find anomalous dimension of the operator $Q_{\ell m}=H_{a\ell}H_{am}-\frac{\delta_{\ell m}}{N}H_{an}H_{an}$
one has to sum only diagrams $F_{1}$, $F_{2}$ and $F_{4}$, $F_{5}$, since all other diagrams are suppressed by $1/N_{h}$ order. Therefore we find 
 \begin{align}
\gamma_{Q1}= \eta_{\textrm{scal},1} (F_{1}+F_{2}+F_{4}+F_{5})  =- \eta_{\textrm{scal},1} \frac{2 (D(D-2)  (M-1)-2)}{4-D}\,,
\end{align}
and thus for $M=3$ we have
 \bea
\Delta_{Q_{\ell m}} &=& D-2 -\frac{\eta_{\textrm{scal},1}}{N_{h}} \frac{4 (D(D-2) -1)}{4-D} \nn
&=& 1 - \frac{64}{3 \pi^2 N_h} \,, \quad D=3 \,. 
\label{dimQN}
\eea
The result in Eq.~(\ref{dimQN}) agrees with Eq.~(\ref{dimQ}) when $\epsilon$ is small.

\section{Impurities and visons}
\label{sec:impvis}
In this section we analyze the spatial texture of the different order parameters defined in \tableref{RepresentationOfSymmetries} in the vicinity of impurities and vortex-like defects in the Higgs field.

\subsection{Impurity-induced order}
\label{sec:imp}
For the analysis of impurities, we will not impose $O(4)$ invariance, {\it i.e.}, relax the assumption $u_1=u_2=u_3$ in the Higgs potential (\ref{HiggsPotential}) which allows to study any of the phases in \figref{MFPhasediagram}. For concreteness, let us focus on phase (C), but the discussion can be easily extended to other phases as well. In a homogeneous system, translational symmetry is preserved in phase (C) and we will analyze how the different translational-symmetry-breaking CDW order parameters emerge locally in the vicinity of a single impurity in an otherwise homogeneous system. 

To this end, we start from the Lagrangian in \equref{FullHiggsAction} with $s<0$ and a combination of $\{u_j\}$ corresponding to phase (C). To parametrize fluctuations of the Higgs field around its saddle-point value in \equref{SaddlePointPhaseC} with $H_0^2 = 2|s|/(4u_0+u_1)$, we first choose a gauge that eliminates $3$ out of the $12$ independent real degrees of freedom and parameterize the remaining $9$ real components, $h_{j}$, $j=0,1,2$ and $H_{a\ell}$, $\ell=3,4$, $a=1,2,3$, according to
\begin{equation}
\vec{\mathcal{H}}_{x} = \frac{1}{\sqrt{2}}\begin{pmatrix} H_0 + h_0 + h_1 & \\ h_2 \\ 0 \end{pmatrix} + \frac{i}{\sqrt{2}} \begin{pmatrix}   h_2 \\ H_0 + h_0 - h_1 \\ 0 \end{pmatrix}, \quad \mathcal{H}_{ay}=H_{a3}+iH_{a4}. \label{ParametrizeHiggs}
\end{equation}
To quadratic order in fields, the Higgs-part of the Lagrangian becomes 
\begin{subequations}
\begin{equation}\begin{split}
\mathcal{L}_{\mathcal{H},\text{Gauss}} = \sum_{j=0}^3\left[ (\partial_\mu h_j)^2 + m_j^2 h_j^2 \right] + \sum_{\ell=3,4} \Bigl[ (\partial_\mu H_{a\ell})^2 + m_{\parallel}^2\left( H_{1\ell}^2+H_{2\ell}^2 \right)  + m_{\perp}^2 H_{3\ell}^2  \Bigr]  
\end{split}\end{equation}
with
\begin{equation}
m^2_0 = 2|s|, \quad m^2_1=m^2_2= 4|s|\frac{u_2}{4u_0 + u_1}, \quad m^2_{\parallel} = 2 |s| \frac{u_3 - u_1}{4u_0 + u_1}, \quad m_{\perp}^2 = 2 |s| \frac{-u_1}{4u_0 + u_1}, \label{MassesPhaseC}
\end{equation}\label{FullQuadraticAction}\end{subequations}
while, as expected from the $\mathbbm{Z}_2$ nature of phase (C), all gauge fields aquire a mass ($m_{A_z}^2=2m_{A_{x}}^2=2m_{A_{y}}^2=H_0^2$). In fact, all bosonic modes are gapped in phase (C) which is physically related to the fact that phase (C) only breaks discrete symmetries and, hence, no Goldstone modes are expected.  It is easily seen that all masses in \equref{MassesPhaseC} are positive in the region of the phase diagram in \figref{MFPhasediagram} where phase (C) emerges. 
A closer comparison with the phase diagram in \figref{MFPhasediagram} reveals that the sign change of $m^2_{1}=m^2_{2}$ is related to the transition (C)$\rightarrow $(A), the sign change of $m^2_\perp$ with (C)$\rightarrow $(F)$_\theta$, and that of $m^2_\parallel$ with (C)$\rightarrow $(D).

We consider an isolated impurity located at the origin $\vec{r}=0$ and invariant under all point symmetries of the system. As is readily seen from \tableref{RepresentationOfSymmetries}, the most general contribution of the impurity to the Lagrangian reads, up to quadratic order in the Higgs fields, as
\begin{align}\begin{split}
    \mathcal{L}_{\text{imp}} = \delta(\vec{r}) \Bigl[ & v_1\left(|\vec{\mathcal{H}}_x|^2 + |\vec{\mathcal{H}}_y|^2\right) + v_2\left(\vec{\mathcal{H}}_x \cdot \vec{\mathcal{H}}_x + \vec{\mathcal{H}}_y \cdot \vec{\mathcal{H}}_y + \text{c.c} \right) \\ & \qquad + v_3 \left( \vec{\mathcal{H}}_x^* \cdot \vec{\mathcal{H}}_y + \vec{\mathcal{H}}_x \cdot \vec{\mathcal{H}}_y + \text{c.c} \right) \Bigr],
\end{split}\end{align}
where $v_j$ are real-valued constants that depend on the microscopic details of the impurity.  

Expressing the different CDW order parameters $\Phi_{x,y}$ and $\Phi_{\pm}$ in terms of the fields in \equref{ParametrizeHiggs} and focusing on the Gaussian fluctuations described by the action in \equref{FullQuadraticAction}, we find 
\begin{subequations}\begin{align}
\braket{\Phi_x(\vec{r})} &\sim \sqrt{\frac{2}{\pi}} v_2H_0^2 \frac{1}{\sqrt{m_1 |\vec{r}|}} e^{-m_1 |\vec{r}|}  \\
\braket{\Phi_y(\vec{r})} &\sim \frac{v_2}{4\pi^{3/2}}\frac{1}{|\vec{r}|^{3/2}}\left(\frac{2}{\sqrt{m_\parallel}}e^{-2 m_\parallel |\vec{r}|} + \frac{1}{\sqrt{m_\perp}}e^{-2 m_\perp |\vec{r}|} \right)  
\\
\braket{\Phi_{+,-}(\vec{r})} &\sim \frac{v_3}{2\sqrt{2\pi}} H_0^2 \frac{1}{\sqrt{m_{\parallel} |\vec{r}|}} e^{-m_{\parallel} |\vec{r}|}
\end{align}\label{NearImpurityBeh}\end{subequations}
in leading (first) order in the impurity strength and in the limit $m_j|\vec{r}| \gg 1$. A few comments on \equref{NearImpurityBeh} are in order. We first note that neither phase (C) nor the impurity itself break the reflection symmetry $\sigma_x$ which is why $\braket{\Phi_+(\vec{r})}=\braket{\Phi_-(\vec{r})}$ is found. Secondly, the qualitatively very different behavior of $\braket{\Phi_x(\vec{r})}$ and $\braket{\Phi_y(\vec{r})}$ is a consequence of the broken $C_4$ rotation symmetry in phase (C) rendering the CDW response with wavevector along $x$ and $y$ inequivalent. We also see that some of the order parameters are induced more strongly than others depending on the value of the masses in \equref{MassesPhaseC}. For instance, if $m_1$ is sufficiently small, phase (C) is expected to be particularly susceptible to impurity-induced CDW order of the type $\Phi_x$. 
This can be understood intuitively from the phase diagram in \figref{MFPhasediagram}: the vanisihing of $m_1$ describes the transition to phase (A) which is characterized by CDW modulations along one of the principle axes. Alternatively, this can be seen in \equref{ParametrizeHiggs} by noticing that fluctuations of $h_1$ induce $\Phi_x$.


\subsection{Visons}
\label{sec:vison}

In the Higgs phases with $\mathbb{Z}_2$ topological order, there are vortex-like solutions corresponding to `visons' \cite{SenthilFisher,NRSS91,Wen91,Bais92}, or the `$m$' particles  of the toric code \cite{Kitaev03}. The visons studied here are physically distinct from those examined earlier \cite{SS92,TSMPAF01}, as they involve gauge rotations
in spin and pseudospin space respectively.
Visons are present both for the Higgs potential in Eq.~(\ref{HiggsPotential}) of Section~\ref{sec:su2}, and also in the O($N_h$) invariant Higgs potential in Eq.~(\ref{HiggsPotential2}). For simplicity, we will only examine these solutions in the O($N_h$) invariant case, although closely related solutions are also present for Eq.~(\ref{HiggsPotential}). We will particularly be interested in the behavior of the $Q_{\ell m}$ order parameters in the presence of such a vortex. Note that the vison has a vortex-like structure in the color SU(2) gauge space, and the gauge-invariant order parameters $Q_{\ell m}$ take a fixed value at a given radial distance from the center of the vortex {\it i.e.\/} there is no rotation of $Q_{\ell m}$ in the flavor O($N_h$) space 
as we go around the center. However, as we move along the radial direction, the $Q_{\ell m}$ can vary, as we will describe below.

For the following discussion it is assumed $u_1>0$, as this is required for the stability of $\mathbb{Z}_2$ topological order and vison solutions. 

\subsubsection{$N_h=3$}
As discussed in Section \ref{sec:mft}, for $N_h \leq 3$ there is no broken symmetry and $\langle Q_{\ell m} \rangle = 0$ in the absence of a vison. It is especially interesting to consider a vison solution for $N_h=3$, and in particular examine the order parameter $\langle Q_{\ell m} \rangle$ in the vicinity of the vison core. Moreover, the results obtained for $N_h=3$ are immediately generalized to any $N_h>3$. We will demonstrate such a generalization.

To begin, we write the Higgs field $H_{bm}$ as an orthogonal matrix where columns correspond to the $N_h$ flavours.  Moreover, we know that the eigenvalues of $(H^0_{bm})$ are triply degenerate and given by $w=\sqrt{|s|/(12u_0)}$ (as per Section \ref{sec:mft}). In the absence of visons, we choose a reference frame whereby,
\begin{align}
\label{H0}
(H^0_{bm})=\left(
\begin{array}{ccc}
 w& 0 & 0 \\
 0& w & 0 \\
 0 & 0 & w\\
\end{array}
\right).
\end{align} 
To account for a vison configuration centered at $r=0$, we demand that the condensate at $r\to\infty$ undergoes a single rotation about the third (color) column of $(H^0_{bm})$, i.e. we add  the rotation matrix $\hat{R}_3(\phi)$, where $\phi$ is the azimuthal angle of the $(x,y)$-plane. This is precisely the winding from the $S^1$ coordinate space to the $S^3/\mathbb{Z}_2$ Higgs order parameter space, which has the homotopy $\pi_1(S^3/\mathbb{Z}_2)=\mathbb{Z}_2$; see Appendix~\ref{app:vison}. Our ansatz for the Higgs field away from the vison core therefore has the following angular dependence,
\begin{align}
\label{Hboundary}
H_{bm}(\phi)&=R_3(\phi)_{bc}H^0_{cm}.
\end{align}
By demanding the vanishing of the covariant derivative at $r\to\infty$, $(\partial_\mu\delta_{ac} -\varepsilon_{abc}A_{b\mu})H_{cm}(\phi)=0$, 
we obtain the asymptotic form of the gauge field (in polar coordinates $\mu=\{t,r,\phi\}$) 
\begin{align}
\label{Aboundary}
\tilde{A}_{1\mu}&=0, \ \tilde{A}_{2\mu}=0, \  \tilde{A}_{3\mu}=\partial_\mu\phi=\frac{1}{r}\delta_{\mu,\phi}.
\end{align}

Having obtained the asymptotic solutions $H_{bm}$ and $\tilde{A}_{b\mu}$ (\ref{Hboundary}) and (\ref{Aboundary}), we now need an ansatz for the solutions for $r\in[0,\infty)$. For the Higgs field we take
\begin{align}
\label{AnsatzH}
H_{bm}(r,\phi)&=H_{bm'}(\phi)(1 + f_{b'}(r))\delta_{bb'}\delta_{mm'}=w\left(
\begin{array}{ccc}
 (1+f_1(r))\cos\phi & -(1+f_2(r))\sin\phi & 0 \\
 (1+f_1(r))\sin\phi&(1+f_2(r))\cos\phi & 0\\
 0& 0 & 1+f_3(r)\\
\end{array}
\right)
\end{align}
where $f_b(r)$ are scalar functions without angular dependence -- this guarantees the boundary conditions. By rotational invariance $f_1(r)=f_2(r)\equiv f(r)$. Now for the gauge field, we consider deviations from the boundary solutions Eq. (\ref{Aboundary}) but maintain the same angular dependence. Explicitly we construct the ansatz 
\begin{align}
\label{AnsatzA}
A_{1\mu}&=\frac{1}{r}h_1(r)\delta_{\mu\phi},\ A_{2\mu}=\frac{1}{r}h_2(r)\delta_{\mu\phi},\
A_{3\mu}=\frac{1}{r}(1+h(r))\delta_{\mu\phi}.
\end{align}
By inspection of the solutions to the field equations, as well as the boundary conditions Eq. (\ref{Aboundary}), we find that $h_1(r)=h_2(r)=0$. Hence the nonabelian gauge field reduces to the single, abelian component $A_{3\mu}$. 

Solutions for the scalar functions, $f(r), f_3(r), h(r)$, must be obtained numerically; we describe such a procedure below. We can, however, anticipate the asymptotic behaviour based on a mean-field analysis of the Lagrangian (\ref{LH2}). At the mean-field level, the Higgs mass is given by $m_H=\sqrt{2|s|}$ and the spontaneously generated gauge boson mass by $m_A=\sqrt{g^2|s|/(6u_0)}=\sqrt{2}gw$. The corresponding length scales, $\xi\sim 1/m_H$ and $\lambda\sim 1/m_A$, set the characteristic scale for the deformations of the Higgs and gauge fields, respectively. Hence we expect $f(r), f_3(r)\sim e^{-r/\xi}$, and $h(r)\sim e^{-r/\lambda}$, far from the vison core. 

To obtain explicit solutions, we use trial functions and minimize the classical energy (obtained from Eq. (\ref{LH2})) with respect to a small set of parameters. A convenient choice is,
\begin{align}
\label{hff3}
h(r)&=\sum_{n=0}^N a_n r^n e^{-a_E r}, \
f(r)=\sum_{n=0}^N b_n r^n e^{-b_E r}, \ 
f_3(r)=\sum_{n=0}^N c_n r^n e^{-c_E r},
\end{align}
where minimization is with respect to the $3N+6$ parameters, $\{a_n, a_E, b_n, b_E, c_n, c_E\}$, and we find sufficient convergence for $N=8$. The number of parameters may be reduced by setting: (i) $b_0=-1$ and (ii) $a_0=-1$. Condition (i) demands that the core expel the in-plane Higgs field. Condition (ii) demands that the flux, $\Phi(r)=2\pi (1+h(r))$, vanishes as $r\to0$. 
\begin{figure}[tb]
\begin{center}
\includegraphics[width=0.45\linewidth]{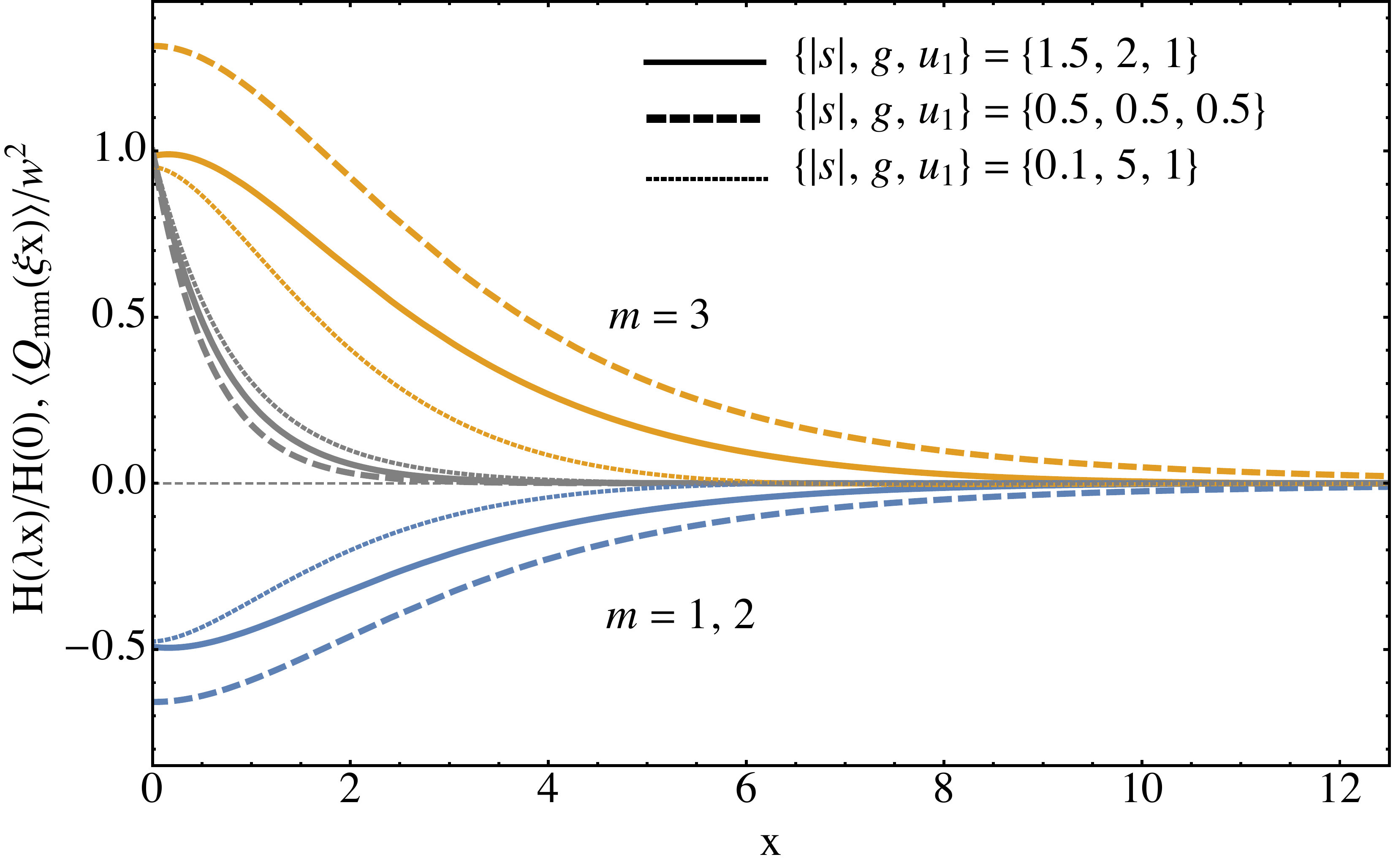}\hspace{0.5cm}
\includegraphics[width=0.45\linewidth]{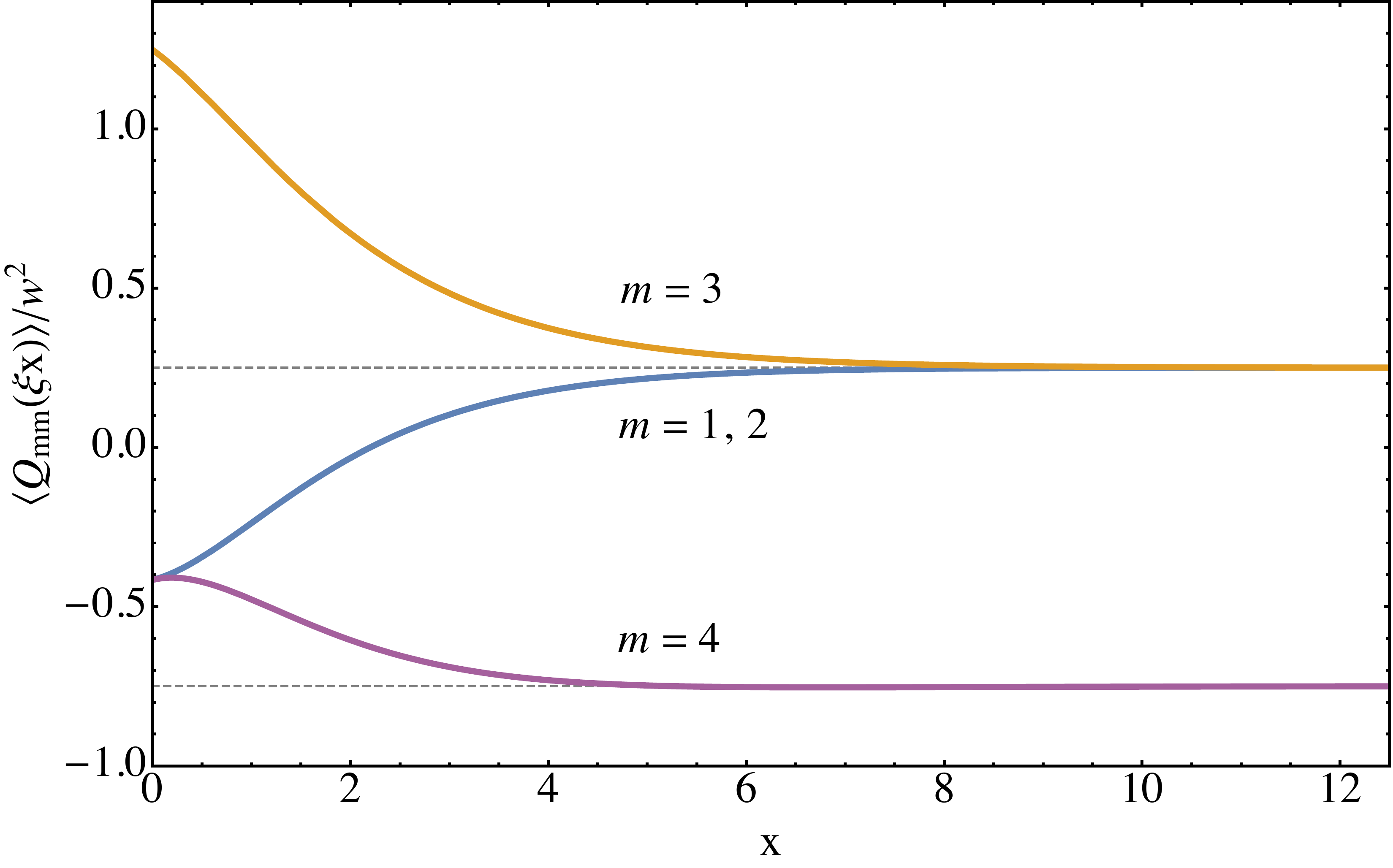}
  \begin{picture}(0,0) 
\put(-465,8){\text{(a)}}
\put(-230,8){\text{(b)}}
\end{picture}
\caption{(a) Nonzero components of the order parameter $Q_{\ell m}$ and magnetic field $H$ in the presence of a vison at the origin $x=0$. Blue lines show $\langle Q_{1 1}\rangle/w^2$, yellow lines show $\langle Q_{3 3}\rangle/w^2$, and grey lines show the normalized magnetic field $H(\lambda x)/H(0)$. $Q_{\ell m}(\xi x)$ and $H(\lambda x)$ have been re-scaled using different length scales $\xi$ and $\lambda$, as discussed in the text. Solutions show various $\{|s|,g,u_1\}$, with fixed $N_h=3, u_0=1$. (b) Nonzero (diagonal) components of the order parameter $Q_{m m}$ for $N_h=4$. The couplings are set to $\{|s|,g,u_0, u_1\}=\{1.5,2,1,1\}$. Recall $w=\sqrt{|s|/(4(3u_0+(1-3/N_h)u_1))}$ is not the same in (a) and (b).} 
\label{fig:visonN3}
\end{center}
\end{figure}
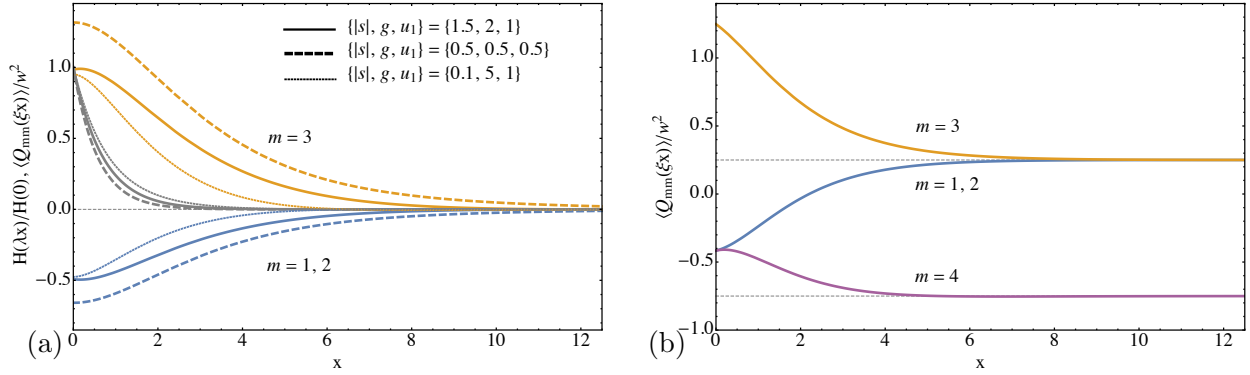

The fields $f(r), f_3(r)$ and $h(r)$ are themselves not gauge invariant. 
From $h(r)$ we can define the gauge invariant magnetic field $H(r)$,
\begin{align}
H(r)&=\frac{1}{r}\frac{d}{dr}(1+h(r)). \label{Hfield}
\end{align}
From $f(r), f_3(r)$, we construct the gauge invariant $Q_{\ell m}$ (CDW) order parameter, which has components,
\begin{align}
\label{CDWN3}
\notag\langle Q_{1 1} \rangle&=\langle Q_{2 2}\rangle=\frac{w^2}{3}(1+f(r))^2-\frac{w^2}{3}(1+f_3(r))^2,\\
\notag\langle Q_{3 3} \rangle&=\frac{2w^2}{3}(1+f_3(r))^2-\frac{2w^2}{3}(1+f(r))^2,\\
\langle Q_{\ell m} \rangle&=0, \ \ \forall \ell\neq m.
\end{align}
Far from the vison core, where $f(r),f_3(r)\to0$, we see from the expression (\ref{CDWN3}) that all components of the $\langle Q_{\ell m} \rangle$ order parameter vanish, as they should. 
Figure \ref{fig:visonN3}(a) plots the nonzero components of the order parameter $\langle Q_{11} \rangle = \langle Q_{22} \rangle$, $\langle Q_{33} \rangle$ and the magnetic field $H(r)$, for three choices of couplings $\{s,u_0,u_1,g\}$.

\subsubsection{$N_h>3$}
For $N_h>3$, the gauge field ansatz Eq. (\ref{Aboundary}) is unaffected, while the Higgs field ansatz Eq. (\ref{Hboundary}) requires the straightforward modification 
\begin{align}
\label{AnsatzH4}
H_{bm}(r,\phi)&=w\left(
\begin{array}{ccc|c}
 (1+f(r))\cos\phi & -(1+f(r))\sin\phi &  0 &\\
 (1+f(r))\sin\phi & (1+f(r))\cos\phi & 0 & \ \  \text{\Large{0}}\\
 0& 0 & 1+f_3(r) & \\
\end{array}
\right),
\end{align}
where now the expectation value reads $w=\sqrt{|s|}/(4(3u_0+(1-3/N_h)u_1))$. The solutions for the scalar functions $f(r), f_3(r), h(r)$, are slightly altered from the $N_h=3$ case due to the change of the expectation value $w$.  
The elements of $\langle Q_{\ell m} \rangle$ change in an important way: For arbitrary $N_h$ they are, 
\begin{align}
\label{CDWN}
\notag\langle Q_{1 1} \rangle&=\langle Q_{2 2}\rangle=w^2\frac{N_h-2}{N_h}(1+f(r))^2-w^2\frac{1}{N_h}(1+f_3(r))^2,\\
\notag\langle Q_{3 3} \rangle&=w^2\frac{N_h-1}{N_h}(1+f_3(r))^2-w^2\frac{2}{N_h}(1+f(r))^2,\\
\notag\langle Q_{\ell \ell} \rangle&=-w^2\frac{2}{N_h}(1+f(r))^2-w^2\frac{1}{N_h}(1+f_3(r))^2, \ \ \ \forall \ell=4,...,N_h\\
\langle Q_{\ell m} \rangle&=0, \ \ \ \forall \ell\neq m,
\end{align}
which shows that even far from the vison core, the diagonal elements are non-vanishing. The components are plotted in Figure \ref{fig:visonN3}(b) for the case $N_h=4$.

\begin{figure}[tb]
\begin{center}
\includegraphics[width=0.4325\linewidth]{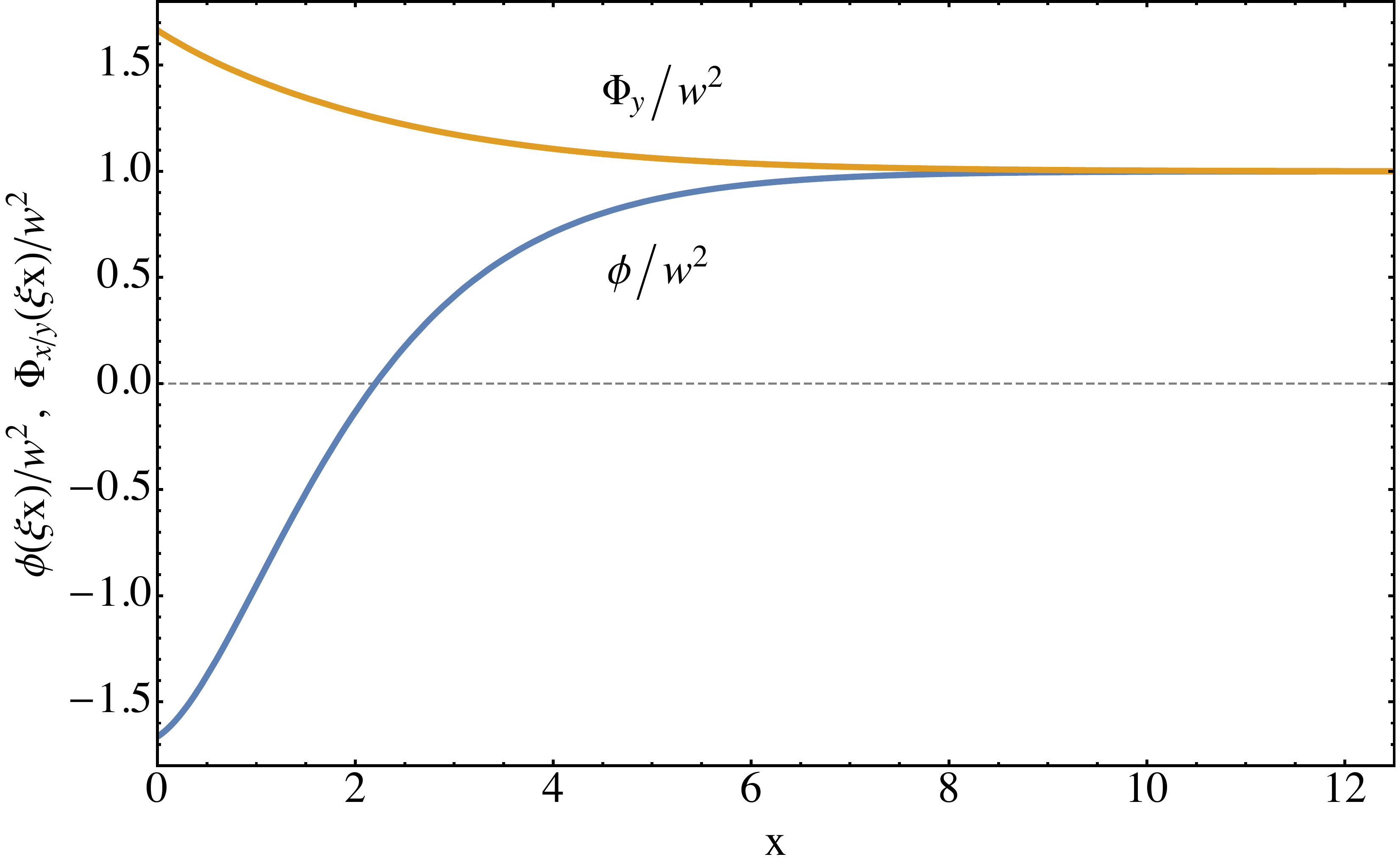}\hspace{0.5cm}
\includegraphics[width=0.431\linewidth]{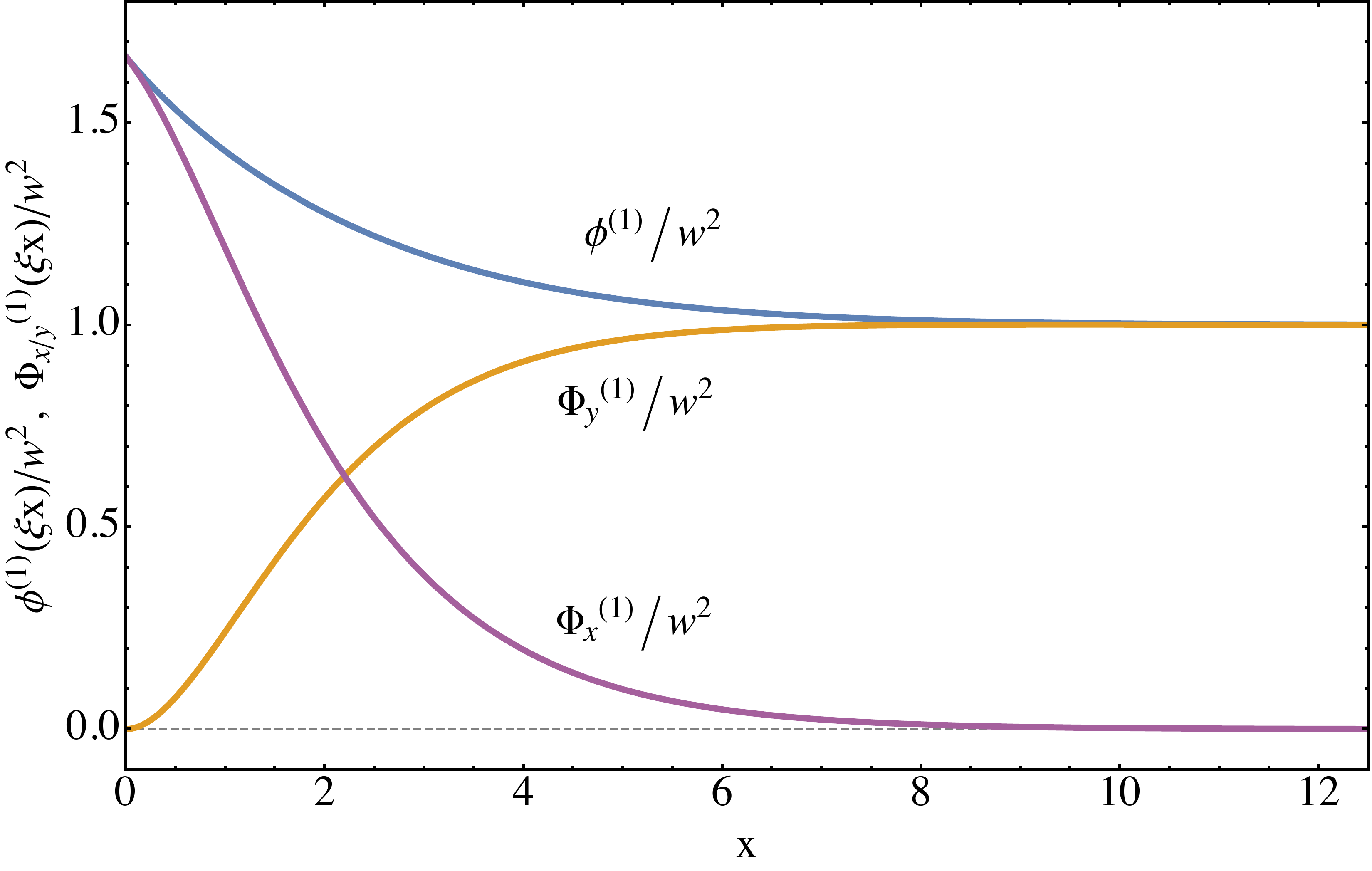}
\includegraphics[width=0.4325\linewidth]{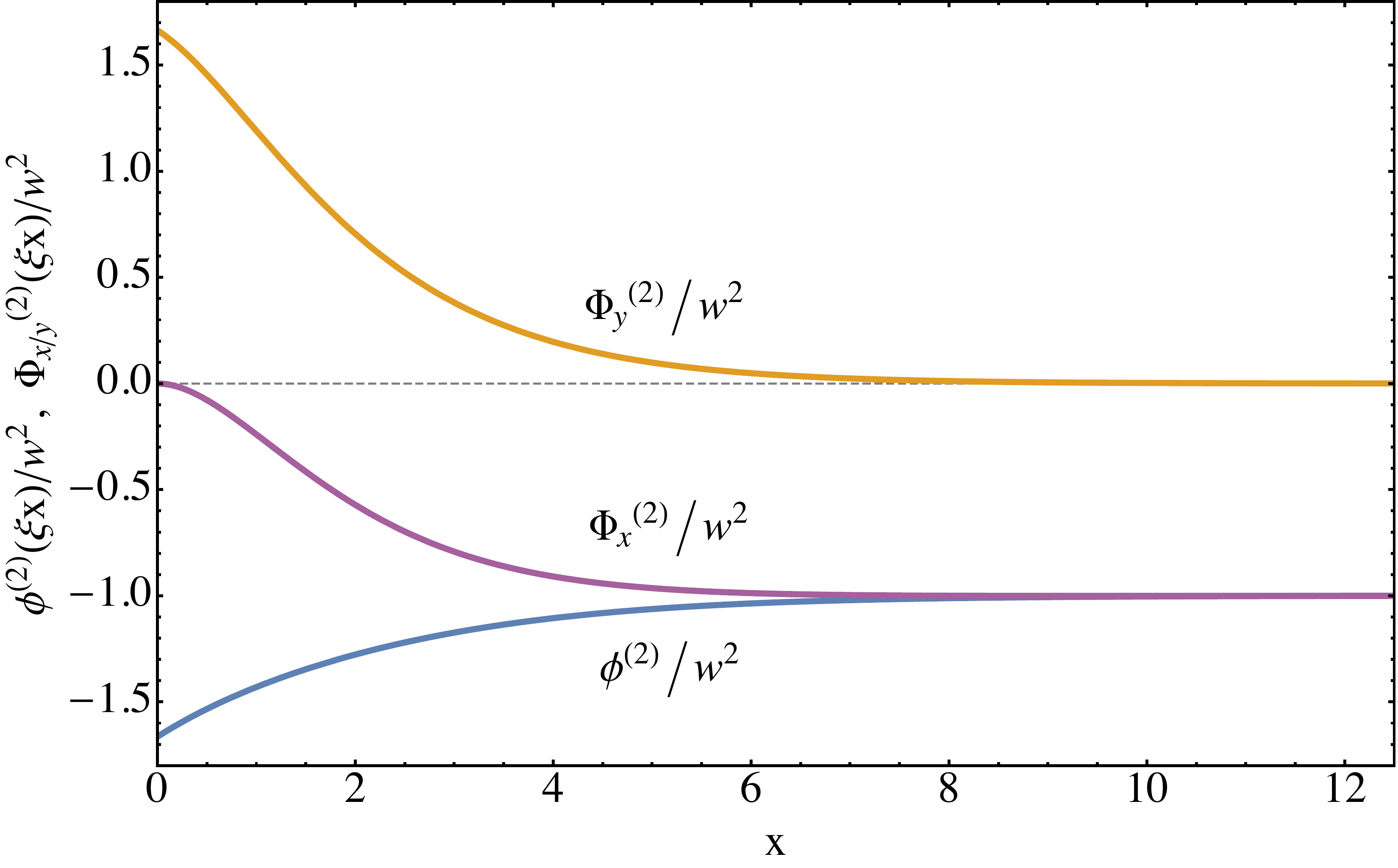}\hspace{0.5cm}
\includegraphics[width=0.4325\linewidth]{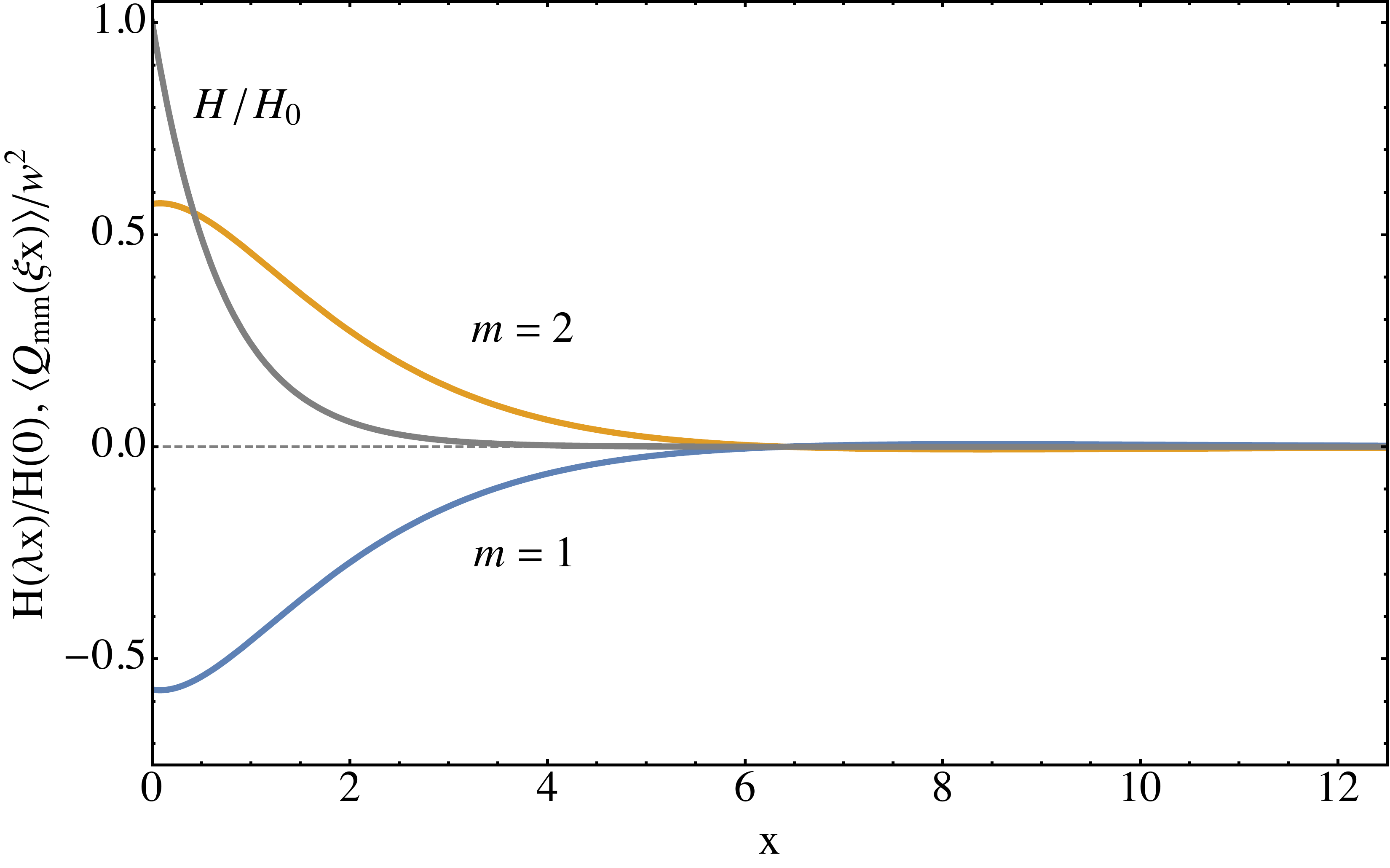}
  \begin{picture}(0,0) 
\put(-458,150){\text{(a)}}
\put(-458,15){\text{(c)}}  
\put(-225,150){\text{(b)}}
\put(-225,15){\text{(d)}}  
\end{picture}
\caption{Order parameter components in the presence of a vison at the origin $x=0$. (a), (b), and (c) consider the order parameters $\phi$, $\Phi_x$ and $\Phi_y$, with $N_h=4$. (a) corresponds to Eq. (\ref{phiQ}), while (b) and (c) correspond to the permutations $R_{1} : Q_{11}\Leftrightarrow Q_{33}$, and $R_{2} : Q_{11}\Leftrightarrow Q_{44}$, which are described in the main text. (d) Considers $N_h=2$, and shows the magnetic field $H(\lambda x)/H_0$ (grey line) and the nonzero order parameter components $\langle Q_{mm}(\xi x)\rangle$ with $m=1,2$, as defined in Eq. (\ref{CDWN2}). In all figures the couplings are set to $\{|s|,g,u_0, u_1\}=\{1.5,2,1,1\}$.} 
\label{fig:RQR}
\end{center}
\end{figure}
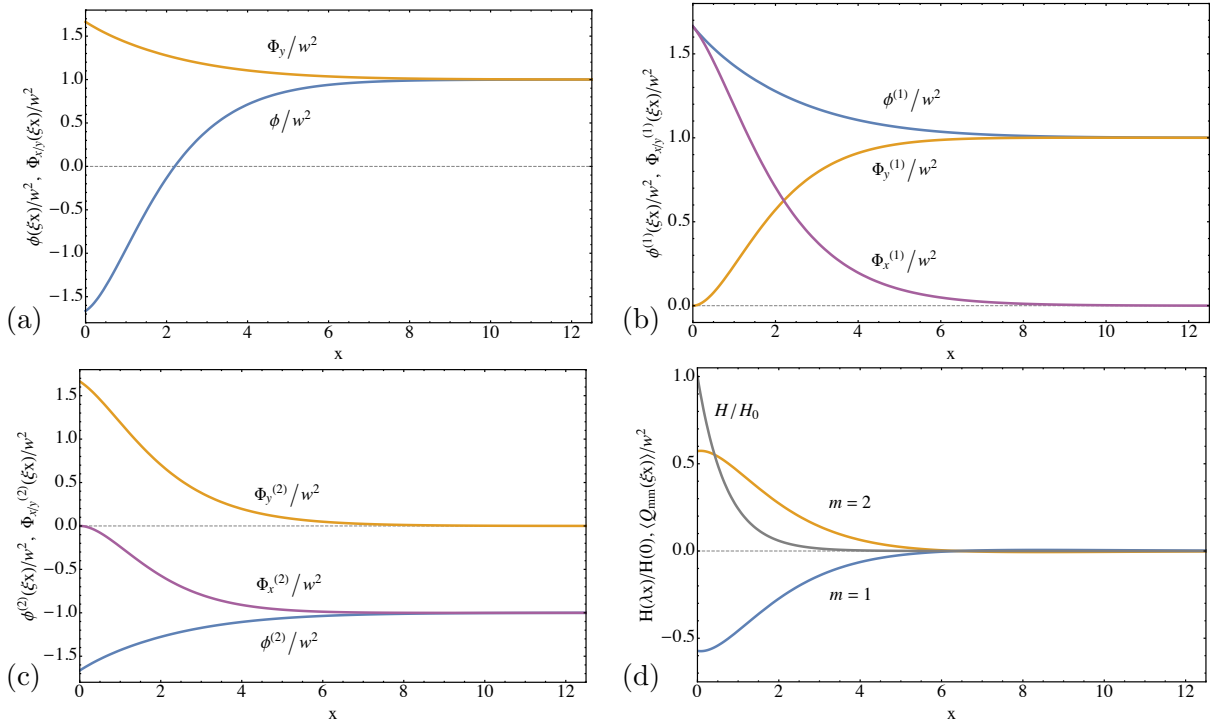

For $N_h=4$, the above solution is of the type in phase $(F)_\theta$ in Fig.~\ref{MFPhasediagram} at all $r$, as is evident from
comparison between Eqs.~(\ref{DefinitionOfFTheta}) and (\ref{AnsatzH4}). We can also compute the $r$  dependence of the order parameters
in Section~\ref{sec:phase} for the non-zero $Q$'s in 
Eq.~(\ref{CDWN}): we use Eqs.~(\ref{ScalarHiggsChriality}), (\ref{HReIm}), and (\ref{defQlm}) to identify
\bea
\phi &=& Q_{11} + Q_{22} - Q_{33} - Q_{44} \nn
\Phi_x &=& Q_{11} - Q_{22} \nn
\Phi_y &=& Q_{33} -Q_{44}\,. \label{phiQ}
\eea
From the solution in Fig.~\ref{fig:visonN3}(b)
we see that only $\phi$ and $\Phi_y$ are non-zero, and we plot their $r$ dependence in Fig.~\ref{fig:RQR}(a).
Far from the vortex, $Q \propto \mbox{diag}(1,1,1,-3)$, and so both $\phi$ and $\Phi_y$ are non-zero.
At the core of the vortex, $Q \propto \mbox{diag}(-1,-1,3,-1)$, and again $\phi$ and $\Phi_y$ are non-zero, but the Ising-nematic order has changed sign.

It is also interesting to consider the transformations of the order parameters $\phi$, $\Phi_x$ and $\Phi_y$, when the diagonal elements of $Q$ are permuted. A set of six orthogonal matrices $\{R_{i} : i=1,...,6\}$ perform all possible pairwise permutations of the diagonal elements of $Q$, via,
\begin{eqnarray}
Q^{R_i}=R_i^TQR_i \label{RQR}. 
\end{eqnarray}
For example, we may define $R_{1}$ as the matrix which acts to exchange $Q_{11}\Leftrightarrow Q_{33}$. Noting that $Q_{11}=Q_{22}$, we now consider all possible permutations. To this end, it is sufficient to identify three distinct combinations: the first is the original basis (\ref{phiQ}); the second is obtained from (\ref{phiQ}) via the transformation $R_{1} : Q_{11}\Leftrightarrow Q_{33}$, and is denote as $\{\phi^{(1)},\Phi_x^{(1)},\Phi_y^{(1)}\}$;  
while the third is obtained from (\ref{phiQ}) via $R_{2} : Q_{11}\Leftrightarrow Q_{44}$, and is denoted $\{\phi^{(2)},\Phi_x^{(2)},\Phi_y^{(2)}\}$.
All other pairwise transformations $R_i$, along with products thereof, can be understood in terms of these three sets. 
Explicitly, the 24 unique combinations are,
\begin{eqnarray}
\pm\phi&,&  \ \pm\Phi_x, \  \pm\Phi_y\nn
\pm\phi^{(1)}&,&\pm\Phi_x^{(1)}, \pm\Phi_y^{(1)}\nn
\pm\phi^{(2)}&,& \pm\Phi_x^{(2)}, \pm\Phi_y^{(2)}. \label{phiQcomb}
\end{eqnarray}
where the sign of each order parameter can be varied independently of the others.
Figure \ref{fig:RQR} plots the set of order parameters (\ref{phiQ}), as well as the two permutations $\{\phi^{(1)},\Phi_x^{(1)},\Phi_y^{(1)}\}$, $\{\phi^{(2)},\Phi_x^{(2)},\Phi_y^{(2)}\}$, from which all other permutations can be immediately understood.

\subsubsection{$N_h<3$}
Only for $N_h>1$ can vison solutions be found; let us consider $N_h=2$. Now there are two distinct topological vison solutions -- the {\it in-plane vison} and the {\it out-of-plane vison} -- which correspond to two distinct mappings to $S^3/{\mathbb Z}_2$, see Eq. (\ref{s3map}) and related discussion. 

Let us orient the Higgs vectors such that the third component lies perpendicular to the two-dimensional configuration space plane. Then, the in-plane vison follows from a simple modification of Higgs field ansatz Eq. (\ref{Hboundary}), which now reads
\begin{align}
\label{H2inplane}
(H_{bm})(r,\phi)&=w\left(
\begin{array}{cc}
 (1+f(r))\cos\phi & -(1+f(r))\sin\phi\\
 (1+f(r))\sin\phi&(1+f(r))\cos\phi\\
 0& 0 \\
\end{array}
\right).
\end{align}
In this case the nonzero eigenvalues of $H_{bm}$ are degenerate, and as a result there are no nonzero elements of the order parameter $\langle Q_{\ell m} \rangle$; we do not pursue explicit solutions for $f(r)$. 

The out-of-plane vison induces the Higgs field,
\begin{align}
\label{H2outplane}
(H_{bm})(r,\phi)&=w\left(
\begin{array}{cc}
 (1+f_1(r))\cos\phi & 0\\
 (1+f_1(r))\sin\phi&0\\
 0& 1+f_2(r) \\
\end{array}
\right),
\end{align}
which subsequently hosts non-degenerate eigenvalues. In this case, the corresponding $\langle Q_{\ell m} \rangle$ order parameter has nonzero components,
\begin{align}
\label{CDWN2}
\langle Q_{1 1} \rangle&=-\langle Q_{2 2}\rangle=\frac{w^2}{2}(1+f_1(r))^2-\frac{w^2}{2}(1+f_2(r))^2.
\end{align}
The gauge field ansatz (\ref{AnsatzA}) remains the same (with $h_1(r)=h_2(r)=0$), and we find solutions for $f_1(r), f_2(r), h(r)$ in the presence of the out-of-plane vison. Considering just gauge invariant quantities, Figure \ref{fig:RQR}(d) plots the order components (\ref{CDWN2}), as well as the magnetic field induced by the gauge field $h(r)$ c.f. Eq. (\ref{Hfield}), for a particular choice of coupling constants.

\section{Conclusions}
\label{sec:conc}

The main claim of our paper (see Fig.~\ref{fig:phasediag_h}) is that near optimal doping, the hole-doped cuprates are described by the SU(2) gauge theory $\mathcal{L}_\mathcal{H}$ in Eq.~(\ref{FullHiggsAction}), coupled via an Ising-nematic order to a large Fermi surface of electrons, as described by $\mathcal{L}_c + \mathcal{L}_\phi$ in Eqs.~(\ref{BareElectronLagr}), (\ref{Lphi}), and (\ref{ScalarHiggsChriality}). The Higgs phase of this theory describes the pseudogap regime, and it has interesting broken symmetries, defect structure, and electronic spectrum, as we review below. Then we discuss the quantum criticality of the Higgs-confinement transition.

\subsection{Broken symmetry in the pseudogap}

We found interesting varieties of broken symmetry in our Higgs phase for the pseudogap state of the hole-doped cuprates. These phases are summarized in Fig.~\ref{MFPhasediagram}, and they involve one or more of Ising-nematic, CDW, and spatially modulated scalar spin chirality orders. Interestingly, these are precisely the orders that have been observed in a variety of experiments over a number of decades (with the exception of the pair density wave \cite{SeamusPDW}, which will be discussed in a forthcoming paper).
In particular, the state $(F)_\theta$ is an excellent fit to many observations, with uni-directional CDW order co-existing with time-reversal breaking, spatially modulated, scalar spin chirality: this combination arises from collinear incommensurate SDW correlations along the CDW direction, and spiral SDW correlations in the other spatial direction
(but note that, as in all of the phases, there is no long range SDW order in either spatial direction).
The phases in Fig.~\ref{MFPhasediagram} are also labeled by $\mathbb{Z}_2$ or U(1): this identifies the residual gauge invariance in the presence of the Higgs condensate. The states labeled $\mathbb{Z}_2$ (including $(F)_\theta$) have $\mathbb{Z}_2$ topological order, and so their FL* states are stable, and they have gapped spinon, chargon, and vison excitations. The states labeled U(1) confine at the longest scales, and so are formally characterized
only by the broken symmetry.

\subsection{Defects}

An important feature of our theory is that all 6 order parameters in Eq.~(\ref{ScalarHiggsChriality}) are essentially equivalent at the quadratic order in the Higgs potential in Eq.~(\ref{HiggsPotential}).
The preferred order parameter(s) are selected only by the quartic terms, and the selection depends sensitively upon the values of the couplings $u_{1,2,3}$, as was shown in Fig.~\ref{MFPhasediagram}. A consequence of this near-equivalence is that an external perturbation from an impurity can induce appearance of a secondary order over significant length scales: this was discussed in Section~\ref{sec:imp}. In Section~\ref{sec:vison} we showed that the topological vison excitations are associated with interesting order parameter profiles (as we noted in Section~\ref{sec:intro}, these visons are distinct from those searched for in earlier experiments \cite{Moler01,Moler02}). Far from the vison core, the dominant order parameter has a uniform condensate. However, within the vortex core, the dominant order parameter can acquire significant radial dependence, and secondary order parameters can appear. Fig.~\ref{fig:RQR} shows sample radial evolutions of the Ising-nematic and CDW orders. The STM observations of Ref.~\onlinecite{Mesaros11} showed correlations between the defects of these orders, and it would be worthwhile to re-examine these results using the framework of the theory in Eq.~(\ref{FullHiggsAction}).

\subsection{Electronic spectrum of the pseudogap}
The nature of the electronic spectrum in the pseudogap was discussed in Section~\ref{sec:flstar}, where we identified it with a fractionlized Fermi liquid (FL*), with large length scale confinement in the U(1) cases. The preferred method of the spectrum computation remains that discussed in other recent papers \cite{WSCSGF,SCWFGS}, which compared favorably with numerics on the Hubbard model, and with photoemission observations on the electron-doped cuprates \cite{Shen18}. These methods include a fractionalization of the electron into spinons and chargons at intermediate length scales in an algebraic charge liquid (ACL) regime (see Fig.~\ref{fig:nambu}). The subsequent binding of the chargons and spinons into electron-like quasiparticles around small Fermi surfaces in a FL* state requires a technically demanding computation, which was only partially 
carried out in Ref.~\onlinecite{SBCS16}; but we note evidence from ultracold atom studies that such bound states do form \cite{Greiner18a,Bloch18,Greiner18b}. Section~\ref{sec:flstar} presented an alternative computation which dispenses with the intermediate scale fractionalization of the electron. 
This is similar in spirit to the `spinon-dopon' approach \cite{Ribeiro06,Punk12}, but it is not clear whether the
suppression of the ACL physics can be quantitatively reliable. The electron-spinon interaction has to be treated in a strong-coupling regime, and in Section~\ref{sec:flstar} we limited ourselves to perturbative considerations. Such computations gave reasonable evidence for the existence of the FL* spectrum of earlier work \cite{Punk15,WSCSGF,SCWFGS}.

\subsection{Quantum criticality}

 We summarize our current understanding of the critical properties of the SU(2) gauge theory in Eqs.~(\ref{FullHiggsAction}), (\ref{BareElectronLagr}), and (\ref{Lphi}) by first describing simpler cases.

\subsubsection{No gauge field, no coupling to fermions}
\label{nogaugefield}
The theory $\mathcal{L}_\mathcal{H}$ in Eq.~(\ref{FullHiggsAction}) with $\vec{A}_\mu = 0$ has been studied in some detail by De Prato {\it et al.\/} \cite{Vicari06}. They employed two different RG schemes to five and six loops, and found that the critical fixed point had the O($N_h=4$) invariant form of Eqs.~(\ref{LH2}) and (\ref{HiggsPotential2}), with $u_1=u_2=u_3$. 
Moreover, the fixed point coupling $u_1 < 0$, which favors collinear spin density wave ordering. We note, however, that large $N_h$ analysis of this case yields $u_1 >0$ at the fixed point, and so there is uncertainty in the sign of $u_1$.

\subsubsection{No coupling to fermions, O(4) symmetry:} 
\label{noFermisurface}
The O(4)-invariant theory $\mathcal{L}_H$ in Eq.~(\ref{LH2}) (this equals $\mathcal{L}_\mathcal{H}$ in Eq.~(\ref{FullHiggsAction}) when
$u_1=u_2=u_3$), with a dynamic SU(2) gauge field $\vec{A}_\mu$, has been studied in several previous works \cite{Hikami:1980qk,Vasiliev:1984bf,Vicari01}, and we reviewed and extended the results to obtain the scaling dimension of the gauge-invariant $Q_{\ell m}$ order parameter in Eq.~(\ref{defQlm}). The different components of this order parameter correspond to the Ising-nematic and CDW orders defined in Eq.~(\ref{ScalarHiggsChriality}) and Table~\ref{RepresentationOfSymmetries}.
We described both the small $\epsilon$ and large $N_h$ expansions, and a suitable fixed-point was obtained only at sufficiently large $N_h$. 

A key property of the critical theory is the sign of the fixed point value of $u_1$. These expansions imply a fixed point value $u_1 > 0$ at large $N_h$, see Eq.~(\ref{FP}). A $u_1>0$ fixed point was also found at large $N_h$ without a gauge field (see Section~\ref{nogaugefield}), and indeed, at leading order in $1/N_h$, the gauge field fluctuations play no role in determining the sign of $u_1$. But, as we noted in Section~\ref{nogaugefield}, five/six loop computations without the gauge field \cite{Vicari06} yielded the opposite sign, $u_1 < 0$.
On the other hand, evidence for a $u_1>0$ fixed point appeared in the Monte Carlo study of Ref.~\onlinecite{Snir18}, which argued for a critical point described by a SU(2) gauge theory at $N_h=3$, coupled also to massless Dirac fermions carrying SU(2) gauge charges. 

As is evident from Fig.~\ref{MFPhasediagram} and Section~\ref{sec:mft}, the sign of $u_1=u_2=u_3$ plays a crucial role in determining the nature of the Higgs phase. For $u_1 < 0$, the Higgs field breaks SU(2) down to U(1), and then the U(1) theory confines with the appearance of CDW and associated Ising/nematic order. So the Higgs phase is ultimately a conventional phase with broken translational symmetry. As the confining phase is also conventional, traditional phase transition theory would state that the critical point should be described by fluctuations of the CDW order parameter. Instead, the critical point could be described by a deconfined SU(2) gauge theory of fluctuating SDW order alone, realizing the novel scenario of `multi-universality' \cite{BiSenthil}.

For $u_1 > 0$, the Higgs phase has $\mathbb{Z}_2$ topological order, accompanied by broken symmetries. Now the large $N_h$ phase transition  is described by deconfined criticality, as expected.

\subsubsection{General case}

We analyzed the influence of the fermions in Section~\ref{sec:CouplingToElectrons}, considering both the cases when the electrons form a large Fermi surface Fermi liquid, and a $d$-wave superconductor with 4 nodal points. The simplest allowed couplings are quartic in the Higgs fields and electrons, and the more disruptive Yukawa-type couplings are not permitted by gauge invariance. In many cases, the quartic couplings are expected to be irrelevant, including those when the electrons form a $d$-wave superconductor. The most significant coupling is that of the large Fermi surface to the Ising-nematic order in Eq.~(\ref{Sd1}) in the metallic case. We estimated its scaling dimension by the large $N_h$ methods reviewed in Section~\ref{noFermisurface}, and found that it is relevant for large $N_h$. The status at $N_h=4$ remains uncertain, given the discussion on the reliability of large $N_h$ results in Section~\ref{nogaugefield}. The ultimate fate of the critical theory in the presence of a relevant coupling to the Fermi surface is not understood, and it would be useful to investigate this question with Monte Carlo simulations.

\subsection*{Acknowledgements}

We thank Zhen Bi, D. Chowdhury,  A.~V.~Chubukov,  J.~C.~Seamus Davis, E. ~Demler, E.~Fradkin, A.~Georges, M.~Greiner, N.~Karthik, Eun-Ah Kim, S.~Kivelson, C.~P\'epin, T.~Senthil, E.~Vicari, A.~Vishwanath, L.~Taillefer, C.~Xu, and J.~Zaanen for valuable discussions.
This research was supported by the National Science Foundation under Grant No. DMR-1664842. Research at Perimeter Institute is supported by the Government of Canada through Industry Canada and by the Province of Ontario through the Ministry of Research and Innovation. SS also acknowledges support from Cenovus Energy at Perimeter Institute. 
MS acknowledges support from the German National Academy of Sciences Leopoldina through grant LPDS 2016-12. HS acknowledges support from the Australian-American Fulbright Commission. G.T.  acknowledges support from the MURI grant W911NF-14-1-0003 from ARO and by DOE grant de-sc0007870.

\appendix

\section{Monopoles}
\label{app:monopole}

In the phases labeled by U(1) in Fig.~\ref{MFPhasediagram}, for the electron-doped case in Section~\ref{sec:electron}, and for the $u_1 <0$ case in the models with O($N_h$) symmetry in Section~\ref{sec:mft}, the condensation of the Higgs field leaves a residual U(1) gauge invariance. 
In these cases there are 't Hooft-Polyakov monopole saddle points \cite{tH74,Polyakov74} of the action, whose proliferation drives confinement. We recall these monopole solutions here for completeness. 

In gauge theories of insulating antiferromagnets \cite{NRSS90,FSX11}, monopole Berry phases play an important role in determining the symmetry breaking in the resulting confining state. In our present study, the fermions are gauge neutral, and their coupling to the monopoles is not via the gauge or Higgs fields, and so it seems reasonable to ignore their Berry phases. In the electron-doped case near optimal doping, 
confinement is expected to restore the large Fermi surface, as in the Landau-damped spin fluctuation computations \cite{Tremblay97,Eremin01,Chubukov10}, and not be associated with any symmetry breaking.

We will limit our monopole analysis to cases in which we can perform a global flavor rotation to a configuration in which $H_{a \ell}$ is non-zero only for a single $\ell$ value, say $\ell = 1$. It is possible that other values of $\ell$ become non-zero near the monopole core, but we will ignore this effect here: including these values will only modify the radial dependence in the core of $f(r)$ and $k(r)$ below. It is convenient to examine the monopole saddle point
of the Lagrangian in Eq.~(\ref{LH2}) after rescaling the fields and length scales via (with $s<0$)
\bea
H_{a\ell} (x) &= & \delta_{\ell 1} \left( \frac{|s|}{4(u_0 + u_1 (N_h-1)/N_h)} \right)^{1/2} h_a (r) \nn
x_\mu &=& \frac{r_\mu}{\sqrt{|s|}} \nn
A_{a\mu} (x) &=& \sqrt{|s|} \, \alpha_{a\mu} (r) 
\eea
Then the action associated with $\mathcal{L}_H$ with spacetime co-ordinate $r_\mu$ is (after subtracting the action of the vacuum)
\bea
\mathcal{S}_H &=& \left[ \frac{\sqrt{|s|}}{4(u_0 + u_1 (N_h-1)/N_h)} \right] \int d^3 r \left( \frac{1}{2} \left( \partial_\mu h_{a} - \varepsilon_{abc} \alpha_{b\mu}
h_{c} \right)^2 + \frac{1}{4} (h_a h_a - 1)^2 \right. \nn
&~&~~~~~~~~~~~~~~~~~~~~~~~~~+\left. \frac{1}{4 \overline{g}^2} \left(\partial_\mu \alpha_{a \nu} - \partial_\nu \alpha_{a \mu} - \varepsilon_{abc} \alpha_{b\mu} \alpha_{c \nu}\right)^2 \right)\,.
\eea
Now the action is characterized by a single dimensionless parameter, $\overline{g}$, which is given by
\beq
\frac{1}{\overline{g}^2} = \frac{4(u_0 + u_1 (N_h-1)/N_h)}{g^2}
\eeq

An important observation can already be made: the only dependence upon $s$, the distance from the critical point, is in the prefactor of the action. So the monopole action $\sim \sqrt{-s}$, and this can become large deep in the Higgs phase, leading to an exponential suppression of the monopole density $\sim \exp(- C \sqrt{-s})$, for some $s$-independent constant $C$.

We now examine saddle points of $\mathcal{S}_H$ of the form
\bea
h_a &=& \frac{r_a}{r} f(r) \nn
\alpha_{b\mu} &=& \varepsilon_{\mu b d} \frac{r_d}{r^2} (1-k(r)) 
\eea
where $r = (r_\mu r_\mu)^{1/2}$. If we choose $f(r \rightarrow \infty) = 1$ and $k(r \rightarrow \infty) = 0$, then the $1/r$ term in the covariant gradient of the Higgs field vanishes as $r \rightarrow \infty$, and we can obtain a monopole saddle point with a finite action. The functions $f(r)$ and $k(r)$ are obtained by minimizing the functional
\beq
I[f,k] = \int_0^\infty r^2 dr \left[ \frac{1}{2} \left( \frac{df}{dr} \right)^2 + \frac{f^2 k^2}{r^2} + \frac{(f^2-1)^2}{4} + \frac{1}{ \overline{g}^2} \left\{ \frac{1}{r^2} \left( \frac{dk}{dr} \right)^2 + \frac{(1-k^2)^2}{2 r^4} \right\} \right] \,.
\eeq
Numerical minimization of $I[f,k]$ then yields a monopole solution with a finite $I$ which depends upon $\overline{g}$.
Consequently the action $\mathcal{S}_H \sim \sqrt{-s}$.

The monopole is associated with a SU(2) gauge flux which obeys
\beq
B_{a\lambda} \equiv \frac{1}{2} \varepsilon_{\lambda \mu\nu} F_{a\mu\nu} = \frac{x_a x_\lambda}{x^4} \quad , \quad x \rightarrow \infty \,.
\eeq
The Higgs condensate reduces the effective gauge theory to U(1), and the corresponding U(1) gauge flux is \cite{tH74,Dunne00}
\bea
B_\lambda &=& \frac{1}{2} \varepsilon_{\lambda\mu\nu} \left( \frac{h_a F_{a \mu\nu}}{h}  - \frac{1}{h^3} \varepsilon_{abc} h_a
(\partial_\mu h_b - \varepsilon_{bde} A_{d\mu} h_e)(\partial_\nu h_c - \varepsilon_{cfg} A_{f\mu} h_g) \right) \nn
&=& \frac{x_\lambda}{x^3} \quad , \quad x \rightarrow \infty \,.
\eea
So the total U(1) gauge flux in this monopole solution is $4 \pi$. In the present reduction from SU(2) to U(1), the spinons, $z$, in Eq.~(\ref{eq:Rz}) transform under a SU(2) fundamental, and consequently they have U(1) charge 1/2.
In the usual compact U(1) gauge theories of interacting spinons \cite{senthil1,senthil2}, the spinons have U(1) charge 1, and a U(1) monopole with gauge flux $2\pi$. Therefore, after appropriate rescaling of the U(1) gauge field, the present SU(2) 't Hooft-Polyakov monopole does indeed reduce to the smallest U(1) Polyakov monopole of compact U(1) gauge theories of interacting spinons \cite{ATSS18}.

\section{Vison topology}
\label{app:vison}

A vison is defined as a configuration of two orthogonal unit vectors $n_{1a}$ and $n_{2a}$ in the SU(2) gauge space. We can parameterize these two vectors in terms of a complex bosonic unit spinor $(z_1, z_2)$ via 
\beq
n_{1a} + i n_{2a} = \varepsilon_{\alpha \gamma} z_\gamma \sigma^a_{\alpha \beta} z_\beta\,, \label{s3map}
\eeq
where $\sigma^a$ are the Pauli matrices.
Then, both $\pm (z_1, z_2)$ map onto the same $n_{1,2a}$, and any vortex configurations which winds from $(z_1, z_2)$ to $- (z_1, z_2)$ is a topologically stable vison associated with $\pi_1 (S_3/\mathbb{Z}_2) = \mathbb{Z}_2$. 
Choosing $z_1 = i e^{-i \phi/2}$, $z_2 = 0$,
we obtain $n_{1a} = (\cos \phi, \sin \phi , 0)$ and $n_{2a} = (-\sin \phi, \cos \phi, 0)$, which is the vortex configuration considered in Section~\ref{sec:vison}. Moreover, choosing $z_1 = e^{i(\phi/2+\pi/2)}/\sqrt{2}$, $z_2  = e^{-i\phi/2}/\sqrt{2}$,
we obtain $n_{1a} = (\cos \phi, -\sin \phi , 0)$ and $n_{2a} = (0, 0, 1)$, which corresponds to the {\it out-of-plane} vortex configuration of Eq. (\ref{H2outplane}), after $\phi\to-\phi$.

\section{$\epsilon$ expansion}
\label{app:epsilon}
We consider the Lagrangian (\ref{LHcounter}) with Higgs potential as defined in Eq. (\ref{VH1}), except here it is convenient to re-scaled the  gauge field $A_{a,\mu}\mapsto g A_{a,\mu}$. 
Our task is to compute the counter-terms $Z_1, Z_2, Z_3, Z_{v_0}, Z_{v_1}, Z_{s}$  (the others are not necessary for the present work) in an $\epsilon$-expansion and obtain the corresponding $\beta$-functions. 

\subsection{Vertices}
\begin{figure}[h]
 {\includegraphics[width=1.0\textwidth,clip]{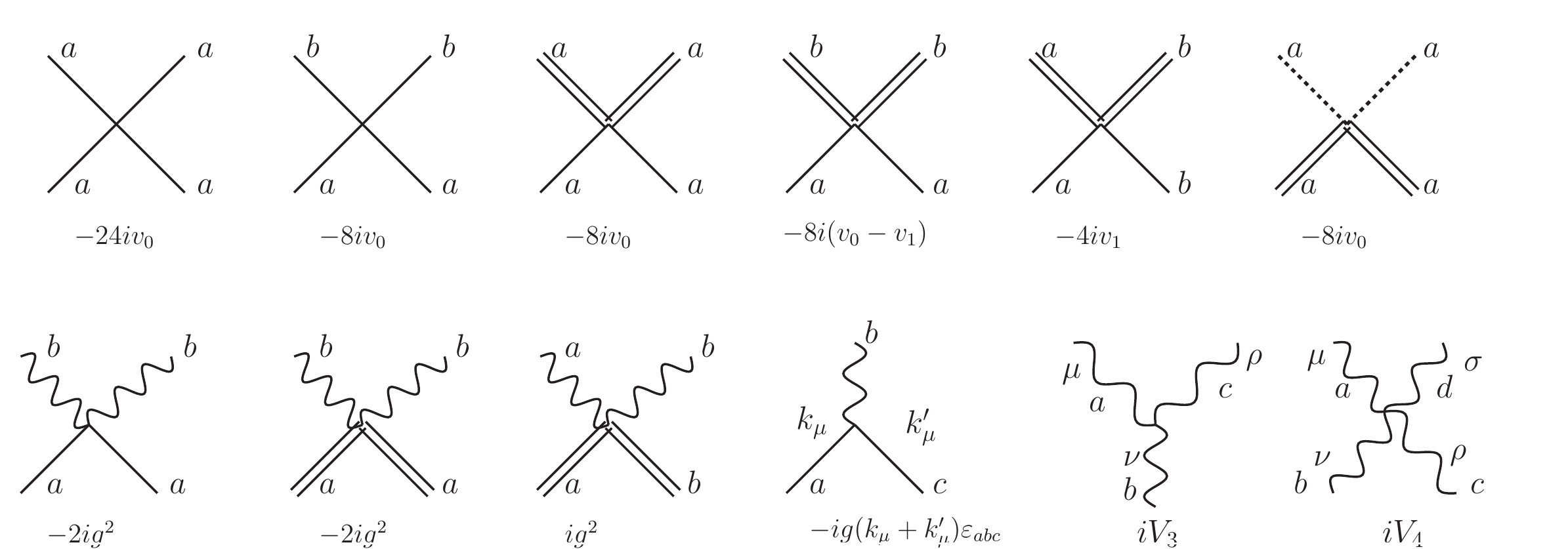}}
 \begin{picture}(0,0) 
\put(-255,110){\text{(a)}}
\put(-255,12){\text{(b)}}  
\put(90,12){\text{(c)}} 
\end{picture}
 \caption{A selection of all possible vertices in the (a) Pure Higgs sector, (b) Higgs-gauge sector, and (c) Pure gauge sector. Single, double and dashed lines refer to $h_a,g_a,H_{al}$ (for $l>2$), while curvy lines refer to the gauge field $A_{a\mu}$. Below each vertex is the associated symmetry factor. The labels $a,b,c$ are indices of the adjoint gauge group. It is assumed that $a\neq b\neq c$ in the above diagrams. Renormalization of $v_0$ and $v_1$ follows from all possible loop insertions built from the above vertices. The pure gauge sector, i.e. ${\cal L}_{YM} + {\cal L}_{\xi} + {\cal L}_{C}$ is renormalized as per textbooks, see e.g. \cite{Peskin} for $V_3$ and $V_4$.}
\label{fig:vertices}
\end{figure}
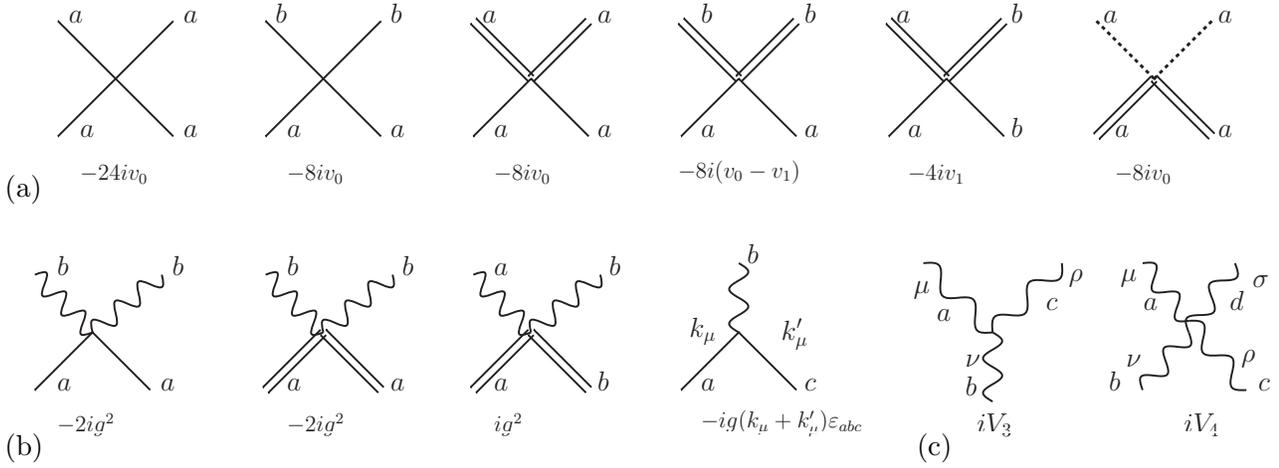

For the purposes of renormalization and diagrammatics, it is convenient to separate two flavour components of $H_{al}$, such that $H_{a1}=h_a$, $H_{a2}=g_a$ and the remaining stay as $H_{al}$ for $l>2$. The pure gauge sector ${\cal L}_{YM} + {\cal L}_{\xi} + {\cal L}_{C}$ is renormalized by standard procedure, and we will at times borrow results from \cite{Peskin}. Figure \ref{fig:vertices} shows the vertices -- and their symmetry factors -- necessary to compute the counterterms. 

\subsection{Table of loop integrals}
For convenience we list and evaluate all necessary integrals. Since we renormalise using the $\overline{MS}$ scheme, it is only necessary to keep the $O \big(\epsilon^{-1}\big)$ terms. All integrals assume an Euclidean metric, and we work in Feynman gauge ($\xi=1$), such that $P^{\mu\nu}(l)=g^{\mu\nu}-(1-\xi^{-1})l^\mu l^\nu/l^2=g^{\mu\nu}$. Let us introduce the following,
\begin{align}
\label{integrals}
\notag I_0&\equiv\int \frac{d^d l}{(2\pi)^d}\frac{1}{(l^2+s)((l+k)^2+s)}= \frac{1}{\epsilon}\frac{1}{8\pi^2},\\
\notag I_1&\equiv\int \frac{d^d l}{(2\pi)^d}\frac{1}{l^2+s}=-s I_0,\\
\notag I_2&\equiv\int \frac{d^d l}{(2\pi)^d}\frac{g^{\mu\nu}(l+2k)^\mu(l+2k)^\nu}{l^2((l+k)^2+s)}=(2k^2-s)I_0\\
\notag I_3&\equiv\int \frac{d^d l}{(2\pi)^d}\frac{(2l+k)^\mu(2l+k)^\nu}{(l^2+s)((l+k)^2+s)} -\int \frac{d^d l}{(2\pi)^d}\frac{2g^{\mu\nu}}{l^2+s}=\frac{1}{3}(k^\mu k^\nu-k^2g^{\mu\nu})I_0\\
\notag I_4&\equiv\int \frac{d^d l}{(2\pi)^d}\frac{g^{\mu\sigma}P^{\sigma\rho}g^{\rho\lambda}P^{\lambda\mu}}{(l^2+\tilde{s})^2}=d I_0,\\
\notag I_5&\equiv\int \frac{d^d l}{(2\pi)^d}\frac{l^\mu l^\sigma P^{\sigma\rho}g^{\rho\lambda}P^{\lambda\mu}}{(l^2+s)l^4}=I_0,\\
\notag I_6&\equiv\int \frac{d^d l}{(2\pi)^d}\frac{l^\mu l^\sigma l^\rho l^\lambda P^{\sigma\rho}P^{\lambda\mu}}{(l^2+s)^2l^4}=I_0,\\
I_7&\equiv\int \frac{d^d l}{(2\pi)^d}\frac{l^\mu l^\nu P^{\mu\nu}}{(l^2+s)^2 l^2}=I_0.
\end{align}
A fictitious mass term, $\tilde{s}$, is introduced into $I_4$ in order to perform the integration -- $\tilde{s}$ does not appear in the final result. $I_3$ arises from diagrams with two gauge field propagators, such as Figure \ref{fig:fourpoint}(a)(iv) and (c)(i). Integrals $I_2$ and $I_3$ require several intermediate steps, one should consult \cite{Peskin} for such details.

\subsection{Two-point functions: Counterterms $Z_s$, $Z_2$ and $Z_3$}
\begin{figure}[h]
 {\includegraphics[width=0.8\textwidth,clip]{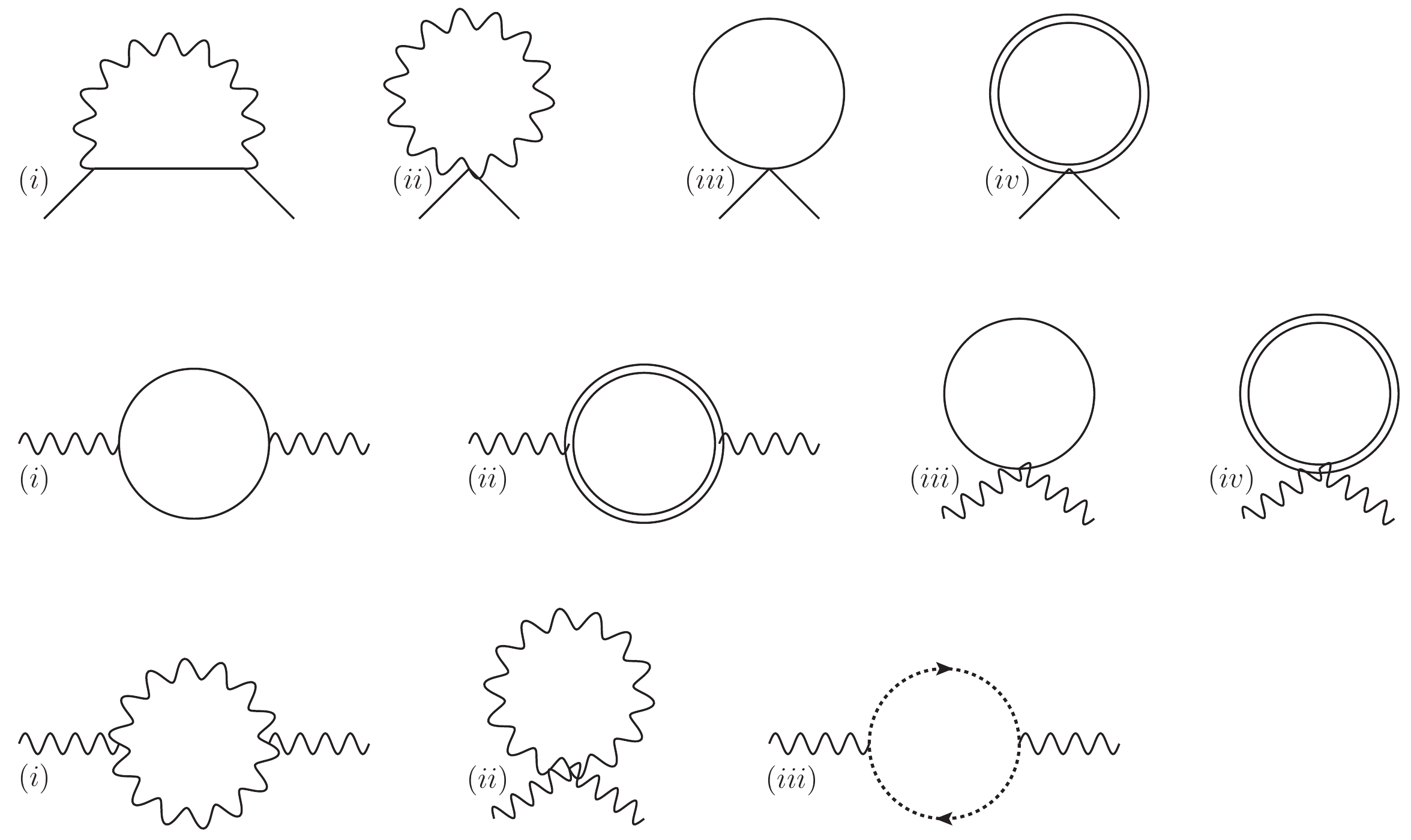}}
   \begin{picture}(0,0) 
\put(-450,165){\text{(a)}}
\put(-450,90){\text{(b)}}  
\put(-450,14){\text{(c)}} 
\end{picture}
 \caption{The one-loop diagrams renormalizing (a) the Higgs mass and field, $Z_s$ and $Z_2$, (b) and (c) the gauge fields, $Z_3$. (b) Shows the Higgs-gauge sector, while (c) shows the pure gauge (and ghost) sector. }
\label{fig:twopoint}
\end{figure}
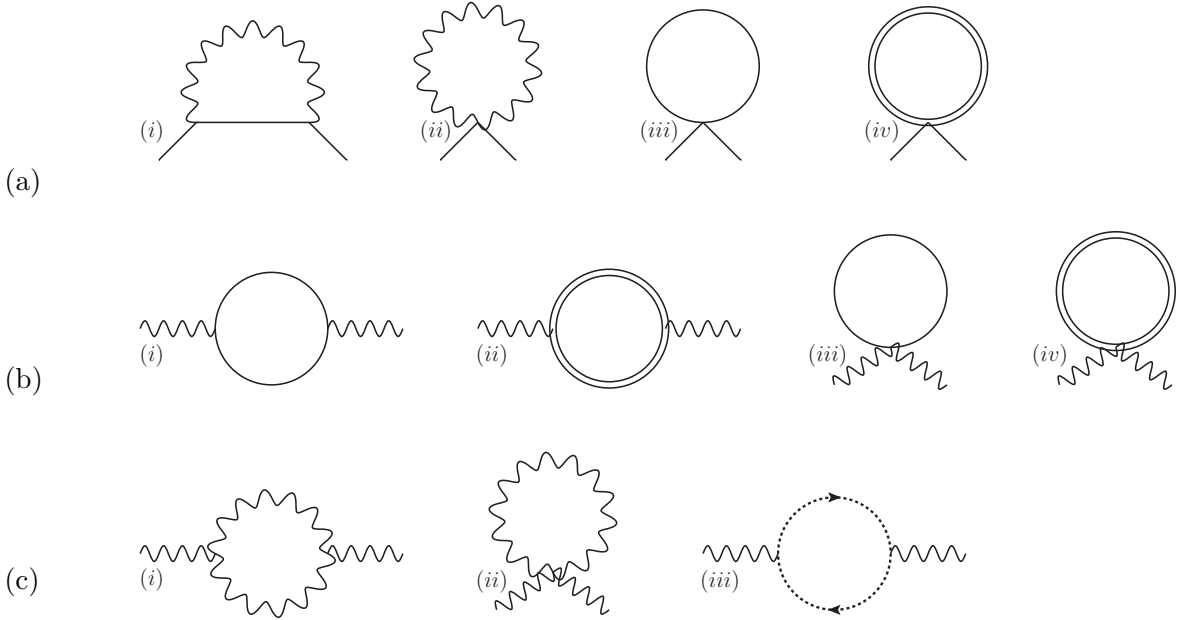
We begin with the two-point functions of Figure \ref{fig:twopoint}. From these diagrams we can compute $Z_s$, $Z_2$ and $Z_3$. Consider first $Z_s$ and $Z_2$, which are computed from Figure \ref{fig:twopoint}a. Fig. \ref{fig:twopoint}a(ii) gives exactly zero, and we set the external momentum of Fig. \ref{fig:twopoint}a(i) to be $k_\mu$, the self-energies are,
\begin{align*}
&i\Pi_{a(i)}+i\Pi_{a(iii)}+i\Pi_{a(iv)}\\
&=-i(Z_2-1)k^2 -i(Z_s-1)s +2g^2iI_2-\left(12v_0+8v_0+4v_0(N_h-1)+8(v_0-v_1)(N_h-1)\right)iI_1\\
&=-i(Z_2-1)k^2 -i(Z_s-1)s + 2g^2(2k^2-s)iI_0+s\left(4v_0(3N_h+2)-8v_1(N_h-1)\right) iI_0.
\end{align*}
The counterterms immediately follow (setting $g^2=\alpha$)
\begin{align}
\label{Z_s}
Z_s&=1+\left(4v_0(3N_h+2)-8v_1(N_h-1) - 2\alpha\right)I_0,\\
\label{Z_2}
Z_2&=1+4\alpha I_0.
\end{align}
$Z_s,Z_2$ are gauge dependent, but the difference $Z_s-Z_2$ is physical and must be gauge independent. To check our computations we have also calculated $Z_s-Z_2$ in Lorenz gauge ($\xi=\infty$) and found it to be invariant (details not shown here).

Consider now $Z_3$, which we write as $Z_3=Z_3^0+\delta Z_3$. The contribution $Z_3^0$ is due to the the pure gauge-ghost sector, shown in Figure \ref{fig:twopoint}(c), for which the result can be extracted from standard textbooks \cite{Peskin}, $Z_3^0=1+10g^2I_0/3$. The contribution $\delta Z_3$ is due to the gauge-Higgs sector, and is shown by the diagrams of Figure \ref{fig:twopoint}(b). Such are diagrams are particularly simple to evaluate by making use of the integral $I_3$, 
\begin{align*}
i\Pi_{b(i)}+i\Pi_{b(ii)}+i\Pi_{b(iii)}+i\Pi_{b(iv)}&=-i\delta Z_3(k^2g^{\mu\nu}-k^\mu k^\nu)+N_h g^2iI_3.
\end{align*}
This gives $\delta Z_3=N_hg^2I_0/3$. Finally, setting $g^2=\alpha$, we obtain
\begin{align}
\label{Z_3}
Z_3=Z_3^0+\delta Z_3&=1+\alpha\left(\frac{10}{3}-\frac{N_h}{3}\right)I_0.
\end{align}
Counterterm $Z_3$ is independent of gauge fixing.

\subsection{Four-point functions: Counterterms $Z_{v_0}$ and $Z_{v_1}$}
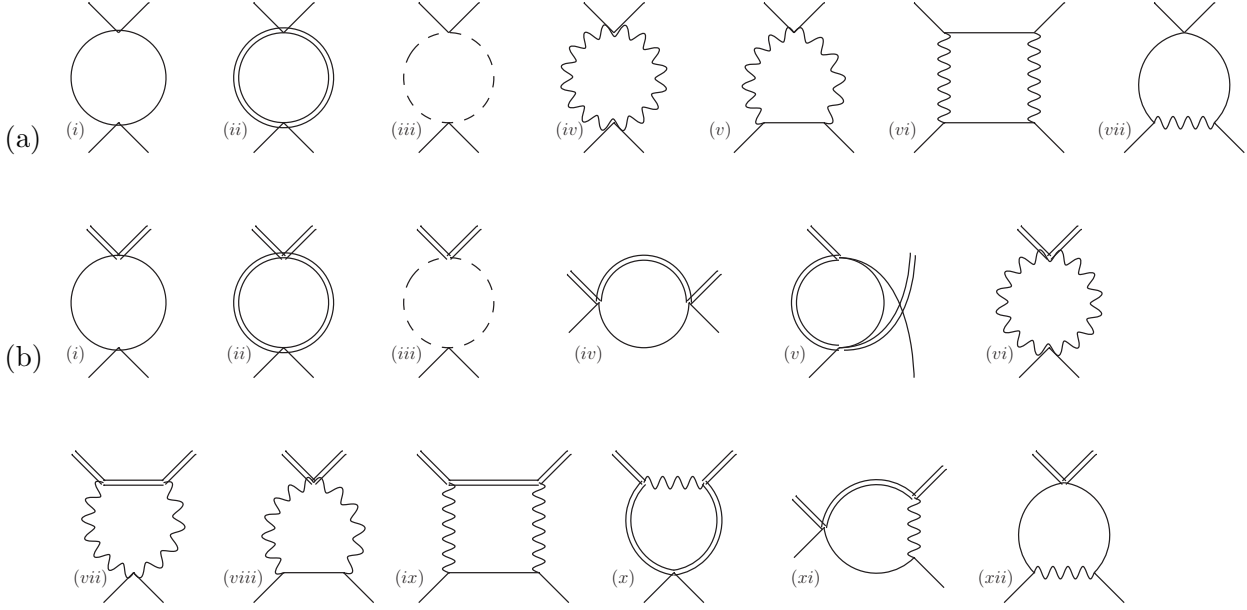
\begin{figure}[h]
 {\includegraphics[width=0.9\textwidth,clip]{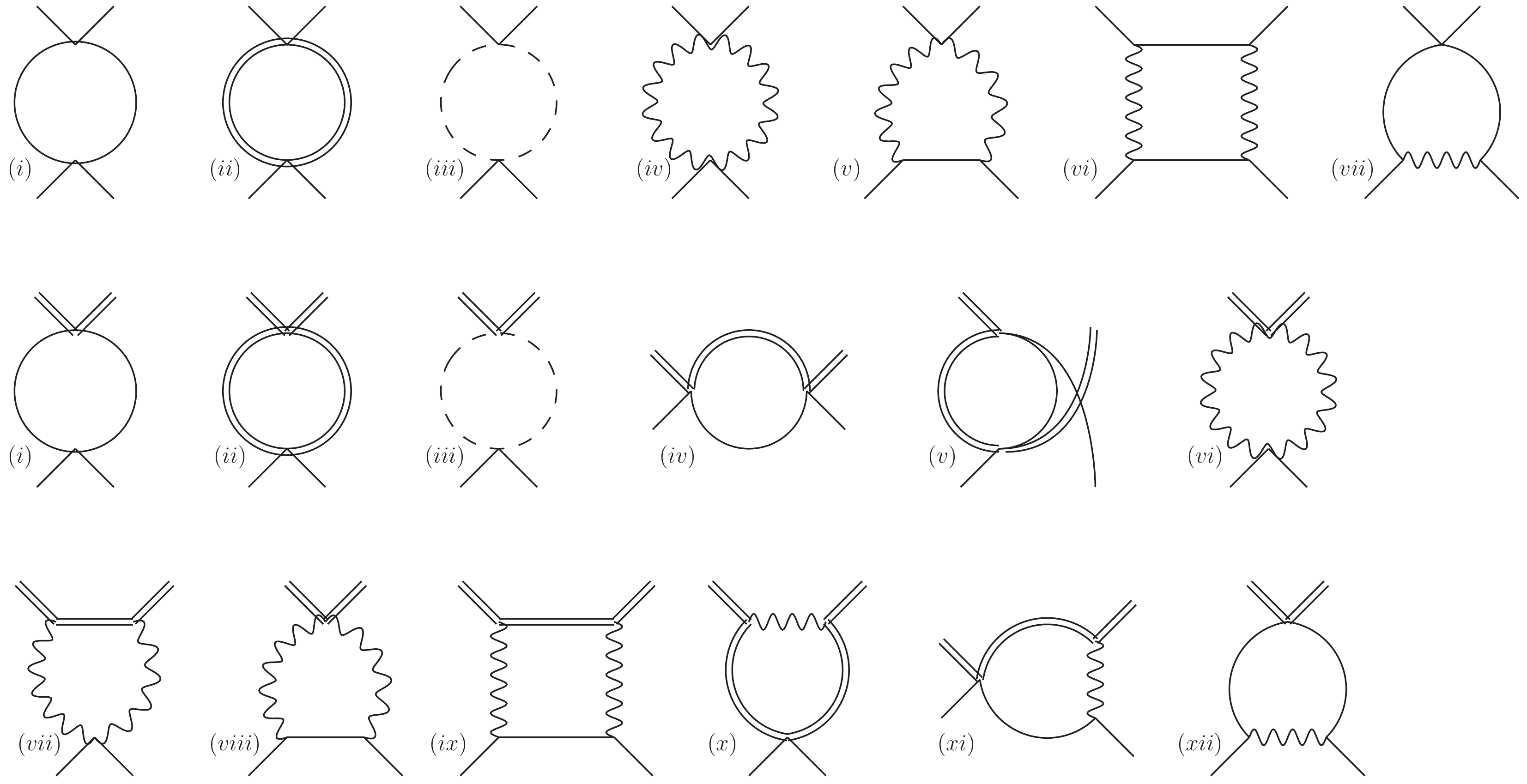}}
  \begin{picture}(0,0) 
\put(-475,175){\text{(a)}}
\put(-475,92){\text{(b)}}  
\end{picture}
 \caption{The one-loop diagrams renormalizing the Higgs quartic couplings $v_0$ and $v_1$ (i.e. the counter terms $Z_{v_0}$ and $Z_{v_1}$). }
\label{fig:fourpoint}
\end{figure}
To find $Z_{v_0}$, we sum the diagrams of Figure \ref{fig:fourpoint}(a) and set all four external legs to be the same, such that the vertex is $H_{11}H_{11}H_{11}H_{11}=h_1h_1h_1h_1$. Following the vertex rules of Figure \ref{fig:vertices}, and counting all possible permutations we obtain
\begin{align*}
iV_{v_0}&=-24i v_0 Z_{v_0}+\left(\frac{1}{2}(24v_0)^2+(8v_0)^2+\frac{1}{2}(8v_0)^2(N_h-1)+8(v_0-v_1)^2(N_h-1)\right)iI_0\\
&\hspace{2.1cm}
 + 12dg^4iI_4-24g^4iI_5+12g^4iI_6 -96g^2v_0 iI_7.
\end{align*}
Substituting the integrals from Eq. (\ref{integrals}), setting $d=4$ and $g^2=\alpha$, we find
\begin{align}
\label{Z_v0}
Z_{v_0}&=1+I_0\left(4(3N_h+8)v_0-16(N_h-1)v_1+8(N_h-1)\frac{v_1^2}{v_0}+\frac{3\alpha^2}{2v_0}-4\alpha\right).
\end{align}
Similarly, to find $Z_{v_1}$ we sum the diagrams of Figure \ref{fig:fourpoint}(b), but now set the external legs to be $H_{11}H_{21}H_{12}H_{22}=h_1h_2g_1g_2$. Following the vertex rules of Figure \ref{fig:vertices}, we find
\begin{align*}
iV_{v_1}&=-4i v_1 Z_{v_1}+(4v_1)^2\left(1+(N_h-2)-4\right)iI_0 - 32v_0 v_1\left(1+1+1+1+1+1\right)iI_0 \\
&\hspace{1.9cm}+ dg^4iI_4-(1+1)g^4iI_5+g^4iI_6 -4(2+1+1)g^2v_1iI_7.
\end{align*}
Upon using Eq. (\ref{integrals}), setting $d=4$ and $g^2=\alpha$, we find
\begin{align}
\label{Z_v1}
Z_{v_1}&=1+I_0\left(4(N_h-5)v_1+48v_0+\frac{3\alpha^2}{4v_1}-4\alpha\right).
\end{align}

\subsection{Three-point function: Counterterm $Z_1$}
The counterterm $Z_1$ can be computed directly by considering the loop corrections to the three-point vertex $Z_1\varepsilon_{abc}A_{a\mu}H_{bm}\partial_\mu H_{cm}$. Alternatively, we note that the difference $Z_1-Z_2$ is independent of the matter content and can be determined from the pure gauge-ghost sector, i.e. from
\begin{align}
Z_1-Z_2=Z_{1c}-Z_{2c}=-\frac{\alpha }{4 \pi ^2 \epsilon},
\end{align} 
where the second equality was extracted from Peskin \& Schroeder \cite{Peskin}. This circumvents the need to directly compute loop corrections to the three-point vertex, instead we use our result for $Z_2$ (\ref{Z_2}), to find
\begin{align}
\label{Z_1}
Z_1&=1+\frac{\alpha }{4 \pi ^2 \epsilon }.
\end{align}

\bibliography{GG}

\end{document}